\documentclass[a4paper,prx,reprint,superscriptaddress,twocolumns,notitlepage,floatfix]{revtex4-2}
\usepackage[utf8]{inputenc}
\usepackage[T1]{fontenc}

\usepackage{amsmath}
\usepackage{amssymb}
\usepackage{stmaryrd}
\usepackage{fourier-orns}
\usepackage{commath}
\usepackage{mathrsfs}
\usepackage{braket}
\usepackage{xcolor}
\usepackage{bm}
\usepackage{csquotes}
\usepackage{graphicx}
\usepackage{booktabs}
\usepackage{enumitem}
\usepackage{siunitx}
\usepackage{microtype}

\usepackage{hyperref}
\usepackage{bookmark}

\definecolor{tab_blue}{HTML}{1F77B4}
\definecolor{tab_orange}{HTML}{FF7F0E}
\definecolor{tab_green}{HTML}{2CA02C}
\definecolor{tab_red}{HTML}{D62728}
\definecolor{tab_purple}{HTML}{9467BD}
\definecolor{tab_brown}{HTML}{8C564B}
\definecolor{tab_pink}{HTML}{E377C2}
\definecolor{tab_gray}{HTML}{7F7F7F}
\definecolor{tab_olive}{HTML}{BCBD22}
\definecolor{tab_cyan}{HTML}{17BECF}

\hypersetup{breaklinks=true,
  colorlinks=true,
  linkcolor=tab_blue,
  filecolor=tab_blue,
  urlcolor=tab_blue,
  citecolor=tab_blue,
}

\def\dd{\text{d}}
\def\ii{\text{i}}
\def\ee{\text{e}}

\def\ZZ{\mathbb{Z}}

\def\vec#1{\bm{#1}}

\def\div{\text{div}}

\def\kB{k_{\text{B}}}

\let\strong\textbf

\definecolor{tab_blue}{HTML}{1F77B4}
\definecolor{tab_orange}{HTML}{FF7F0E}
\definecolor{tab_green}{HTML}{2CA02C}
\definecolor{tab_red}{HTML}{D62728}
\definecolor{tab_purple}{HTML}{9467BD}
\definecolor{tab_brown}{HTML}{8C564B}
\definecolor{tab_pink}{HTML}{E377C2}
\definecolor{tab_gray}{HTML}{7F7F7F}
\definecolor{tab_olive}{HTML}{BCBD22}
\definecolor{tab_cyan}{HTML}{17BECF}

\def\header#1{\emph{#1.}} 
\makeatletter\begin{document}

\title{The odd ideal gas: Hall viscosity and thermal conductivity from non-Hermitian kinetic theory}
\author{Michel Fruchart}
\email{fruchart@uchicago.edu}
\affiliation{James Franck Institute, University of Chicago, Chicago, Illinois 60637, USA}
\affiliation{Department of Physics, University of Chicago, Chicago, IL 60637, USA}

\author{Ming Han}
\affiliation{James Franck Institute, University of Chicago, Chicago, Illinois 60637, USA}
\affiliation{Pritzker School of Molecular Engineering, University of Chicago, Chicago, Illinois 60637, USA}

\author{Colin Scheibner}
\affiliation{James Franck Institute, University of Chicago, Chicago, Illinois 60637, USA}
\affiliation{Department of Physics, University of Chicago, Chicago, IL 60637, USA}

\author{Vincenzo Vitelli}
\email{vitelli@uchicago.edu}
\affiliation{James Franck Institute, University of Chicago, Chicago, Illinois 60637, USA}
\affiliation{Department of Physics, University of Chicago, Chicago, IL 60637, USA}
\affiliation{Kadanoff Center for Theoretical Physics, University of Chicago, Chicago, IL 60637, USA} 

\begin{abstract}
The flow of momentum and energy in a fluid is typically associated with dissipative transport coefficients: viscosity and thermal conductivity. 
Fluids that break certain symmetries such as mirror symmetry and time-reversal invariance can display non-dissipative transport coefficients called odd (or Hall) viscosities and thermal conductivities.
The goal of this paper is to elucidate the microscopic origin of these dissipationless transport coefficients using kinetic theory. 
We show that odd viscosity and odd thermal conductivity arise when the linearized collision operator is not Hermitian. 
This symmetry breaking occurs when collisions are chiral, i.e. not mirror symmetric.
To capture this feature in a minimalistic way, we introduce a modified relaxation time approximation in which the distribution function is rotated by an angle characterizing the average chirality of the collisions. In the limit of an infinitesimal rotation, the effect of the parity-violating collisions can be described as an emergent effective magnetic field.  \end{abstract}

\maketitle

 \header{Introduction} Hydrodynamics describes a wide range of phenomena across scales, from the motion of cells in biological tissues~\cite{Marchetti2013,Doostmohammadi2018,Prost2015,Shankar2020,Colen2021} to the collective motion of electrons in solid-state systems~\cite{Lucas2018,Cook2019,Sulpizio2019,Ku2020} and from the dynamics of quark-gluon plasma~\cite{Policastro2001,Kovtun2005} to cosmology~\cite{Cen1992,Andersson2021} 
-- or more modestly, the twirl of coffee in an espresso cup.
Hydrodynamic theories are effective theories built upon a separation of time scales between fast degrees of freedom and slowly relaxing hydrodynamic variables arising, for instance, from the existence of conserved quantities~\cite{Forster1975,Chaikin2000,Schafer2014,LandauVI,Dubovsky2012,Haehl2016,Liu2018}. 
As the total value of conserved quantities is fixed, the dynamics of the corresponding fields at large length scales is slow enough to warrant a hydrodynamic description.
In a simple fluid such as water, there are only three independent conserved quantities: mass, linear momentum, and energy.
Their conservation is expressed by the Navier-Stokes equations
\begin{subequations}
\label{NS}
\begin{align}
	D_t \rho &= - \rho \nabla \vec{v} \\
	\rho \, D_t \vec{v} &= \nabla \cdot \sigma^{\text{tot}} + \vec{f} \\
	n c_{\text{v}} \, D_t T &= - \nabla \cdot \vec{Q} - \sigma_{i j}^{\text{tot}} \partial_j v_i
\end{align}
\end{subequations}
that describe the evolution in time of the mass density $\rho(t, \vec{r})$, the velocity field $\vec{v}(t, \vec{r})$ and the temperature field $T(t, \vec{r})$.
Here, $D_t = \partial_t + \vec{v} \cdot \nabla$ is the convective derivative, $n = \rho / m$ the number density of particles with mass $m$, $\sigma^{\text{tot}}$ the stress tensor, $\vec{f}$ the density of body forces, $\vec{Q}$ the heat flux, and $c_{\text{v}}$ the specific heat at constant volume.

\begin{figure}
    \includegraphics{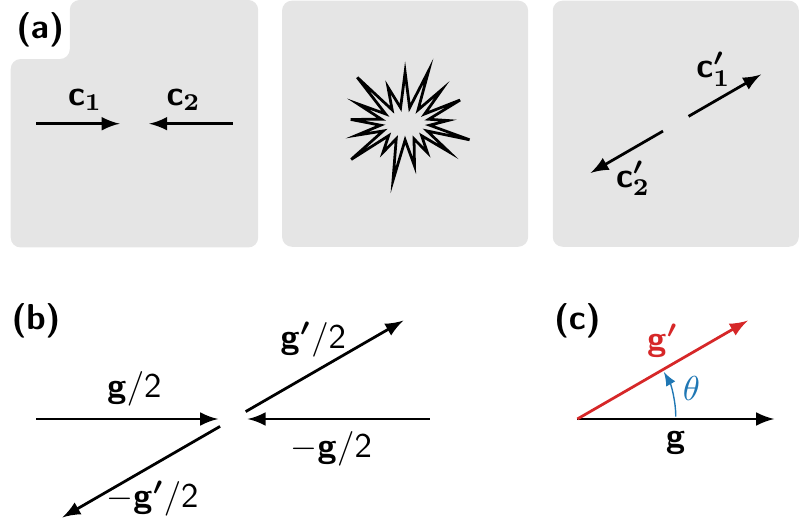}
    \caption{\label{figure_collision}
    \textbf{Two-body collision.}
    (a) Two particles (1) and (2) collide. Their incoming velocities are $\vec{c_1}$ and $\vec{c_2}$, and their outgoing velocities are $\vec{c_1'}$ and $\vec{c_2'}$.
    (b) In the center of mass reference frame, the incoming velocities become $\pm \vec{g}/2$, 
	where $\vec{g} = \vec{c_2} - \vec{c_1}$.
	When momentum is conserved, the outgoing velocities are $\pm \vec{g'}/2$ where $\vec{g'} = \vec{c_2'} - \vec{c_1'}$ (the particles have identical masses).
    By rotating the system of coordinate (this is only possible in an isotropic system), we can set $\vec{g}$ to be along the x axis.
    (c) The angle between $\vec{g}$ and $\vec{g'}$ (in the plane defined by these vectors) is called $\theta$. As the particles are identical, it is always possible to choose $-\pi < \theta \leq \pi$ by exchanging $\vec{c_1'} \leftrightarrow \vec{c_2'}$.
    Collisions are parity-violating when there is, on average, an imbalance between collisions with $\pm \theta$.
    }
\end{figure}

In Newtonian fluids, the stress tensor is
\begin{equation}
	\sigma_{i j}^{\text{tot}} = - p \, \delta_{i j} + \eta_{i j k \ell} \, \dot{e}_{k \ell}
\end{equation}
in which $p$ is the hydrostatic pressure, $\dot{e}_{k \ell} = \partial_\ell v_k$ are velocity gradients, and $\eta_{i j k \ell}$ is the viscosity tensor.
Similarly, the thermal conductivity tensor relates the heat flux to temperature gradients through Fourier's law
\begin{equation}
	Q_i = - \lambda_{i j} \partial_j T.
\end{equation}
Viscosity and thermal conductivity are usually associated with dissipation.  
However, only their symmetric parts $\eta^{\text{sym}}_{i j k \ell} = [\eta_{i j k \ell} + \eta_{k \ell i j}]/2$ and 
$\lambda^{\text{sym}}_{i j} = [\lambda_{i j} + \lambda_{j i}]/2$ are dissipative.
The antisymmetric parts $\eta^{\text{anti}}_{i j k \ell} = [\eta_{i j k \ell} - \eta_{k \ell i j}]/2$ and $\lambda^{\text{anti}}_{i j} = [\lambda_{i j} - \lambda_{j i}]/2$ contain, instead, non-dissipative transport coefficients.
In usual fluids such as water or coffee, these non-dissipative transport coefficients vanish because of the presence of certain symmetries, such as time-reversal invariance and mirror reflection.
When these symmetries are broken, non-dissipative transverse viscosities (variously called odd viscosity, Hall viscosity, or gyroviscosity) and heat conductivities (called Hall thermal conductivity or Righi-Leduc coefficient) appear. 
Examples of such situations include magnetized plasma and fluids under rotation~\cite{Kaufman1960,ChapmanCowling,Braginskii1958,Braginskii1965}, neutral polyatomic gases under magnetic field~\cite{Korving1966,Korving1967,Kagan1962,Kagan1962b,Kagan1967,Moraal1969,McCourt1969,Hulsman1970,Knaap1967,McCourt1967,Levi1968,Waldmann1958b,Beenakker1970,Mccourt1990,Hess2003} (under the name of Senftleben—Beenakker effect), quantum fluids such as liquid helium~\cite{Furusawa2021,Fujii2018}, vortex matter~\cite{Wiegmann2014,Bogatskiy2019}, or electron fluids in magnetized solids~\cite{Tokatly2007,Alekseev2016,Scaffidi2017,Berdyugin2019},
and active matter~\cite{vanZuiden2016,Tsai2005,Banerjee2017,Soni2019,Han2020}. 
Experimental observations of odd viscosities have been reported in polyatomic gases~\cite{Beenakker1970,Mccourt1990}, active rotating colloids~\cite{Soni2019,Bililign2021}, graphene~\cite{Berdyugin2019}, and it has been suggested that they can occur in neural progenitor cells~\cite{Yamauchi2020}. 
These lines of inquiry have led to a surge of interest in the topic \cite{Hosaka2021,Reichhardt2021,Kogan2016,Khain2022,Hosaka2021b,Yang2021,Hargus2020,Epstein2020,Lapa2014,Jensen2012,Kaminski2014,Tauber2019,Souslov2019,Bao2021,Kirkinis2019,Ganeshan2017,Monteiro2021,Abanov2018,Bogatskiy2019,Abanov2019,Landsteiner2016,Lucas2014,Holder2019,Markovich2021,Barabanov2015,Maksimov2017}.
In magnetized plasma and fluids under rotation, an external field acts on the trajectories of individual particles through the Lorentz or Coriolis force. In these systems, odd viscosity is primarily a single-particle effect persisting even in the collisionless limit~\cite{ChapmanCowling,Braginskii1958,Braginskii1965,Kaufman1960,Nakagawa1956}, that can be captured within a Hamiltonian structure~\cite{Morrison2014,Morrison1984,Lingam2020,Markovich2021,Monteiro2021b}. The physical mechanism behind odd viscosity can be traced down to the fact that the particles follow circular orbits.

\medskip

Our goal is to construct a minimal kinetic theory describing how odd viscosity and thermal conductivity can arise from collisions. 
Viscosity is ultimately about linear momentum transport. Hence, it should be possible to understand its origin in a gas of point particles where all internal degrees of freedom are neglected.
As we shall see, we will also obtain an odd thermal conductivity in this way.
In order to deviate as little as possible from an ideal gas, we will assume that both energy and linear momentum are conserved, and we will ignore correlations between the particles by using the Boltzmann equation for a dilute gas.
These assumptions are not crucial, and are indeed violated in typical active matter systems.
For concreteness, we focus on isotropic gases in dimension $d=2$. Three-dimensional systems can be treated in a similar way.

\begin{figure*}
    \includegraphics[width=\textwidth]{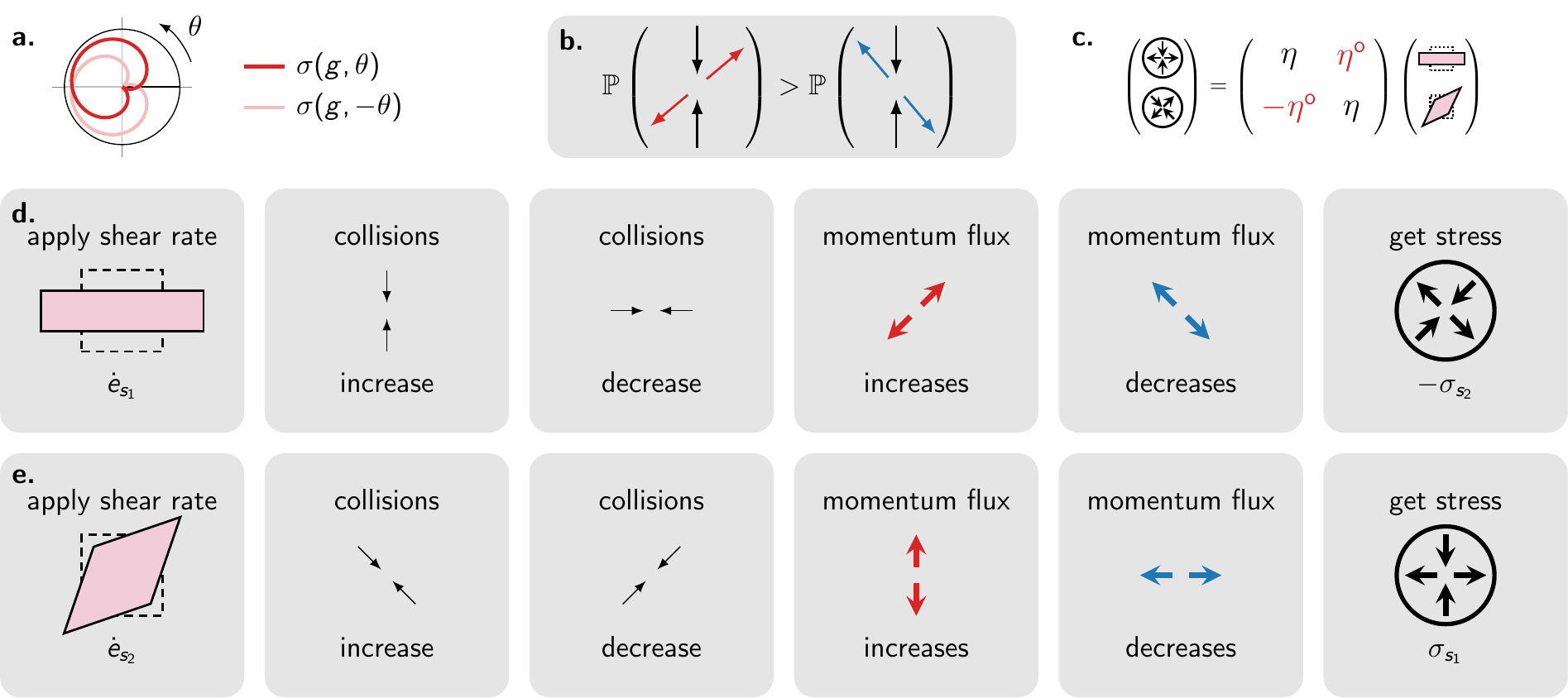}
    \caption{\label{figure_mechanism_odd_viscosity}
    \textbf{Schematic mechanism leading to odd viscosity from parity-violating collisions.}
    (a) Consider parity-violating collisions: the cross-section $\sigma(g, \theta) \neq \sigma(g, -\theta)$ (see Fig.~\ref{figure_collision} for definitions), as pictorially represented in (b). Note that the collisions in panel (b) can be globally rotated.
    (c) Shear viscosities relate shear stress and rate of shear deformation. In a 2D isotropic system, only a normal shear viscosity $\eta$ and an odd shear viscosity $\eta^{\text{o}}$ are possible.
    (d) In line d, the fluid is subjected to a constant shear rate $\dot{e}_{s_1}$ (defined below). It undergoes a rate of vertical compression combined with a rate of horizontal expansion.  As a consequence, there are more vertical collisions and less horizontal collisions, compared to the fluid at rest.
    The change in the momentum flux (compared to the fluid at rest) is qualitatively obtained by looking at where particles go after collision.
    As the collisions are asymmetric (panel b), there is an increase of the momentum flux at \ang{45} (in red) and a decrease of the momentum flux at \ang{-45} (in blue).
    Combining these, we find the resulting viscous stress $-\sigma_{s_2}$ in the last panel.
    Note that the momentum flux tensor (pressure tensor) $P_{i j}$ is the opposite of the stress tensor $\sigma_{i j} = - P_{i j}$.
    There is also a change in the horizontal and vertical momentum fluxes, not pictured there. It corresponds to normal shear viscosity (see SI Fig.~\ref{figure_mechanism_normal_viscosity}). 
(e) We follow the same reasoning as in (d) when the the fluid is subjected to a constant shear rate $\dot{e}_{s_2}$ (obtained by rotating $\dot{e}_{s_1}$ by \ang{45}, and orthogonal to it in the sense of linear algebra). The result is a viscous stress $\sigma_{s_1}$.
    Hence, we have found that $\sigma_{s_1} \propto \dot{e}_{s_2}$ and $\sigma_{s_2} \propto - \dot{e}_{s_1}$ (ignoring normal shear viscosity), which is indeed the effect of odd viscosity, represented by the pink coefficient $\eta^{\text{o}}$ in panel c.
 	We have defined $\dot{e}_{s_1} = \dot{e}_{x x} - \dot{e}_{y y}$, $\dot{e}_{s_2} = \dot{e}_{x y} + \dot{e}_{y x}$ and $\sigma_{s_1} = (\sigma_{x x} - \sigma_{y y})/2$, $\sigma_{s_2} = (\sigma_{x y} + \sigma_{y x})/2$, in which $\dot{e}_{i j} = \partial_j e_i$. 
    }
\end{figure*}

\header{Parity-violating collisions} From the point of view of symmetries, the only requirement for collision-generated odd viscosity is that collisions violate parity (Figs.~\ref{figure_collision} and \ref{figure_mechanism_odd_viscosity}a-c).
The occurrence of odd viscosity from this asymmetry can be understood pictorially: see Fig.~\ref{figure_mechanism_odd_viscosity}.
We subject the fluid to a constant shear rate in which it undergoes vertical compression combined with horizontal expansion (panel d).
As a consequence, there are more vertical collisions and less horizontal collisions compared to the fluid at rest.
The change in the momentum flux compared to the fluid at rest is qualitatively obtained by looking at where particles go after collision.
As the collisions are asymmetric (panel b, rotated as needed), there is an increase of the momentum flux at \ang{45} and a decrease of the momentum flux at \ang{-45}.
The corresponding viscous stress is the opposite of this momentum flux.
Repeating the same argument for a constant shear rate at \ang{45} (panel e), we find an increase in the vertical momentum flux and a decrease in the horizontal momentum flux. Note that, crucially, the relations between the resulting stresses and the strain rates we apply in (d) and (e) are antisymmetric (panel c): this is odd viscosity. 

Mathematically, the requirement that collisions violate parity is expressed as
\begin{equation}
	\label{asymmetry_cross_section}
	\sigma(g,\theta) \neq \sigma(g,-\theta)
\end{equation}
where $\sigma(g,\theta)$ is the differential scattering cross-section.
Here, $g$ is the relative momentum in the center of mass reference frame and $\theta$ is the angle between the incoming and outgoing momenta, see Fig.~\ref{figure_collision}. The asymmetry in Eq.~\eqref{asymmetry_cross_section} can be observed in Fig.~\ref{figure_mechanism_odd_viscosity}a. 

In our minimal model, parity-violating collisions originate from effective parity-violating  forces~\cite{Han2020}.
In actual physical systems, they arise from internal degrees of freedoms. 
This occurs in systems ranging from rotating colloids~\cite{Soni2019,Bililign2021} and rotating frictional grains~\cite{Tsai2005} to
starfish embryos~\cite{Tan2021}.
Realistic microscopic descriptions of magnetized neutral polyatomic gases taking into account internal degrees of freedom indeed capture the measured non-dissipative transport coefficients~\cite{Korving1966,Korving1967,Kagan1962,Kagan1962b,Kagan1967,Moraal1969,McCourt1969,Hulsman1970,Knaap1967,McCourt1967,Levi1968,Waldmann1958b,Hess2003,Beenakker1970,Mccourt1990}.

\header{Non-Hermitian kinetic theory} We now show how the intuitive explanations in Fig.~\ref{figure_mechanism_odd_viscosity} can be put on a firmer mathematical ground using kinetic theory~\cite{ChapmanCowling,Hirschfelder1964,Waldmann1958,Grad1958,Harris2004,Dorfman2021,Reif2009}.
Consider a dilute gas of point particles with masses $m$, that collide with each other according to a cross-section $\sigma(g,\theta)$.
The distribution function $f(t,\vec{r},\vec{c})$ gives the probability $f(t,\vec{r},\vec{c}) \dd^d \vec{r} \dd^d \vec{c}$ to find a particle in an infinitesimal volume centered at $(\vec{r}, \vec{c})$ in phase space at time $t$, where $\vec{r}$ is the position and $\vec{c}$ is the velocity.
At equilibrium, the distribution function is the Boltzmann distribution
$f^\circ(\vec{c}) = n \, (\beta m/2 \pi)^{d/2} \, \ee^{- \beta m c^2/2}$
in which $n$ is the number density, $\beta = (k_{\text{B}} T)^{-1}$ is the inverse temperature and $k_{\text{B}}$ is the Boltzmann constant.

The fate of perturbations $f = f^\circ (1 + \phi)$ on top of the equilibrium distribution is described by the linearized Boltzmann equation
\begin{equation}
	\label{LBE_main}
	\frac{\partial \phi}{\partial t} 
	+ c_i \frac{\partial \phi}{\partial r_i}
	+ b_i \frac{\partial \phi}{\partial c_i}
	= L \phi
\end{equation}
in which $m \vec{b}$ represents external forces, $b_i$ and $c_i$ are Cartesian components of $\vec{b}$ and $\vec{c}$, and
\begin{equation}
	\label{linearized_collision_operator_cons_main}
	L \phi = 
	\int 
	  g \sigma(g, \theta') f^{\circ}_2 \big[ \phi_1' + \phi_2' - \phi_1 - \phi_2 \big] 
	  \dd^d \vec{c_2} \, \dd^{d-1} \Omega'
\end{equation}
is the linearized collision operator, which captures the redistribution of probabilities following a collision.
Here, $\phi_1'$ is shorthand for $\phi(\vec{c_1'})$ (and similarly for the other terms, associated with the incoming and outgoing velocities represented in Fig.~\ref{figure_collision}). Velocities are constrained by conservation laws (see SI), and $\dd^{d-1} \Omega'$ is the differential solid angle associated with $\vec{g'} = \vec{c_2'} - \vec{c_1'}$. 
The equilibrium distribution allows us to define the scalar product 
\begin{equation}
	\label{scalar_product_main}
	(\chi, \phi) = \frac{1}{n} \int f^{\circ}(\vec{c_1}) \, \overline{\chi(\vec{c_1})} \, \phi(\vec{c_1}) \dd^d\vec{c_1}.
\end{equation}
on functions $\vec{c} \mapsto \phi(\vec{c})$ of the velocity $\vec{c}$.
A direct computation shows that the condition \eqref{asymmetry_cross_section} is satisfied when
\begin{equation}
	L \neq L^\dagger,
\end{equation}
namely when the linearized collision operator is not Hermitian with respect to the scalar product in Eq.~\eqref{scalar_product_main} (see Ref.~\cite{Ashida2020} for a review of non-Hermitian physics and Refs.~\cite{Lhuillier1982,AhSam1971,Resibois1970} for applications to kinetic theory).

\header{From non-Hermitian kinetic theory to parity-violating hydrodynamics}
To obtain hydrodynamic equations, we use a projection operator method~\cite{Zwanzig2001,Mori1965,Nakajima1958,Zwanzig1960,Resibois1970,Balescu1975,Ernst1970,Bixon1971,Hauge1970}. 
Hydrodynamic modes correspond to conserved quantities that span the nullspace of $L$.
Hydrodynamic equations of the form of Eq.~\eqref{NS} are obtained by projecting Eq.~\eqref{LBE_main} onto this nullspace (i.e. the long wavelength modes) under the assumption that the remaining non-hydrodynamic modes are stationary, see SI.

Inspection of the explicit form of the Navier-Stokes equations obtained in this way shows that the viscosity and thermal conductivity tensors of the dilute gas are
\begin{align}
	\label{viscosity_main}
 	\frac{\eta_{i j k \ell}}{\rho} &= - \frac{m}{k_{\text{B}} T} \, 
 	\left(c_i c_j, L^{-1} \, c_k c_\ell\right) \\
\intertext{and}
	\label{thermal_conducitivity_main}
	\lambda_{i j} &= - \frac{m^2 n}{4 \kB T^2} \left(c^2 \, c_i, L^{-1} c^2 \, c_j \right)
\end{align}
in which $c_i c_j$ are shorthands for functions $\vec{c} \to c_i c_j$ of the velocities of the particles, and $L^{-1}$ is restricted to non-hydrodynamic modes, i.e. the orthogonal complement of the nullspace of $L$ (SI).
Equations~\eqref{viscosity_main} and \eqref{thermal_conducitivity_main} can also be obtained using the Chapman-Enskog method~\cite{ChapmanCowling} (SI).

\header{Non-dissipative response coefficients and Onsager-Casimir relations} 
In standard kinetic theory, the linearized collision operator $L$ is Hermitian ($L^\dagger = L$, where $\dagger$ represents the transpose of the conjugate). Under this hypothesis, the viscosity tensor in Eq.~\eqref{viscosity_main} is symmetric:
\begin{equation}
	\label{symmetry_eta}
\begin{split}
	\eta_{i j k \ell} 
	&= - \frac{\rho \, m}{k_{\text{B}} T} \, \left({c_i c_j, L^{-1} c_k c_\ell}\right)
	\\
	&= - \frac{\rho \, m}{k_{\text{B}} T} \, \left({L^{-1} c_i c_j,  c_k c_\ell}\right)
	= \overline{\eta_{k \ell i j}}
	= \eta_{k \ell i j}
\end{split}
\end{equation}
as $\eta_{i j k \ell}$ is real-valued. Equation~\eqref{symmetry_eta} is known as Onsager reciprocity~\cite{DeGrootMazur,DeGroot1973}.
Conversely, when the linearized collision operator $L$ is not Hermitian, then it is possible that $\eta_{i j k \ell} \neq \eta_{k \ell i j}$.
A simple extension of Eq.~\eqref{symmetry_eta} yields the Onsager-Casimir relations $\eta_{i j k \ell}(B) = \eta_{k \ell i j}(-B)$ and $\lambda_{i j}(B) = \lambda_{j i}(-B)$ from the assumption that $L(B) = L^\dagger(-B)$, in which $B$ is an arbitrary parameter (see SI).

\begin{figure}
	\hspace{-0.5cm}\includegraphics{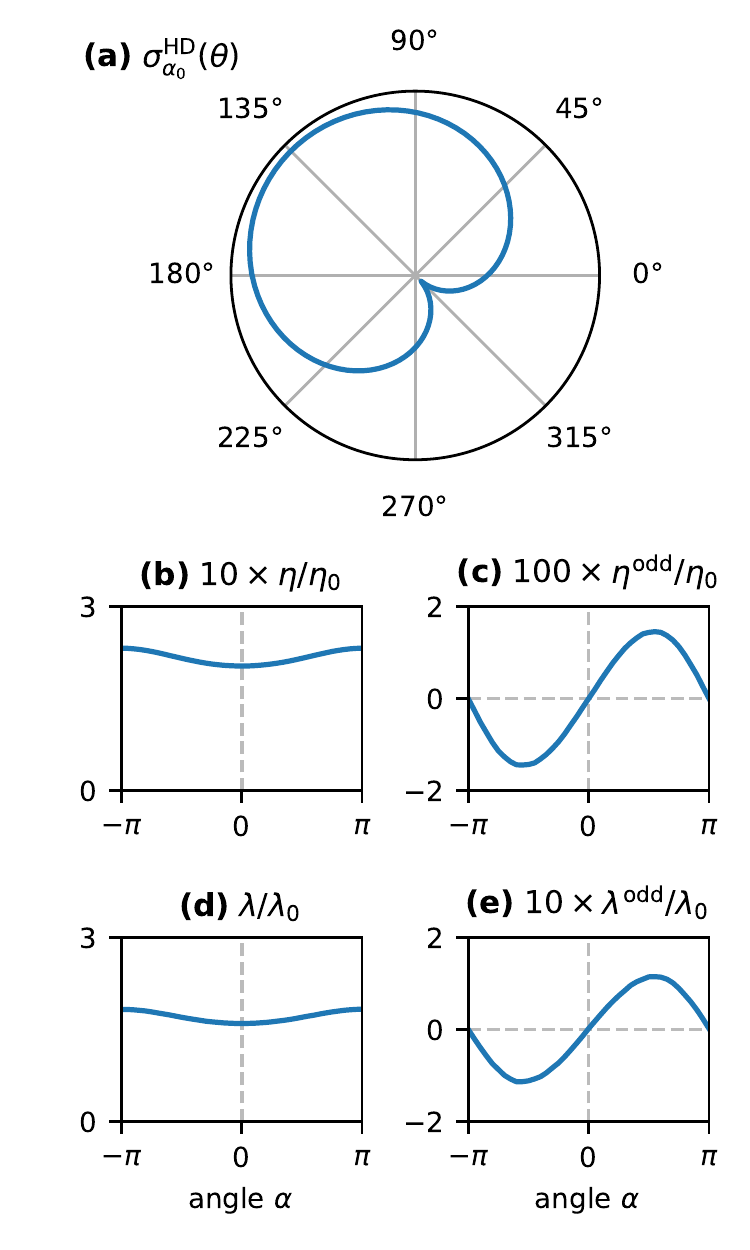}
    \caption{\label{figure_transport_coefficients_hard_disks_main}
    \textbf{Transport coefficients for a rotated hard disk cross-section.}
    We plot (b) the shear viscosity $\eta$, (c) the odd (or Hall) viscosity $\eta^{\text{o}}$, (d) the heat conductivity $\lambda$ and (e) the odd (or Hall) heat conductivity as a function of the angle $\alpha$, for the rotated hard-disk cross-section of Eq.~\eqref{cross_section_HD_main} represented in (a) for $\alpha = \pi/4$.
    The usual transport coefficients $\lambda$ and $\eta$ are even functions of $\alpha$, while the non-dissipative transport coefficients
    $\eta^{\text{o}}$ and $\lambda^{\text{o}}$ are odd functions of $\alpha$.
    These coefficients are the only ones present in an isotropic system (as considered here). They enter the heat conductivity tensor as
    $\lambda_{ij} = \lambda \delta_{ij} +  \lambda^{\text{o}} \epsilon_{ij}$ 
    and the viscosity tensor as 
    $\eta_{i j k \ell} = \eta [\delta_{ik} \delta_{j\ell} + \delta_{i\ell} \delta_{jk} - \delta_{ij} \delta_{k\ell}]
	  + \eta^{\text{o}} [\epsilon_{ik} \delta_{j\ell}+\epsilon_{j\ell} \delta_{ik}]$.
	All quantities are nondimensionalized (see SI).
    We have used $\num{5}$ Hermite polynomials in each direction. 
    The integrals are truncated up to $\lvert \tilde{c} \rvert \sim \num{5}$ and computed using the quasi-Monte-Carlo method Vegas of the Cuba library~\cite{Hahn2005} with $\num{500000}$ points.
    }
\end{figure}

\header{Example} 
We now evaluate Eqs.~\eqref{viscosity_main} and \eqref{thermal_conducitivity_main} numerically in a toy model obtained by rotating the standard hard-disk cross-section by an arbitrary angle $\alpha$ (Fig.~\ref{figure_transport_coefficients_hard_disks_main}a), yielding the modified hard-disk differential cross-section
\begin{equation}
	\label{cross_section_HD_main}
	\sigma_{\alpha}^{\text{HD}}(g, \theta) = \frac{\sigma_0}{4} \, \left| \sin\left(\frac{\theta + \alpha}{2}\right) \right|.
\end{equation}
When $\alpha = 0$, we recover the usual hard-disk cross-section and $\sigma_{0}^{\text{HD}}(g, \theta) = \sigma_{0}^{\text{HD}}(g, -\theta)$. This symmetry is broken when $\alpha \neq 0$.
The results are presented in Fig.~\ref{figure_transport_coefficients_hard_disks_main}. 
The usual transport coefficients $\lambda$ and $\eta$ are even functions of $\alpha$, while the non-dissipative transport coefficients $\eta^{\text{o}}$ and $\lambda^{\text{o}}$ are odd functions of $\alpha$.

A concrete microscopic model is considered in Ref.~\cite{Han2020}, in which an actively rotating granular gas is analyzed in detail. The collision cross-section is indeed asymmetric, as evidenced by Fig.~2d of Ref.~\cite{Han2020}, and Fig.~1f of Ref.~\cite{Han2020} demonstrates the existence of odd viscosity from molecular dynamics simulations.  
In the SI, we consider another example cross-section to illustrate how the relative magnitude of odd versus normal shear viscosities and thermal conductivities can be increased by tuning microscopic interactions. We emphasize that Eqs.~\eqref{viscosity_main} and \eqref{thermal_conducitivity_main} allow us to determine the viscosity and thermal conductivity from experimental measurements of the cross-section obtained in two-body collisions, as long as the gas is dilute.

\header{Green-Kubo relations} 
Equations~\eqref{viscosity_main} and \eqref{thermal_conducitivity_main} can be cast in the form of a Green-Kubo relation for the whole viscosity and thermal conductivity tensors~\cite{Balescu1975,Han2020}. After a few algebraic manipulations (see SI), we find that
\begin{equation}
	\label{green_kubo}
	\frac{\eta_{i j k \ell}}{\rho} = \frac{m}{k_{\text{B}} T} \int_{0}^{\infty} \braket{ \hat{\sigma}_{i j}(t) \, \hat{\sigma}_{k \ell}(0)}_{\text{eq}} \dd t
\end{equation}
in which $\hat{\sigma}_{i j}(\vec{c}) = c_i c_j - (1/d) c^2 \delta_{i j}$ is the microscopic stress tensor (projected on the nullspace of $L$), $\braket{\cdot}_\text{eq}$ is an equilibrium average weighted by the Boltzmann distribution, and the time evolution is generated by Eq.~\eqref{LBE_main} (see SI).
The Green-Kubo formula \eqref{green_kubo} relates the integrated time correlations of the fluctuating stress tensor with the viscosity tensor.
Note that the microscopic stress tensor of the dilute gas is symmetric by construction, as it does not contain the virial part of the Irving-Kirkwood formula~\cite{Han2020}. A similar result can be obtained for $\lambda_{i j}$.

\header{Relaxation time approximation} Finally, we introduce an analytically tractable collision operator
\begin{equation}
	\label{SRTA_main}
	[L^{\text{S}} \phi](\vec{c}) = - \frac{1}{\tau} \phi(R^{-1} \vec{c}).
\end{equation}
in which $R = R(\alpha) = \ee^{- \alpha \epsilon}$ is a rotation matrix by the angle $\alpha$, and $\epsilon$ is the Levi-Civita symbol.
Equation~\eqref{SRTA_main} is a generalized version of the relaxation time approximation (RTA) in which $L \phi \simeq - \phi/\tau$~\cite{Bhatnagar1954}, that is recovered when $\alpha = 0$. 
Essentially, it captures the most relevant eigenvalues of the linearized collision operator.
We can directly compute the viscosity and thermal conductivity tensors in this modified RTA by first determining the inverse $[(L^{\text{S}})^{-1} \phi](\vec{c}) = - \tau \phi(R \vec{c})$, and computing the integrals in Eqs.~\eqref{viscosity_main} and \eqref{thermal_conducitivity_main}. We find
\begin{equation}
\begin{split}
	\frac{\eta_{i j k \ell}}{\rho} = \frac{\tau}{\beta m} \Big( 
	  &\cos(2 \alpha) \, [\delta_{ik} \delta_{j\ell} + \delta_{i\ell} \delta_{jk} - \delta_{ij} \delta_{k\ell}] \\
	+ &\sin(2 \alpha) \, [\epsilon_{ik} \delta_{j\ell}+\epsilon_{j\ell} \delta_{ik}]
	\Big)
\end{split}
\end{equation}
and
\begin{equation}
	\lambda = 
	\frac{2 n k_{\text{B}} \tau}{\beta m}
	\begin{pmatrix}
		\cos \alpha & \sin \alpha \\
		- \sin \alpha & \cos \alpha
	\end{pmatrix}.
\end{equation}
When $\alpha = 0$, we recover the usual expression of the viscosity and thermal conductivity tensors of a dilute monoatomic gas in the relaxation time approximation. 
When $\alpha \neq 0$, a non-vanishing odd viscosity appears. Its magnitude $(\tau/\beta m) \, \sin(2 \alpha)$ is odd with the angle $\alpha$.
Microscopically, the phenomenological angle $\alpha$ can be interpreted as an average twisting angle characterizing the parity-violating nature of the collisions.
Similarly, a non-vanishing odd thermal conductivity appears, whose magnitude $(2 k \tau/\beta m) \, \cos \alpha$ is again odd with the angle $\alpha$.
In the limit of small angles, \begin{equation}
    L^{\text{S}} \simeq - \frac{1}{\tau} [ 1 + \alpha \, \vec{c} \times \nabla_{\vec{c}}] + \mathcal{O}(\alpha^2)
\end{equation}
in which we recognize the out-of-plane component $\ii \ell_z = \vec{c} \times \nabla_{\vec{c}}$ of the infinitesimal generator of rotations in velocity space.
This expression is formally equivalent to the contribution of
an out-of-plane magnetic field proportional to $\alpha$ in the last term in the LHS of the linearized Boltzmann equation \eqref{LBE_main}.
In this limit, the effect of collisions can therefore be accounted for by an effective magnetic field that depends on their average chirality.

To sum up, we have shown how odd viscosity and thermal conductivity arise from parity-violating collisions in an dilute gas through the non-Hermiticity of the linearized collision operator.
 
\clearpage

\begin{center}
  \textbf{Supplemental Material}
\end{center}

\def\mainTextFigureMechanismOddViscosity{Fig.~2 of the main text}
\def\mainTextFigureCollision{Fig.~1 of the main text}

\setcounter{equation}{0}
\let\oldtheequation\theequation
\renewcommand{\theequation}{S\arabic{equation}}

\setcounter{figure}{0}
\let\oldthefigure\thefigure
\renewcommand\thefigure{S{\arabic{figure}}}
\renewcommand{\theHfigure}{S\arabic{figure}}

In this SI, we give a summary of the aspects of kinetic theory relevant to our analysis, and prove the results quoted in the main text. 
For more details on kinetic theory, we refer to the literature, including the monographs by Chapman and Cowling~\cite{ChapmanCowling} and by Hirschfelder, Curtiss and Bird~\cite{Hirschfelder1964}
and the articles by Waldmann~\cite{Waldmann1958} (in German) and by Grad~\cite{Grad1958}. For readable first introductions, we suggest the textbook by Reif~\cite{Reif2009} and the introductory book by Harris~\cite{Harris2004}. 
The recent textbook by Dorfman, van Beijeren and Kirkpatrick~\cite{Dorfman2021} contains both introductory chapters and discussions on recent developments.

\section{The Boltzmann equation}
\label{sec:boltzmann_equation}

Statistical mechanics trades the exact deterministic description of a many-body system for a probabilistic description, more tractable because it is more amenable to certain approximations. In the case of a dilute gas, the relevant approximation is the molecular chaos hypothesis (\emph{Stosszahlansatz}) that particles about to collide are completely uncorrelated before the collision. Under this hypothesis, it is possible to obtain a closed form dynamical equation for the one-particle probability distribution function, called the Boltzmann equation, which is the basis of the kinetic theory of dilute gases.

The distribution function $f(t,\vec{r},\vec{c})$ gives the probability $f(t,\vec{r},\vec{c}) \dd^d \vec{r} \dd^d \vec{c}$ to find a particle in an infinitesimal volume centered at $(\vec{r}, \vec{c})$ in phase space at time $t$, where $\vec{r}$ is the position and $\vec{c}$ is the velocity (so $m \vec{c}$ is the momentum).
The Boltzmann equation describes the evolution of the distribution function $f(t,\vec{r},\vec{c})$ as
\begin{equation}
	\label{BE}
	\frac{\partial f}{\partial t} 
	+ c_\mu \frac{\partial f}{\partial r_\mu}
	+ b_\mu \frac{\partial f}{\partial c_\mu}
	\equiv \mathscr{D}(f) 
	= \mathscr{C}(f)
\end{equation}
in which $m \vec{b}$ is the external force applied to the system.
We will write $\vec{c} \equiv \vec{c_1}$ interchangeably in the following.
The right-hand side is the collision integral
\begin{equation}
\label{collision_integral}
\begin{split}
	\mathscr{C}(f) \equiv \iiint \big[
	 &W(\vec{c_1'}, \vec{c_2'} \mid \vec{c_1}, \vec{c_2}) f_1' f_2' \\
	-&W(\vec{c_1}, \vec{c_2} \mid \vec{c_1'}, \vec{c_2'}) f_1 f_2
	\big] \dd \vec{c_2} \, \dd \vec{c_1'} \, \dd \vec{c_2'}
\end{split}
\end{equation}
in which $W(\vec{c_1}, \vec{c_2} | \vec{c_1'}, \vec{c_2'}) \geq 0$ is the (conditional) transition probability (rate) for two particles with incoming velocities $\vec{c_1}$ and $\vec{c_2}$ to have outgoing velocities $\vec{c_1'}$ and $\vec{c_2'}$ after collision (i.e., for the binary collision $(\vec{c_1}, \vec{c_2}) \to (\vec{c_1'}, \vec{c_2'})$).
We have followed the notations of Ref.~\cite{Waldmann1958}, in which the arguments of $W(b|a)$ are permuted compared to the usual notation for conditional probabilities (to recover the habitual order, one can define $w(a|b)=W(b|a)$ as in Ref.~\cite{Hoffman1969}).
We have used the notations
\begin{subequations}
\begin{align}
	f_1 &= f(t,\vec{r},\vec{c_1}) 
	&
	\quad
	f_2 &= f(t,\vec{r},\vec{c_2}) 
	&
	\\
	f_1' &= f(t,\vec{r},\vec{c_1'}) 
	&
	\quad
	f_2' &= f(t,\vec{r},\vec{c_2'}) 
\end{align}
\end{subequations}
Similarly, we wrote $\mathscr{C}(f)$ as a shortcut for the function applied to the arguments $[\mathscr{C}(f)](t, \vec{r}, \vec{c_1})$.

Several assumptions underlie the Boltzmann equation for a dilute gas:
\begin{itemize}[nosep,label=--,wide, labelwidth=!, labelindent=0pt]
	\item There are well-defined collisions, localized in space and time, outside of which the interaction is negligible. For instance, this would  not be true with long-range interactions~\cite{Alexandre2001,Campa2009}. The fluid of vortices analysed by Abanov and Wiegmann in Ref.~\cite{Wiegmann2014} is an example of system with long-range interactions that displays odd viscosity. This is outside of the range of this work, as drastic modifications are required for kinetic theory to handle these systems. 
	\item Just before a collision occurs, the two-point distribution function factorizes as 
	$f^{[2]}(t, \vec{r_1}, \vec{c_1}, \vec{r_2}, \vec{c_2}) = f^{[1]}(t, \vec{r_1}, \vec{c_1}) \, f^{[1]}(t, \vec{r_2}, \vec{c_2})$
	in which we have temporarily used $[n]$ to label the $n$-particle distribution function. 
	This is the molecular chaos hypothesis. Systems in which correlations before collisions cannot be neglected are called \enquote{liquids} in kinetic theory, and require a more elaborate theory. We refer to Ref.~\cite{Liao2019} for the effect of correlations on odd viscosity, in which it was shown that the distortion of the pair correlation function can indeed lead to non-dissipative transverse responses.
\end{itemize}

\subsection{Linearized Boltzmann equation}

Starting from a known solution $f^{(0)} \equiv f^\circ$ of the Boltzmann equation, we can linearise it to understand the fate of perturbations by writing
\begin{equation}
	f = f^{(0)} + f^{(1)} = f^{(0)} (1 + \phi) 
\end{equation}
in which $f^{(1)} = f^{(0)} \, \phi$ is a small perturbation.
We assume that the unperturbed solution satisfies both $\mathscr{C}(f^\circ) = 0$ and $\mathscr{D}(f^\circ) = 0$ (this is indeed the case of the Boltzmann distribution describing a uniform fluid).
We obtain the linearized Boltzmann equation
\begin{equation}
	\label{lbe}
	\mathscr{D}(\phi) = L \phi
\end{equation}
in which $L$ is the linearized collision operator defined by~\footnote{In Eq.~\eqref{linearized_collision_operator}, the bilinear function $\mathscr{C}(f,g)$ is defined by Eq.~\eqref{collision_integral} in which $f_2$ is replaced by $g_2$ and $f_2'$ by $g_2'$.}
\begin{equation}
	\label{linearized_collision_operator}
	L \phi = \frac{1}{f^\circ} \left[ \mathscr{C}(f^\circ, f^\circ \phi) + \mathscr{C}(f^\circ \phi, f^\circ) \right].
\end{equation}

\subsection{Hermiticity of the linearized collision operator}

Equipped with a stationary distribution $f^\circ$, we can define a scalar product on the functions of the velocity $\vec{c} \mapsto \phi(\vec{c})$ by
\begin{equation}
	\label{scalar_product}
	(\chi, \phi) = \frac{1}{n} \int f^{\circ}(\vec{c_1}) \, \overline{\chi(\vec{c_1})} \, \phi(\vec{c_1}) \dd^d\vec{c_1}.
\end{equation}

We can then compute
\begin{equation}
\begin{aligned}
	(\chi, L \phi) - (L \chi, \phi) \\
	=
	\frac{1}{2} \int \Delta W \, \overline{(\chi_1' + \chi_2')} \, (\phi_1 + \phi_2)
\end{aligned}
\end{equation}
in which we have assumed that $W$ and $f^\circ$ are real-valued, and have defined
\begin{equation}
	\label{DeltaW}
	\Delta W = W(1 2 | 1' 2') f_1^\circ f_2^\circ - W(1' 2' | 1 2) f_1^{\circ \prime} f_2^{\circ \prime}.
\end{equation}
in which we abbreviated $\vec{c}_1$ as $1$, etc. 

Hence, the linearized collision operator is Hermitian (symmetric) provided that $\Delta W = 0$, namely
\begin{equation}
	\label{LeqLdagger_DeltaW}
	L = L^\dagger \; \Leftrightarrow \; \Delta W = 0.
\end{equation}

Non-Hermitian collision operators arise in spin-polarized gases~\cite{Lhuillier1982} and systems under magnetic field (Ref.~\cite{AhSam1971} and references therein). They also occur in the theory of dense gases~\cite{Resibois1970}.
Here, we will assume in section~\ref{collision_symmetries} that the collisions conserve energy and momentum, and that they are isotropic (invariant under $SO(d)$).
In space dimension $d=3$, this is enough to guarantee that $L$ is Hermitian. In dimension $d=2$, under these assumptions, $L = L^\dagger$ is equivalent to $\sigma(g,\theta)=\sigma(g,-\theta)$ in which $\sigma$ is the scattering cross-section [see Eq.~\eqref{Lhermitian2d}].
This constraint can be a consequence of time-reversal or/and mirror reflection about one axis.

\subsection{Conserved quantities and symmetries}
\label{collision_symmetries}

In this work, we assume that the collisions conserve mass, energy and momentum.
Note that this is not guaranteed in active systems: energy is not conserved in driven-dissipative systems; momentum is not conserved in collections of self-propelled particles, nor in the case of particles on a substrate; and mass can be lost and gained through chemical reactions.
All of these constraints can be lifted. The kinetic theory of granular gases provides an example in which energy is not conserved\cite{Garzo1999,Lutsko2005,Garzo2019,Brilliantov2004} (in active systems, losses can be balanced by active forces to produce an effective equilibrium). The kinetic theory of self-propelled particles provides an example in which momentum is not conserved~\cite{Bertin2006,Bertin2009,Ihle2011}. Chemically reacting gas mixtures provide an example in which the number of particles in each species is not conserved~\cite{Kremer2010}.
Nevertheless, we can ignore these additional complexities to get a zeroth-order picture of collision-generated non-dissipative transport coefficients.
In addition, we assume that the system is invariant under proper rotations (i.e., under the group $SO(d)$).
Under theses assumptions, the collision integral Eq.~\eqref{collision_integral} can be written in the form (see Appendix~\ref{symmetries_collisions})
\begin{equation}
\label{simplified_collision_integral}
	\mathscr{C}(f) = \int g \, \sigma(g, \theta') 
	\big[ f_1' f_2' -  f_1 f_2 \big] \dd^d \vec{c_2} \, \dd^{d-1} \Omega'
\end{equation}
in which $\sigma(g, \theta')$ is the differential cross-section, that characterizes the two-body collisions.
We have introduced $\vec{g} = \vec{c_2} - \vec{c_1}$ and $\vec{g'} = \vec{c_2'} - \vec{c_1'}$ and the angle $\theta'$ between $\vec{g}$ and $\vec{g}'$ (see \mainTextFigureCollision{}). The differential solid angle $\dd^{d-1} \Omega'$ is defined such that $\dd^d \vec{g}' = g^{d-1} \dd g \dd^{d-1} \Omega'$. Note that all the variables in the equation should be seen as functions of $(t, \vec{r}, \vec{c_1})$ (arguments of the LHS) and $(\vec{c_2}, \Omega')$ (integration variables) defined by the conservation laws (see Appendix~\ref{appendix_dependent_variables} for an explicit version in the 2D case).

The linearized collision operator Eq.~\eqref{linearized_collision_operator} then becomes
\begin{equation}
	\label{linearized_collision_operator_cons}
	L \phi = 
	\int 
	  g \sigma(g, \theta') f^{\circ}_2 \big[ \phi_1' + \phi_2' - \phi_1 - \phi_2 \big] 
	  \dd^d \vec{c_2} \, \dd^{d-1} \Omega'
\end{equation}
In this case, the stationary distribution of the dilute gas is the Boltzmann distribution
\begin{equation}
\label{boltzmann_distribution}
\begin{split}
	f^\circ(\vec{c}) = n \, \left( \frac{m \beta}{2 \pi} \right)^{\frac{d}{2}} 
	\, \exp\left( - \frac{1}{2} \, \beta m \lVert \vec{c} - \vec{v} \rVert^2 \right)
\end{split}
\end{equation}
in which $n$ is the number density and $\beta = (\kB T)^{-1}$ is the inverse temperature.

The nullspace of $L$ is spanned by collisional invariants, whose averages are conserved quantities. 
Here, they are (by construction) respectively the mass, the linear momentum, and the energy, 
\begin{equation}
	\label{collision_invariants}
	m, \; m c_\mu, \; m \frac{\lvert \vec{c} \rvert^2}{2}.
\end{equation}

In two dimensions, the linearized collision operator Eq.~\eqref{linearized_collision_operator_cons} is Hermitian provided that $\sigma(g, \theta')$ is an even function of $\theta'$, namely
\begin{equation}
    \label{Lhermitian2d}
	L = L^\dagger \; \Leftrightarrow \; \sigma(g, \theta') = \sigma(g, - \theta').
\end{equation}
This can be seen from Eq.~\eqref{LeqLdagger_DeltaW} as follows. First, the Boltzmann distribution Eq.~\eqref{boltzmann_distribution} satisfies the local equilibrium condition $f_1^{\circ \prime} f_2^{\circ \prime} =  f_1^{\circ} f_2^{\circ}$, so the symmetry condition Eq.~\eqref{LeqLdagger_DeltaW} becomes $W(1 2 | 1' 2') = W(1' 2' | 1 2)$. Second, permuting $(\vec{c_1}, \vec{c_2}) \leftrightarrow (\vec{c_1'}, \vec{c_2'})$ corresponds to permuting $\vec{g} \leftrightarrow \vec{g'}$.
As $\theta$ is the angle from $\vec{g}$ to $\vec{g'}$ in the plane (see \mainTextFigureCollision{}), this corresponds to $\theta' \to -\theta'$.

\section{From the Boltzmann equation to hydrodynamics}
\label{sec:boltzmann_to_hydro}

\subsection{Hydrodynamic variables}

We define the local mass and number density \begin{equation}
\begin{split}
	\rho(t,\vec{r}) &= m \, n(t, \vec{r}) = m \int f(\vec{c}) \; \dd^{d}\vec{c} \\  \end{split}
\end{equation}
and the momentum and energy densities
\begin{equation}
\label{momentum_energy_densities}
\begin{split}
	\rho_{v_i}(t,\vec{r}) &=  \rho \, \braket{c_i} = \rho \, v_i(t, \vec{r}) \\
	\rho_{\text{e}}(t,\vec{r}) &= \rho \braket{c^2/2} = \rho \, v^2/2 + n \, e_{\text{kin}}(t, \vec{r})
\end{split}
\end{equation}
in which
\begin{equation}
	\label{average_def}
	\braket{A} = \frac{1}{n} \, \int A f \dd^d \vec{c}
\end{equation}
These equations also define the hydrodynamic velocity $\vec{v}$, and the local kinetic energy $e_{\text{kin}}$.
We also define the local temperature $T$ such that 
\begin{equation}
	e_{\text{kin}}(t,\vec{r}) = \frac{n_{\text{dof}}}{2} \, k T(t,\vec{r}).
\end{equation}
In a monoatomic gas, $n_{\text{dof}} = d$.

\subsection{Balance equations}

Starting from the Boltzmann equation \eqref{BE}, and integrating over velocities, we obtain for any quantity $A$ the integral balance equation
\begin{equation}
	 \int A(t, \vec{r}, \vec{c}) \mathscr{D}(f) \dd \vec{c} = \int A(t, \vec{r}, \vec{c}) \mathscr{C}(f) \dd \vec{c}.
\end{equation}
The left-hand side of this equation can be converted to a local balance equation~\cite[\S~3.13]{ChapmanCowling}.
This is particularly simple in the case of conserved quantities $A$ for which the right-hand side of this equation vanishes.

With the collisional invariants in Eq.~\eqref{collision_invariants}, we find the usual local conservation equations
\begin{subequations}
\label{conservation_eqs}
\begin{align}
	&\frac{D \rho}{D t} + \rho \, \div(\vec{v}) = 0 \\
	\rho &\frac{D v_\mu}{D t} = \rho b_\mu - \partial_\nu P_{\mu \nu} \\
	n &\frac{D e_{\text{kin}}}{D t} + \div(\vec{Q}) + P_{\mu \nu} \partial_\nu  v_\mu = 0
\end{align}
\end{subequations}
in which we have defined the pressure tensor
\begin{equation}
	\label{pressuredef}
	P_{\mu \nu} = \rho \, \Braket{U_\mu \, U_\nu}
\end{equation}
and the heat current
\begin{equation}
	\label{eqdef}
	\vec{Q} = n \Braket{\frac{1}{2} m U^2 \; \vec{U}}
\end{equation}
where $\vec{U} = \vec{c} - \vec{v}$.
It can also be convenient to work with the stress tensor
\begin{equation}
    \sigma_{\mu \nu} = - P_{\mu \nu}.
\end{equation}
We note that the pressure tensor (and the stress tensor) of a dilute gas of point particles is symmetric by construction: $P_{\mu \nu} = P_{\nu \mu}$.

Computing the zeroth-order pressure tensor gives the ideal gas law
\begin{equation}
	\label{ideal_gas_law}
	P^{(0)}_{\mu \nu} = n \kB T \, \delta_{\mu \nu}.
\end{equation}
Similarly, we compute the zeroth-order heat flux
\begin{equation}
	\label{q_zeroth_order}
	Q_\mu^{(0)} = 0.
\end{equation}
and the zeroth-order kinetic energy
\begin{equation}
	\label{ekin_zeroth_order}	
	e_{\text{kin}}^{(0)} = \frac{d}{2} k_{\text{B}} T^{(0)} 
	\equiv c_{\text{v}} T^{(0)}
\end{equation}
in which $c_{\text{v}} = (d/2) \kB$ (we have set $n_\text{dof}=d$).

The energy conservation is equivalent to the temperature equation
\begin{equation}
	n c_{\text{v}} \frac{D T}{D t} = - \left[ \div(\vec{Q}) + P_{\mu \nu} \partial_\nu  v_\mu \right]
\end{equation}

\subsection{Linearized hydrodynamic equations}

We linearize the balance equations \eqref{conservation_eqs} by considering the departures from equilibrium
\begin{align}
	\label{delta_hydro_variables}
	\delta \rho &= \rho - \rho^{(0)} = m \, n \, (1, \phi) \\
	\delta v_\mu &= v_\mu - {v}^{(0)}_\mu = (c_\mu, \phi) \\
	\delta T &= T - T^{(0)} = T \, \left(m \frac{c^2}{n_{\text{dof}} \, k_{\text{B} T}} - 1, \phi\right)
\end{align}
that we have expressed as scalar products $(\cdot, \phi)$ with the perturbation $\phi$ of the distribution function, and in which we have used that $\vec{v}^{(0)} = \vec{0}$.
We obtain
\begin{align}
	\label{linearized_NS}
	\partial_t \delta \rho &= - \rho^{(0)} \div(\vec{\delta v}) = 0 \\
	\rho^{(0)} \partial_t \delta v_\mu &= - \partial_\nu P^{(1)}_{\mu \nu} \\
	n^{(0)} c_{\text{v}} \partial_t \delta T &= - \div(\vec{Q}^{(1)}) - n^{(0)}  k_{\text{B}} T^{(0)} \div(\vec{\delta v})
\end{align}
in which $c_{\text{v}} = (d/2) \kB$ is the specific heat at constant volume.
Note that the zeroth-order quantities are given by Eqs.~(\ref{ideal_gas_law},\ref{q_zeroth_order},\ref{ekin_zeroth_order}) and that their the space and time derivatives vanish.

Here, $P^{(1)}_{\mu \nu} = [\kB n^{(0)} \delta T + \kB T^{(0)} \delta n] \delta_{\mu \nu} + P^{(1), \text{vis}}_{\mu \nu}$ in which $P^{(1), \text{vis}}$ is the viscous stress tensor.

\subsection{Transport coefficients}
\label{transport_coefficients}

Following Refs.~\cite{Resibois1970,Ernst1970,Balescu1975,Dorfman2021}, we will now determine the transport coefficients by matching the linearized Boltzmann equation projected onto hydrodynamic variables with the linearized balance equations.

Applying a Fourier transform to the linearized Boltzmann equation \eqref{lbe}, we find in the absence of external forces ($\vec{b} = 0$)
\begin{equation}
	\label{eom_distribution}
	\partial_t \phi + \ii q_\mu c_\mu \, \phi = L \, \phi
\end{equation}
in which $\phi = \phi(t,\vec{q},\vec{c})$ is the Fourier transform of the physical-space perturbation of the distribution function $\phi(t,\vec{r},\vec{c})$.

\medskip

Our goal is to obtain a hydrodynamic theory, which describes slow long-wavelength ($q \to 0$) modes. Starting from the full dynamics at finite $q$, an effective dynamics for the hydrodynamic modes is obtained by integrating out the fast degrees of freedom, namely, by solving for their value (under the assumption that they quickly relax compared to the hydrodynamic modes) and replacing it in the equation.
The hydrodynamic modes are then obtained by projecting the state of the system on the linear space spanned by conserved quantities (i.e. on the nullspace of $L(\vec{q} = \vec{0})$). 
The corresponding vectors relax slowly when $q \to 0$ (typically as $q$ or $q^2$), while non-hydrodynamic variables are assumed to relax at a finite rate in the limit $q \to 0$.
Hence, we decompose $\phi(\vec{q})$ into two parts $\phi=(\phi_\sslash, \phi_\bot)$ in which $\phi_\sslash$ is the projection of $\phi(\vec{q})$ on the nullspace of $L$ (i.e., hydrodynamic variables) and $\phi_\bot$ is the projection on the orthogonal complement (i.e., fast non-hydrodynamic modes).
Pictorially, this will produce two equations
\begin{subequations}
\label{split_linearized_boltzmann}
\begin{align}
	\partial_t \phi_\sslash &= a \, \phi_\sslash + b \, \phi_\bot \\
	\partial_t \phi_\bot &= c \, \phi_\sslash + d \, \phi_\bot
\end{align}
\end{subequations}
in which $a$, $b$, $c$, $d$ are linear operators (this notation is used only in this paragraph; here, $d$ is not the space dimension). We assume that the non-hydrodynamic modes are stationary: $\partial_t \phi_\bot \approx 0$ (please note that this is different from assuming that $\phi_\bot \approx 0$). 
Solving for $\phi_\bot$ and replacing, we obtain
\begin{equation}\label{compact_NS}
	\partial_t \phi_\sslash = a \, \phi_\sslash - b d^{-1} c \, \phi_\sslash
\end{equation}
which is a compact version of the Navier-Stokes equations.

The method above is known as a \enquote{projection operator method}, because we will use a projection operator to project things.
We refer to Refs.~\cite{Zwanzig2001,Mori1965,Nakajima1958,Zwanzig1960} for general discussions, and to Refs.~\cite{Resibois1970,Balescu1975,Dorfman2021} for the application to kinetic theory (see also Refs.~\cite{Ernst1970,Bixon1971,Hauge1970}).
Here, we only use a poor man's projection operator method in which the fluctuations are ignored, and we refer to Ref.~\cite{Han2020} in which a fluctuating hydrodynamic description of chiral active fluids with odd viscosity was developed.

\medskip

As promised, we now define the projector $P$ on the nullspace of the linearized collision operator $L$, and the projector $Q = 1 - P$ on its orthogonal complement. 
To do so, we orthonormalize the collisional invariants \eqref{collision_invariants} with respect to the scalar product \eqref{scalar_product} to get the functions
\begin{equation}
	\label{orthonormalized_basis_kernel}
	\Phi(\vec{c}) = \begin{pmatrix}
		\Phi_{\text{m}} \\
		\Phi_{\text{v},\mu} \\
		\Phi_{\text{e}} 
	\end{pmatrix}
	= 
	\begin{pmatrix}
		1 \\
		\sqrt{\beta m} \; c_\mu \\
		\left( \frac{2}{d} \right)^{1/2} \, \left[ \beta m \frac{\lvert \vec{c} \rvert^2}{2} - \frac{d}{2} \right]
	\end{pmatrix}
\end{equation}
that span the nullspace of $L$ and satisfy~\footnote{
	In general, some care is needed when working with non-normal operators, because they cannot be diagonalized in an orthonormal basis. 
	Instead, one can use a biorthonormal system composed of right (usual) eigenvectors $\Phi_i$ and left eigenvectors $\tilde{\Phi}_i$ that can be chosen so that $(\tilde{\Phi}_i, \Phi_j) = \delta_{i j}$ provided that the matrix is diagonalizable. The standard functional calculus of normal matrices is then extended in a simple way (e.g., by replacing $|\Phi_i)(\Phi_i|$ with $|\Phi_i)(\tilde{\Phi}_i|$ in Eq.~\eqref{projector_kernel_L}, see Refs.~\cite{Curtright2007,Brody2013,Riechers2018} and references therein. 
	The (right) eigenvectors of $L$ in Eq.~\eqref{orthonormalized_basis_kernel} are also the corresponding left eigenvectors ($\tilde{\Phi}_i = \Phi_i$, $i=1,\dots,n_{\text{cons}}$), so we do not need to distinguish them.
}
\begin{equation}
	(\Phi_i,\Phi_j) = \delta_{i j}.
\end{equation}
The projector on the nullspace is then
\begin{equation}
	\label{projector_kernel_L}
	P = \sum_{i=1}^{n_{\text{cons}}} |\Phi_i)(\Phi_i|
\end{equation}
and one can check by an explicit computation that it is orthogonal ($P^\dagger = P$).

Any operator can then be written as a block operator
\begin{equation}
	A \simeq \begin{pmatrix} 
	A_{Z Z} & A_{Z F} \\
	A_{F Z} & A_{F F}
	\end{pmatrix}
\end{equation}
in which we have defined $A_{Z Z} = P A P$, $A_{Z F} = P A Q$, $A_{F Z} = Q A P$, $A_{F F} = Q A Q$.
Depending on context, these operators are to be interpreted as restricted to the range of the projection (i.e., only the block without all the zeros) when needed. For instance, $A_{Z Z}$ can be interpreted either as a $n_{\text{cons}} \times n_{\text{cons}}$ matrix, or as an infinite matrix with zeros everywhere except in a block on size $n_{\text{cons}} \times n_{\text{cons}}$, and we use the same notation for both.
Similarly, a vector $\psi$ is decomposed as $\psi = (\psi_Z, \psi_F)$ in which $\psi_Z = P \psi$ and $\psi_F = Q \psi$ (again, $\psi_Z$ and $\psi_F$ should be interpreted as restricted to the ranges of the projector when needed).

After performing this block decomposition, Eq.~\eqref{eom_distribution} becomes
\begin{equation}
	\partial_t \begin{pmatrix} \phi_Z \\ \phi_F \end{pmatrix}
	=
	L(\vec{q})
	\begin{pmatrix} \phi_Z \\ \phi_F \end{pmatrix}
\end{equation}
where
\begin{equation}
	L(\vec{q}) = 
	\begin{pmatrix}
	- \ii q_\mu \hat{c}^{\,\mu}_{Z Z} & - \ii q_\mu \hat{c}^{\,\mu}_{Z F}  \\
	- \ii q_\mu \hat{c}^{\,\mu}_{F Z}  & L_{F F} 
	\end{pmatrix}
\end{equation}
at lowest order in $q$ in each block, in which the operator $\hat{c}^{\,\mu}$ multiplies functions of $\vec{c}$ by $c_\mu$.
Assuming that $\partial_t \phi_F \approx 0$, we obtain $\phi_F = \ii q_\mu (L_{FF})^{-1} \hat{c}^{\,\mu}_{F Z} \phi_Z$
and hence
\begin{equation}
	\partial_t \phi_Z = L_{\text{eff}}(\vec{q}) \, \phi_Z
\end{equation}
where
\begin{equation}
	L_{\text{eff}}(\vec{q}) = - \ii q_\mu \hat{c}^{\,\mu}_{Z Z} + q_\mu q_\nu \, \hat{c}^{\,\mu}_{Z F} \, (L_{FF})^{-1} \, \hat{c}^{\,\nu}_{F Z}
\end{equation}

The last equation can be written more explicitly as
\begin{equation}
	\label{eom_projected_phi}
	\partial_t (\Phi_i, \phi) = \sum_j [L_{\text{eff}}(\vec{q})]_{i j} (\Phi_j, \phi).
\end{equation}

Using $P^\dagger = P$, $Q^\dagger = Q$, $(\hat{c}_\mu)^\dagger = \hat{c}_\mu$ and $P \Phi_i = \Phi_i$, the matrix elements of $L_{\text{eff}}(q)$ are expressed as
\begin{equation}
\begin{split}
	&[L_{\text{eff}}(q)]_{i j}
	= - \ii q_\mu (\Phi_i, \hat{c}^{\,\mu}_{Z Z} \Phi_j) \\ 
	&+ q_\mu q_\nu \, (Q \hat{c}_\mu \Phi_i, (L_{FF})^{-1} \, Q  \hat{c}_\nu \Phi_j) 
\end{split}
\end{equation}
in which $i,j = 1, \dots, n_{\text{cons}}$. 

The matrix elements of $(\hat{c}_\rho)_{Z Z}$ are obtained by computing the moments of a Gaussian distribution (see for instance Ref.~\cite{ZinnJustin2021}) and we get
\begin{subequations}
\label{matrix_elements_UZZ}
\begin{align}
	(\Phi_{\text{m}}, \hat{c}_\rho \, \Phi_{\text{m}}) &= 0 \\
	(\Phi_{\text{m}}, \hat{c}_\rho \, \Phi_{\text{v},\nu}) &= \frac{1}{\sqrt{\beta m}} \, \delta_{\rho \nu} \\
	(\Phi_{\text{m}}, \hat{c}_\rho \, \Phi_{\text{e}}) &= 0 \\
(\Phi_{\text{v},\mu}, \hat{c}_\rho \, \Phi_{\text{m}}) &= \frac{1}{\sqrt{\beta m}} \, \delta_{\rho \mu} \\
	(\Phi_{\text{v},\mu}, \hat{c}_\rho \, \Phi_{\text{v},\nu}) &= 0 \\
	(\Phi_{\text{v},\mu}, \hat{c}_\rho \, \Phi_{\text{e}}) &= \sqrt{\frac{2}{d}} \, \frac{1}{\sqrt{\beta m}} \, \delta_{\mu \rho} \\
(\Phi_{\text{e}}, \hat{c}_\rho \, \Phi_{\text{m}}) &= 0 \\
	(\Phi_{\text{e}}, \hat{c}_\rho \, \Phi_{\text{v},\nu}) &=  \sqrt{\frac{2}{d}} \, \frac{1}{\sqrt{\beta m}} \, \delta_{\rho \nu} \\ 
	(\Phi_{\text{e}}, \hat{c}_\rho \, \Phi_{\text{e}}) &= 0.
\end{align}
\end{subequations}

We can then compute
\begin{equation}
	Q \hat{c}_\nu \Phi_i = \hat{c}_\nu \Phi_i - \sum_{j=1}^{n_{\text{cons}}} (\Phi_j, \hat{c}_\nu \Phi_i) \Phi_j
\end{equation}
and using Eqs.~\eqref{matrix_elements_UZZ}, we find
\begin{subequations}
\label{QPhis}
\begin{align}
	Q  \hat{c}_\nu \Phi_{\text{m}} &= 0 \label{Q_c_Phi_m} \\
	Q  \hat{c}_\nu \Phi_{\text{v},\mu} &= \sqrt{\beta m} \left( c_\mu c_\nu - \frac{1}{d} c^2 \delta_{\mu \nu} \right) \\
	Q  \hat{c}_\nu \Phi_{\text{e}} &= \sqrt{\frac{2}{d}} \, \left(\beta \, \frac{1}{2} m c^2 - \frac{d+2}{2} \right) c_\nu.
\end{align}
\end{subequations}

Let us now express the projection $\phi_Z$ on the nullspace of the linearized collision operator of the perturbation $\phi$ of the distribution function in terms of the departures of the hydrodynamic variables from equilibrium defined in Eq.~\eqref{delta_hydro_variables}. 
We find
\begin{equation}
	\phi_Z = \begin{pmatrix}
		(\Phi_{\text{m}}, \phi) \\
		(\Phi_{\text{v},\mu}, \phi) \\
		(\Phi_{\text{e}}, \phi) \\
	\end{pmatrix}
	=
	\begin{pmatrix}
		\dfrac{1}{m \, n} \; \delta \rho \\
		\phantom{\dfrac{1}{m}} \sqrt{\beta m} \; \delta v_{\mu} \\
		\sqrt{\dfrac{d}{2}} \, \dfrac{\delta T}{T}
	\end{pmatrix}.
\end{equation}

We can then rewrite Eqs.~\eqref{eom_projected_phi} explicitly as
\begin{widetext}
\begin{subequations}
\label{hydro_from_boltzmann}
\begin{align}
	\partial_t \delta \rho = & m n (- \ii q_\mu) \delta v_\mu \\
	\rho \partial_t \delta v_\mu =& - \ii q_\mu ( \kB T \delta n + n \kB \delta T) 
	- q_\rho q_\sigma \eta_{\mu \rho \nu \sigma} \delta v_\nu
	- q_\rho q_\sigma \alpha_{\mu \rho \sigma} \delta T   \\
	n c_{\text{v}} \partial_t \delta T
	=& -\ii q_\mu n \kB T \delta v_\mu 
	- q_\rho q_\sigma \alpha'_{\rho \sigma \mu} \delta v_\mu
	- q_\rho q_\sigma \lambda_{\rho \sigma} \delta T  
\end{align}
\end{subequations}
\end{widetext}
in which we have defined the transport tensors
\begin{widetext}
\begin{align}
	\label{transport_from_boltzmann}
	\eta_{\mu \rho \nu \sigma} &= - \rho \beta m (Q c_\mu c_\rho, L_{FF}^{-1} Q c_\nu c_\sigma) \\
	\alpha_{\mu \rho \sigma} &= - \frac{\rho}{\kB T^2} \left(Q c_\mu c_\rho, L_{FF}^{-1} \left[ \frac{m c^2}{2} - \frac{d+2}{2 \beta}\right] c_\sigma\right) \\
\lambda_{\rho \sigma} &= - \frac{n}{\kB T^2} \left(\left[ \frac{m c^2}{2} - \frac{d+2}{2 \beta}\right] c_\rho, L_{FF}^{-1} \left[ \frac{m c^2}{2} - \frac{d+2}{2 \beta}\right] c_\sigma \right) \\
\alpha'_{\rho \sigma \mu} &= - n \beta m \left(\left[ \frac{m c^2}{2} - \frac{d+2}{2 \beta}\right] c_\rho, L_{FF}^{-1} Q c_\sigma c_\mu \right)
\end{align}
\end{widetext}
and where $c_{\text{v}} = (d/2) \kB$ and $\delta n = \delta \rho/m$.
Here, $\eta_{\mu \rho \nu \sigma}$ is the viscosity tensor, and $\lambda_{\rho \sigma}$ is the heat conductivity tensor.
The tensors $\alpha$ and $\alpha'$ described cross-responses between momentum and energy. 
We expect that these will be a generic feature of anisotropic systems, such as the 3D chiral active fluids analysed in Refs.~\cite{Khain2022,Markovich2021}.
However, they vanish in a rotation-invariant 2D system (see App.~\ref{symmetries_tensors}) and we will therefore ignore them in the following. 
Equations~\eqref{hydro_from_boltzmann} and \eqref{transport_from_boltzmann} give an explicit form of the viscous stress tensor $P_{\mu \nu}^{(1), \text{vis}}$ and the first-order heat flux $Q_\mu^{(1)}$ in the linearized Navier-Stokes Eq.~\eqref{linearized_NS}.

\medskip
\medskip

Note that Eq.~\eqref{hydro_from_boltzmann} only determines the products $q_\rho q_\sigma \eta_{\mu \rho \nu \sigma}$ (not directly $\eta_{\mu \rho \nu \sigma}$). Hence, we have implicitly made a choice in defining $\eta_{\mu \rho \nu \sigma}$ in Eq.~\eqref{transport_from_boltzmann}.
Indeed, any tensor antisymmetric in $\nu \leftrightarrow \sigma$ can be added to $\eta_{\mu \nu \rho \sigma}$ in Eq.~\eqref{transport_from_boltzmann} without changing Eq.~\eqref{hydro_from_boltzmann}. 
This ambiguity is related to the ambiguity in defining the stress tensor in continuum mechanics, that we discuss in Appendix~\ref{bulk_ambiguity}.
In the case of dilute gases, however, there is a good reason to prefer Eq.~\eqref{transport_from_boltzmann} to any other choice, because the microscopic expression of the (purely kinetic) stress tensor of the dilute gas is unambiguous. 
Unfortunately, this choice cannot be justified within the projection operator formalism used here. 
Instead, we have to go back to the standard Chapman-Enskog theory: this is done in Appendix~\ref{chapman_enskog}.
There, the linear response approximation is performed at the level of the distribution function, and the stress tensor is then computed from the perturbed distribution function. Hence, we can obtain a unique stress tensor under the condition that the microscopic expression of the stress tensor has no ambiguity (which is the case for dilute gases, but not for liquids, see Appendix~\ref{bulk_ambiguity} and references therein).

\subsection{Onsager-Casimir relations}

Under certain assumptions, we can obtain Onsager-Casimir-like relations from Eq.~\eqref{transport_from_boltzmann}.
Consider a family of linearized collision operators $L(B)$ depending on a parameter $B$, that measures how much parity and time-reversal are broken.
We assume that
\begin{equation}
	\label{L_duality}
	L(B) = L^\dagger(-B).
\end{equation}
The parameter $B$ might be a magnetic field, an angular velocity, or another parameter. 
As an example, $B$ can represent the angle $\alpha$ in Eqs.~\eqref{cross_section_HD} and \eqref{cross_section_KC}.
The property Eq.~\eqref{L_duality} is inherited by $[L_{FF}(B)]^{-1}$.

Starting with Eq.~\eqref{transport_from_boltzmann}, we find
\begin{subequations}
\begin{align}
	\eta_{\mu \rho \nu \sigma}(B)
	 &= - \rho \beta m (Q c_\mu c_\rho, L_{FF}^{-1}(B) Q c_\nu c_\sigma) \\
	 &= - \rho \beta m ([L_{FF}^{-1}(B)]^\dagger Q c_\mu c_\rho,  Q c_\nu c_\sigma) \\
	 &= - \rho \beta m (Q c_\nu c_\sigma, [L_{FF}^{-1}(B)]^\dagger Q c_\mu c_\rho) \\
	 &= - \rho \beta m (Q c_\nu c_\sigma, L_{FF}^{-1}(-B) Q c_\mu c_\rho) \\
	 &= \eta_{\nu \sigma \mu \rho}(-B)
\end{align}
\end{subequations}
in which we have assumed that $\eta_{\mu \rho \nu \sigma}$ is real. 
Hence, we obtain the Onsager-Casimir relation
\begin{align}
	\eta_{\mu \rho \nu \sigma}(B) = \eta_{\nu \sigma \mu \rho}(-B)
\end{align}
Similarly, we find
\begin{equation}
	\lambda_{\mu \nu}(B) = \lambda_{\nu \mu}(-B).
\end{equation}
The particular case of a Hermitian collision operator corresponds to $B=0$, in which the viscosity and heat conductivity are symmetric.

\subsection{Green-Kubo formulas}

The transport coefficients obtained in Sec.~\ref{transport_coefficients} can be rewritten as the integrals of correlation functions called Green-Kubo formulas~\cite[\S~13.4]{Balescu1975}. In this paragraph, we temporarily omit the subscript \enquote{FF} in $L_{FF}$.

The perturbation to the distribution function $\phi$ follows the equation of motion \eqref{eom_distribution}. Hence, any initial perturbation $\phi_0$ is evolved in time into $\phi(t) = U(t) \phi_0$ by the evolution operator
\begin{equation}
	U(t) = \ee^{t L}.
\end{equation}
To compute time-correlation functions, it is more convenient to use the Heisenberg picture, in which the observables $A$ (here, these are functions $\vec{c} \mapsto A(\vec{c})$) evolve according to
\begin{equation}
	A(t) = U^\dagger(t) A
\end{equation}
in which $U^\dagger(t) = \ee^{t L^\dagger}$ (see \cite[\S~2.2]{Balescu1975} for details).

Let us now consider two observables $A$ and $B$, and compute
\begin{equation}
\begin{split}
	\int_{0}^{\infty} \braket{A(t) B(0)}_0 \dd t 
	=
	&\int_{0}^{\infty} \braket{(\ee^{t L^\dagger} A) B}_0 \dd t \\
	=
	&-\braket{([L^\dagger]^{-1} A) B}_0 \\
\end{split}
\end{equation}
in which
\begin{equation}
	\braket{A}_0 = \frac{1}{n} \, \int A f^\circ \dd^d \vec{c}
\end{equation}
where $\braket{\cdot}_0$ is the average defined by Eq.~\eqref{average_def} in which $f$ is taken to be the equilibrium distribution $f^{(0)}$.
Using the definition of the scalar projected Eq.~\eqref{scalar_product}, we identify
\begin{equation}
	\braket{([L^\dagger]^{-1} A) B}_0 = ([L^\dagger]^{-1} A, B) = (A, L^{-1} B)
\end{equation}
to obtain
\begin{equation}
	(A, L^{-1} B) = - \int_{0}^{\infty} \braket{A(t) B(0)}_0 \dd t.
\end{equation}

We have assumed that $L^\dagger$ is invertible (it has been projected on the orthogonal complement to its nullspace), and that the real part of its eigenvalues (the relaxation times) are all positive (so that $\ee^{t L^\dagger} \to 0$ when $t \to \infty$).

Equipped with this equality, we can rewrite the expression of the viscosity Eq.~\eqref{transport_from_boltzmann} as the correlation function
\begin{equation}
	\frac{\eta_{\mu \nu \rho \sigma}}{\rho} = \beta m \int_{0}^{\infty} \braket{ j_{\mu \nu}(t) j_{\rho \sigma}(0)}_0 \dd t
\end{equation}
in which we have introduced the dynamical functions $j_{\mu \nu}(\vec{c}) = c_\mu c_\nu - (1/d) c^2 \delta_{\mu \nu}$ and $j_{\mu \nu}(t) = U^\dagger(t) j_{\mu \nu}$ (note that in this equation, $j_{\mu \nu}(t)$ and $j_{\mu \nu}$ are vectors in the linear space of functions of the velocity, on which $U^\dagger(t)$ acts linearly).
Up to a constant prefactor, $j_{\mu \nu}$ is the projection on the nullspace of $L$ of the stress tensor.
Hence, we obtain a Green-Kubo formula for the viscosity tensor. A similar procedure can be applied to other transport coefficients.

\section{Nondimensionalized equations}

To compute the transport coefficients numerically, it is convenient to first nondimensionalize all the expressions.
(In this section, nondimensionalized quantities are decorated with a tilde.)
To do so, we introduce
\begin{equation}
	c_\mu = \sqrt{\frac{2}{\beta m}} \, \tilde{c}_\mu
	\quad
	\text{and}
	\quad
	\sigma(g,\chi) = \sigma_0 \, \tilde{\sigma}(\tilde{g},\chi)
\end{equation}
in which $\sigma_0$ is a characteristic cross-section and $\tilde{\sigma}$ is dimensionless.

We can then write
\begin{equation}
	L \phi =
	\frac{n \sigma_0}{\pi^{d/2}} \, \sqrt{\frac{2}{\beta m}}
	\,
	\tilde{L} \phi
\end{equation}
in which 
\begin{equation}
	\tilde{L} \phi =
	\,
	\int 
	  \tilde{g} \tilde{\sigma}(\tilde{g}, \chi') \ee^{-\tilde{c}_2^2} \big[ \phi_1' + \phi_2' - \phi_1 - \phi_2 \big] 
	  \dd^d \vec{\tilde{c}_2} \, \dd^{d-1} \Omega'
\end{equation}
in which the prefactor is the inverse of a characteristic time.

We also define the nondimensionalized inner product
\begin{equation}
	\label{scalar_product_nondim}
	(\tilde{\chi}, \tilde{\phi})_{\text{ND}} 
	=  \int \ee^{-\tilde{c}^2} \overline{\tilde{\chi}(\vec{\tilde{c}})} \, \tilde{\phi}(\vec{\tilde{c}}) \dd^d \vec{c}.
\end{equation}
that satisfies
\begin{equation}
	(\chi, \phi) = \frac{1}{\pi^{d/2}} \, (\tilde{\chi}, \tilde{\phi})_{\text{ND}} 
\end{equation}
provided that $\tilde{\chi}(\vec{\tilde{c}}) = \chi(\vec{c})$ and $\tilde{\phi}(\vec{\tilde{c}}) = \phi(\vec{c})$.
For instance,
\begin{equation}
	(c_\mu c_\nu, c_\rho c_\sigma) = \frac{4}{(\beta m)^2} \, \frac{1}{\pi^{d/2}} \, (\tilde{c}_\mu \tilde{c}_\nu, \tilde{c}_\rho \tilde{c}_\sigma)_{\text{ND}}.
\end{equation}

All the transport tensors in Eq.~\eqref{transport_from_boltzmann} can then be expressed in terms of non-dimensionalized inner products that can be computed numerically or analytically.
Explicitly, we obtain
\begin{subequations}
\begin{align}
	\label{transport_from_boltzmann_nondim}
	&\frac{\eta_{\mu \rho \nu \sigma}}{\eta_0} = - 
(Q \tilde{c}_\mu \tilde{c}_\rho, \tilde{L}_{FF}^{-1} Q \tilde{c}_\nu \tilde{c}_\sigma)_{\text{ND}}  \\
&\frac{\lambda_{\rho \sigma}}{\lambda_0} = - \left(\left[ \tilde{c}^2 - \frac{d+2}{2}\right] \tilde{c}_\rho, \tilde{L}_{FF}^{-1} \left[ \tilde{c}^2 - \frac{d+2}{2}\right] \tilde{c}_\sigma \right)_{\text{ND}}
\end{align}
\end{subequations}
in which
\begin{equation}
\label{ref_for_nondim}
	\eta_0 = \rho \, \frac{2 \sqrt{2}}{\sqrt{\beta m}} \, \frac{1}{n \sigma_0}
	\qquad
	\text{and}
	\qquad
	\lambda_0 = \frac{\sqrt{2} \kB}{\sigma_0 \sqrt{\beta m}}.
\end{equation}
Similar expressions can be obtained for $\alpha$ and $\alpha'$, but we don't include them as both coefficients vanish from symmetry in the cases we consider.

\section{Approximation scheme}

It is now incumbent upon us to compute the scalar products of the form $(\tilde{A}, \tilde{L}^{-1} \tilde{B})_{\text{ND}}$ in Eq.~\eqref{transport_from_boltzmann_nondim}. 
This is not entirely straightforward as $\tilde{L}$ acts on an infinite-dimensional space. 
Luckily, we can get a reasonable approximation $[\tilde{L}]$ of $\tilde{L}$ by computing matrix elements on appropriate (truncated) basis functions to reduce the problem to a finite matrix. Traditionally, the basis of functions consists of orthogonal polynomials such as Sonine polynomials (also called associated Laguerre polynomials) or Hermite polynomials~\cite{Kumar1966,Grad1958}.
Here, we will use normalized Hermite polynomials
\begin{equation}
	H_n(x) = \frac{1}{\sqrt{n! \; 2^n \sqrt{\pi}}} \, (-1)^{n} \ee^{x^2} \frac{\dd^n}{\dd x^n} \ee^{-x^2}
\end{equation}
with $n \in \mathbb{N}$. They satisfy
\begin{equation}
	(H_m, H_n)_{\text{1D}} \equiv \int_{-\infty}^{\infty} \ee^{-x^2} H_m(x) H_n(x) \dd x = \delta_{m n}
\end{equation}
and form an orthonormal basis of $L^2(\mathbb{R}, \ee^{x^2} \dd x)$.
We further define the multidimensional Hermite polynomials as products of 1D Hermite polynomials
\begin{equation}
	H_{n_1 \dots n_d}(x_1, \dots, x_d) = H_{n_1}(x_1) \cdots H_{n_d}(x_d)
\end{equation}
and these satisfy
\begin{equation}
	(H_M, H_N) = \delta_{M N}
\end{equation}
in which the inner product is induced by the 1D version, we have collected the indices in a single symbol $N=(n_1 \, n_2 \, \dots n_d)$, and the Kronecker symbol $\delta_{M N}$ is one iff $m_i = n_i$ for all $i$.

To approximate the linearized collision operator, we compute
\begin{equation}
	[L]_{M N} = (H_M, \tilde{L} \, H_N)
\end{equation}
for $n_i, m_i = 0, \dots, n_{\text{max}}$ to obtain a finite matrix $[L]$ with matrix elements $[L]_{M N}$.

We want to compute a scalar product of the form $(A, \tilde{L}^{-1} B)$ in which $A$ and $B$ are functions of $\vec{\tilde{c}}$.
To do so, we also compute
\begin{equation}
	[A]_M = (H_M, A)
	\quad
	\text{and}
	\quad
	[B]_M = (H_M, B).
\end{equation}
Then, diagonalize $[L]$ to project it (as well as $[A]$ and $[B]$) on the orthogonal complement of its nullspace.
 We call $[L]_{\text{red}}$, $[A]_{\text{red}}$, $[B]_{\text{red}}$ the projections (restricted to the range of the projector, i.e. we extract the non-zero block of the block matrix).
As $[L]_{\text{red}}$ is now invertible, we finally compute
\begin{equation}
	(A, \tilde{L}^{-1} B) \approx \braket{[A]_{\text{red}}, [L]_{\text{red}}^{-1} [B]_{\text{red}}}.
\end{equation}
in which $\braket{}$ is the standard scalar product on $\mathbb{C}^{n_{\text{max}}^d}$.

\section{Numerical examples in two dimensions}

\begin{figure}
	\hspace{-0.5cm}\includegraphics{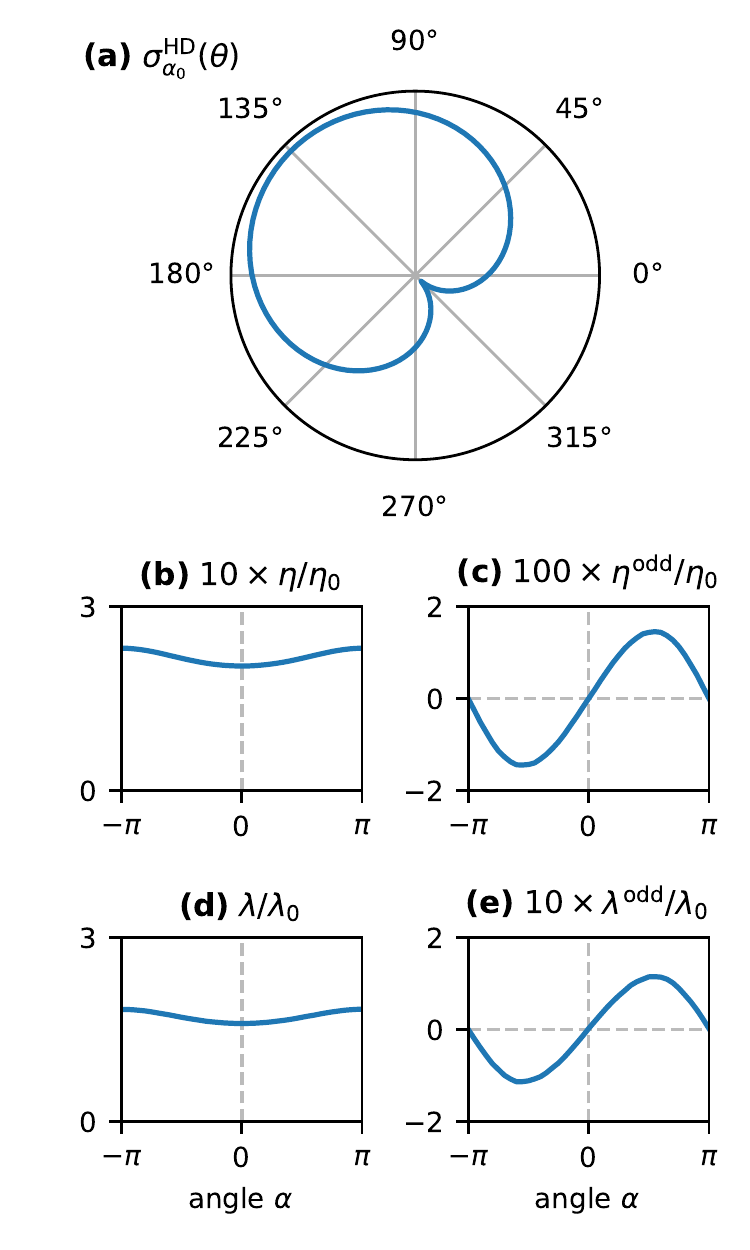}
    \caption{\label{figure_transport_coefficients_hard_disks}
    \textbf{Transport coefficients for a rotated hard disk cross-section.}
    We plot (b) the shear viscosity $\eta$, (c) the odd viscosity $\eta^{\text{o}}$, (d) the (standard) heat conductivity $\lambda$ and (e) the odd heat conductivity (as defined by defined in Eqs.~\eqref{decomposition_lambda} and \eqref{decomposition_eta}) as a function of the angle $\alpha$ determining the chirality of the collisions, for the rotated hard-disk cross-section of Eq.~\eqref{cross_section_HD} represented in (a) for $\alpha = \pi/4$.
    The usual transport coefficients $\lambda$ and $\eta$ are even functions of $\alpha$, while the non-dissipative transport coefficients
    $\eta^{\text{o}}$ and $\lambda^{\text{o}}$ are odd functions of $\alpha$.
    All quantities are nondimensionalized.
    We have used $n_{\text{max}} + 1$ with $n_{\text{max}} = \num{4}$ Hermite polynomials in each Cartesian direction. 
    The integrals are truncated up to $\lvert \tilde{c} \rvert \sim \num{5}$, and are computed using the quasi-Monte-Carlo method Vegas of the Cuba library~\cite{Hahn2005} with $n=\num{500000}$ points.
    }
\end{figure}

\begin{figure}
	\hspace{-0.5cm}\includegraphics{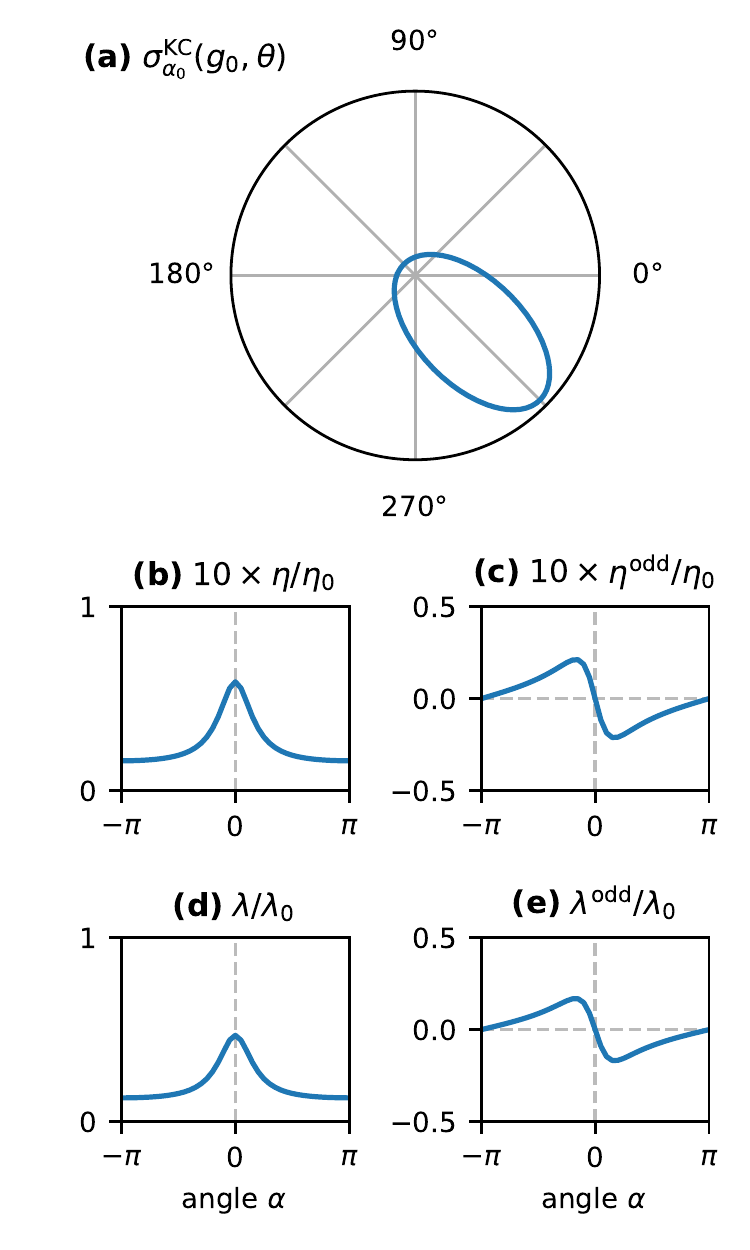}
    \caption{\label{figure_transport_coefficients_regularized_coulomb}
    \textbf{Transport coefficients for a regularized Kepler–Coulomb cross-section.}
    We plot (b) the shear viscosity $\eta$, (c) the odd viscosity $\eta^{\text{o}}$, (d) the (standard) heat conductivity $\lambda$ and (e) the odd heat conductivity (as defined by defined in Eqs.~\eqref{decomposition_lambda} and \eqref{decomposition_eta}) as a function of the angle $\alpha$ determining the chirality of the collisions, for the rotated (regularized) Kepler–Coulomb cross-section of Eq.~\eqref{cross_section_KC} with $\epsilon = \num{0.1}$ represented in (a) for $\alpha = \pi/4$ and $\tilde{g} = 1$.
    The usual transport coefficients $\lambda$ and $\eta$ are even functions of $\alpha$, while the non-dissipative transport coefficients
    $\eta^{\text{o}}$ and $\lambda^{\text{o}}$ are odd functions of $\alpha$.
    All quantities are nondimensionalized.
    See Fig.~\ref{figure_transport_coefficients_hard_disks} for details about the numerical computation.
}
\end{figure}

In this section, we compute numerically the viscosity and heat conductivity tensors for two families of scattering cross-sections in order to illustrate that $\sigma(\theta) \neq \sigma(-\theta)$ indeed leads to odd viscosity and heat conductivity.
To do so, we start with familiar examples of scattering cross-sections~\cite{Friedrich2015}, and rotate them by an angle $\alpha$.
We will therefore consider the modified hard-disk differential cross-section
\begin{equation}
	\label{cross_section_HD}
	\sigma_{\alpha}^{\text{HD}}(g, \theta) = \frac{\sigma_0}{4} \, \left| \sin\left(\frac{\theta + \alpha}{2}\right) \right|.
\end{equation}
as well as the cross-section
\begin{equation}
	\label{cross_section_KC}
	\sigma_{\alpha}^{\text{KC}}(g, \theta) =  \frac{\sigma_0}{\epsilon + \tilde{g}^2 \, \sin^2\left(\dfrac{\theta + \alpha}{2}\right)}
\end{equation}
in which $\tilde{g} = \sqrt{\beta m/2} \, g$.
The cross-section Eq.~\eqref{cross_section_KC} qualitatively describes the scattering by a screened Coulomb interaction, where the non-dimensional parameter $\epsilon$ is inversely proportional to the screening length~\cite{Monceau2002,Everhart1955,Lane1960}. 
Typical cross-sections are plotted in Figs.~\ref{figure_transport_coefficients_hard_disks}a and \ref{figure_transport_coefficients_regularized_coulomb}b for $\alpha = \pi/4$.
In both cases, a nonzero $\alpha$ introduces a chirality in the system, and we recover the standard non-chiral cross-section when $\alpha = 0$.

As a result of this procedure, we obtain the transport coefficients: these are tensors, and it is convenient to decompose them into physically meaningful components. 
In an isotropic system (such as the ones we consider), the heat conductivity tensor can be decomposed as
\begin{equation}
	\label{decomposition_lambda}
	\lambda_{a b} = \lambda \delta_{ab} +  \lambda^{\text{o}} \epsilon_{ab}
\end{equation}
in which $\delta$ is the Kronecker symbol and $\epsilon$ the Levi-Civita symbol.
The coefficient $\lambda$ is the standard heat conductivity, and $\lambda^{\text{o}}$ is the odd heat conductivity (also called thermal Hall conductivity; its existence is also known as the Righi-Leduc effect).
In a system of point particles, there is no bulk viscosity and no antisymmetric stress (see Refs.~\cite{Han2020,Khain2022,Scheibner2020} for discussions on the general case), so the viscosity tensor can be decomposed as
\begin{equation*}
\begin{split}
	\label{decomposition_eta}
	\eta_{abcd} = \eta [\delta_{ac} \delta_{bd} + \delta_{ad} \delta_{bc} - \delta_{ab} \delta_{cd}]
	+ \eta^{\text{o}} [\epsilon_{ac} \delta_{bd}+\epsilon_{bd} \delta_{ac}].
\end{split}
\end{equation*}
The coefficient $\eta$ is the usual shear viscosity, and $\eta^{\text{o}}$ is the odd shear viscosity (or Hall viscosity).

The results are presented in Fig.~\ref{figure_transport_coefficients_hard_disks} for the cross-section of Eq.~\eqref{cross_section_HD}, and in Fig.~\ref{figure_transport_coefficients_regularized_coulomb} for the cross-section of Eq.~\eqref{cross_section_KC}. 
Qualitatively, both cases are similar: the usual transport coefficients $\lambda$ and $\eta$ are even functions of $\alpha$, while the non-dissipative transport coefficients $\eta^{\text{o}}$ and $\lambda^{\text{o}}$ are odd functions of $\alpha$.
We note that in the case of the rotated hard-disk cross-section, the highest value of $\eta^{\text{o}}/\eta$ (or $\lambda^{\text{o}}/\lambda$) is approximately \num[parse-numbers = false]{1/15}, while it is approximately \num[parse-numbers = false]{1/2} for the rotated regularized Kepler–Coulomb cross-section with $\epsilon = \num{0.1}$.

\subsection*{Comparison with the literature}

In this paragraph, we compare our results for $\alpha = 0$ (i.e., for a parity-preserving gas) with values of the viscosity and thermal conductivity known in the literature.

The viscosity and thermal conductivity of a dilute gas of hard disks with diameter $D$ are~\cite{Sengers1969,Gass1971}
\begin{equation}
	\lambda_0^{\text{HD}} = \frac{2}{D} \, \sqrt{\frac{k^3 T}{m \pi}} \, a_1(N)
\end{equation}
and
\begin{equation}
	\eta_0^{\text{HD}} = \frac{1}{2 D} \, \sqrt{\frac{m k T}{\pi}} \, b_0(N)	
\end{equation}
in which $a_1(N) \simeq 1$ and $b_0(N) \simeq 1$ are Sonine polynomial correction factors ($N$ is the order of the Enskog approximation, and $a_1(1) = b_0(1) = 1$).

The corresponding non-dimensional values (obtained using Eq.~\eqref{ref_for_nondim} with $\sigma_0 = 2 D$, $D$ being the diameter of the hard disks) are
\begin{equation}
	\frac{\lambda_0^{\text{HD}}}{\lambda_0} = \frac{2 \sqrt{2}}{\sqrt{\pi}} \simeq \num{1.596}
\end{equation}
and
\begin{equation}
	\frac{\eta_0^{\text{HD}}}{\eta_0} = \frac{1}{2 \sqrt{2 \pi}} \simeq \num{0.199}
\end{equation}
compatible within numerical uncertainty with the values $\lambda/\lambda_0 \simeq \num{1.60}$ and $\eta/\eta_0 \simeq \num{0.20}$ obtained in Fig.~\ref{figure_transport_coefficients_hard_disks}.

\section{Relaxation time approximation}

The relaxation time approximation (RTA) consists in replacing the linearized collision operator $L$ by an approximate version~\cite{Bhatnagar1954}
\begin{equation}
	L^{\text{RTA}} \phi = - \frac{1}{\tau} \phi.
\end{equation}

Here, we introduce a generalization of the RTA designed to reproduce the main features of time-reversal- and parity-violating fluids.
We define the skewed relaxation time approximation (SRTA)
\begin{equation}
	\label{SRTA}
	[L^{\text{SRTA}} \phi](\vec{c}) = - \frac{1}{\tau} \phi(R^{-1} \vec{c}).
\end{equation}
in which $R$ is a rotation matrix.

Let us focus on two-dimensional systems for simplicity, in which we can parameterize rotation matrices
\begin{equation}
	R(\alpha) = \ee^{-\alpha \epsilon} = \begin{pmatrix}
		\cos \alpha & - \sin \alpha \\
		\sin \alpha & \cos \alpha 
	\end{pmatrix}
\end{equation}
by an angle $\alpha$. 
Up to the constant multiplicative factor $-1/\tau$, the operator $L^{\text{SRTA}}$ is a representation of the rotation group $U(1)$ on functions, so we can directly obtain the eigenvalues $-1/\tau_n = -\ee^{-\ii n \alpha}/\tau$ ($n \in \ZZ$) of $L^{\text{SRTA}}$ from the characters of the irreducible representations (which are Fourier modes on the circle). 
In this very simple case, we can obtain the inverse of $L^{\text{SRTA}}$ exactly as
\begin{equation}
 	[(L^{\text{SRTA}})^{-1} \phi](\vec{c}) = - \tau \phi(R \vec{c}).
\end{equation} 
This allows us to compute the transport tensors exactly. Using Eqs.~\eqref{transport_from_boltzmann}, \eqref{QPhis}, \eqref{orthonormalized_basis_kernel}, \eqref{scalar_product}, \eqref{boltzmann_distribution},
we find that
\begin{align}
    \eta_{i j k \ell} 
    &= - \rho \beta m (Q c_i c_j, L^{-1} Q c_k c_\ell) \\
    &= \rho \beta m \tau R_{k k'} R_{\ell \ell'} (Q c_i c_j,  Q c_{k'} c_{\ell'}).
\end{align}
Using the identity
\begin{equation}
\begin{split}
	\int \frac{a}{\pi} \ee^{-a c^2} \left(c_i c_j - \frac{1}{2} c^2 \delta_{i j} \right) \, \left(c_k c_\ell - \frac{1}{2} c^2 \delta_{k \ell} \right) \dd^2 \vec{c} \\
	=  \frac{1}{4 \, a^2} \left( \delta_{i k} \delta_{j \ell} + \delta_{i \ell} \delta_{j k} -  \delta_{i j} \delta_{k \ell} \right)
\end{split}
\end{equation}
we find that
\begin{equation}
	\frac{\eta_{i j k \ell}}{\rho} = \frac{\tau}{\beta m} R_{k k'} R_{\ell \ell'} \left( \delta_{i k'} \delta_{j \ell'} + \delta_{i \ell'} \delta_{j k'} -  \delta_{i j} \delta_{k' \ell'} \right).
\end{equation}
Equivalently,
\begin{equation}
\begin{split}
	\frac{\eta_{i j k \ell}}{\rho} = \frac{\tau}{\beta m} \Big( 
	  &\cos(2 \alpha) \, [\delta_{ik} \delta_{j\ell} + \delta_{i\ell} \delta_{jk} - \delta_{ij} \delta_{k\ell}] \\
	+ &\sin(2 \alpha) \, [\epsilon_{ik} \delta_{j\ell}+\epsilon_{j\ell} \delta_{ik}]
	\Big)
\end{split}
\end{equation}
The viscosity tensor expresses a linear relationship between stress and strain. 
Hence, it can be expressed in matrix form as
\begin{equation}
\label{viscosity_matrix_def}
\begin{pmatrix}
	\sigma_{s_1} \\ \sigma_{s_2}
\end{pmatrix}
= 
\begin{pmatrix} 
	\eta & \eta^{\text{odd}} \\
	-\eta^{\text{odd}} & \eta
\end{pmatrix}
\begin{pmatrix}
	\dot{e}_{s_1} \\ \dot{e}_{s_2}
\end{pmatrix}
\end{equation}
in which the irreducible components of the shear stress and deformation rate are defined in the caption of \mainTextFigureMechanismOddViscosity{}. 
We have assumed that the system is isotropic, and we have kept only the shear block (as other viscosities vanish in the current model). 
We refer to Ref.~\cite{Han2020} for more details on this notation (see also Ref.~\cite{Khain2022} for a 3D system in which full rotation invariance is not assumed).
Here, the viscosity matrix takes the form
\begin{equation}
	\label{viscosity_matrix_srta}
	\eta = 
	\begin{pmatrix} 
	\eta & \eta^{\text{odd}} \\
	-\eta^{\text{odd}} & \eta
	\end{pmatrix}
	= 
	\frac{\rho \tau}{\beta m} \,
	\begin{pmatrix}
		\cos 2 \alpha & \sin 2 \alpha \\
		- \sin 2 \alpha & \cos 2 \alpha
	\end{pmatrix}.
\end{equation}

The same calculation can be performed for the thermal conductivity tensor. Using the identity
\begin{equation}
\begin{split}
	\int \frac{a}{\pi} \ee^{-a c^2} \left(\frac{c^2}{2} - \frac{1}{a} \right)^2 c_i c_j \dd^2 \vec{c}
	=  \frac{1}{4 \, a^3} \delta_{i j}
\end{split}
\end{equation}
we find that (in $d=2$)
\begin{equation}
	\lambda_{i j} = \frac{2 k \tau}{\beta m} \, R_{j j'} \delta_{i j'}
\end{equation}
Following Eq.~\eqref{decomposition_lambda}, we write the thermal conductivity tensor as
\begin{equation}
	\lambda
	=
	\begin{pmatrix}
	\lambda & \lambda^{\text{odd}} \\
	- \lambda^{\text{odd}} & \lambda
	\end{pmatrix}
	=
	\frac{2 k_{\text{B}} \tau}{\beta m}
	\begin{pmatrix}
		\cos \alpha & \sin \alpha \\
		-\sin \alpha & \cos \alpha
	\end{pmatrix}.
\end{equation}

While it captures the essential features of the fluid, the collision operator in Eq.~\eqref{SRTA} is slightly too simple. For any $\alpha \neq 0$, there is a $n$ high enough that the eigenvalue $-1/\tau_n$ has a strictly positive real part (meaning that the system is unstable).
To solve this issue, one can consider multiple relaxation times and rotation matrices with a combination of the form
\begin{equation}
	\label{MSRTA}
	[L^{\text{MSRTA}} \phi](\vec{c}) = - \sum_n \frac{1}{\tau_n} \, \phi(R_n^{-1} \vec{c}).
\end{equation}
This expression is less amenable to exact analytical calculations because the inverse of the linearized collision operator is not easily expressed, but we expect that approximate analytical calculations in the first Sonine polynomial approximation should be enough to capture relevant features. 
Both Eqs.~\eqref{SRTA} and \eqref{MSRTA} can in principle be applied to higher-dimensional (e.g. 3D) systems.

\subsection*{The limit of infinitesimal rotations}

Consider now the case of an infinitesimal rotation (again in 2D for simplicity)
\begin{equation}
	R(\alpha) = 1 - \alpha \epsilon + \mathcal{O}(\alpha^2)
\end{equation}
in which $\epsilon$ is the Levi-Civita symbol. When $\alpha$ is small, we therefore have
\begin{equation}
	\phi(R(\alpha)^{-1} \vec{c}) 
= \phi(\vec{c}) + \alpha \, \frac{\partial \phi}{\partial c_i} \,  \epsilon_{i j} c_j + \mathcal{O}(\alpha^2)
\end{equation}
Going back to Eq.~\eqref{SRTA} in this limit, we find
\begin{equation}
	\label{SRTA_infinitesimal}
	[L^{\text{SRTA}} \phi](\vec{c}) = - \frac{1}{\tau} \phi(\vec{c}) - b^{\text{eff}}_i \; \frac{\partial \phi}{\partial c_i} + \mathcal{O}(\alpha^2)
\end{equation}
in which we have defined $b^{\text{eff}}_i = (\alpha/\tau) \, \epsilon_{i j} c_j$. 
We recognize the Lorentz force
\begin{equation}
	m \vec{b}^{\text{eff}} = q \vec{c} \times \vec{B}
\end{equation}
due to an out-of-plane magnetic field 
\begin{equation}
	\vec{B} = \frac{m}{q} \, \frac{\alpha}{\tau} \, \vec{e}_z
\end{equation}
in which $\vec{e}_z=(0,0,1)^T$ in Cartesian coordinates $(x,y,z)$.
The same term would arise from the Boltzmann equation \eqref{BE} with a constant magnetic field (and only normal parity-preserving collisions).
Upon defining the infinitesimal generator of rotations (\enquote{angular momentum} operator) in velocity space $\hat{\ell}_z$ through
\begin{equation}
	\ii \hat{\ell}_z = c_x \frac{\partial}{\partial c_y} - c_y \frac{\partial}{\partial c_x}
\end{equation}
we also have
\begin{equation}
	[L^{\text{SRTA}} \phi] = - \frac{1}{\tau} \phi(\vec{c}) + \frac{\alpha}{\tau} \, \ii \hat{\ell}_z \phi + \mathcal{O}(\alpha^2)
\end{equation}
Note that $\hat{\ell}_z = - \ii \partial_\theta$ in polar coordinates, and its eigenvalues are $m \in \ZZ$.

\subsection*{Viscosity in the limit of infinitesimal rotations}

In this paragraph, we compute the viscosity of a two-dimensional dilute gas subject to a Lorentz-like force in the relaxation time approximation, following Ref.~\cite{Kaufman1960}.
We refer to \cite{ChapmanCowling,Braginskii1958,Braginskii1965} for a more detailed calculations.
The Lorentz-like force can arise from an external magnetic field, but can also occur from a series expansion of the rotated RTA, as shown in the previous paragraph (see Eq.~\eqref{SRTA_infinitesimal}). 

We start with the Boltzmann equation \eqref{BE} in the relaxation time approximation
\begin{equation}
	\label{BE_RTA_magnetic}
	\frac{\partial f}{\partial t} 
	+ c_\mu \frac{\partial f}{\partial r_\mu}
	+ b_\mu \frac{\partial f}{\partial c_\mu}
	= - \frac{1}{\tau} \left( f - f^\circ \right)
\end{equation}
and consider the force 
\begin{equation}
	b_i(\vec{c}) = \omega_{\text{B}} \epsilon_{i j} c_j
\end{equation}
due to an out-of-plane magnetic field $B$, where $\omega_{\text{B}} = (q/m) B$ ($q$ is the charge) and $\epsilon$ is the Levi-Civita symbol.

Multiplying both sides of Eq.~\eqref{BE_RTA_magnetic} with $U_k U_\ell$ and integrating over velocities $\vec{c}$, we find
\begin{equation}
\begin{split}
	\partial_t P_{k \ell} + \partial_i\left( q_{i k \ell} + v_i P_{k \ell} \right) + P_{i k} \partial_i v_\ell + P_{i \ell} \partial_i v_k \\
	=
	\omega_{\text{B}} \left( \epsilon_{k j} P_{j \ell} + \epsilon_{\ell j} P_{j k} \right)
	- \frac{1}{\tau} \left( P_{k \ell} - p \delta_{k \ell} \right).
	\end{split}
\end{equation}
Here, $\vec{U} = \vec{c} - \vec{v}$, $P_{i j}$ is the pressure tensor defined by Eq.~\eqref{pressuredef}, $p \delta_{k \ell}$ is its equilibrium value given by Eq.~\eqref{ideal_gas_law}, and $q_{i j k} = \rho \, \Braket{U_i U_j U_k}$. 
The trace of this equation expresses the conservation of energy (see Eq.~\eqref{momentum_energy_densities}). 
After removing the trace, the time derivative, and keeping only first order quantities in deviations from equilibrium and gradients, we get
\begin{equation*}
\begin{split}
	p \left( 
	\partial_k u_\ell + \partial_\ell u_k - \tfrac{2}{d} (\partial_i u_i) \delta_{k \ell}
	\right) \\
	= \omega_{\text{B}} \left( \epsilon_{k m} P^{(1)}_{m \ell} + \epsilon_{\ell m} P^{(1)}_{m k} \right)
	- \frac{1}{\tau} \left[ P^{(1)}_{k \ell} - \tfrac{1}{d} P^{(1)}_{i i} \delta_{k \ell} \right].
\end{split}
\end{equation*}
Projecting on the shear degrees of freedom, we obtain 
\begin{equation}
\label{viscosity_magnetic_field}
\begin{pmatrix}
	P_{s_1} \\ P_{s_2}
\end{pmatrix}
= 
- 
\frac{p \tau}{1+ 4 \tau^2 \omega_{\text{B}}^2}
\begin{pmatrix}
	1 &  2 \omega_{\text{B}} \tau \\
  - 2 \omega_{\text{B}} \tau & 1
\end{pmatrix}
\begin{pmatrix}
	\dot{e}_{s_1} \\ \dot{e}_{s_2}
\end{pmatrix}
\end{equation}
using the notation defined in Eq.~\eqref{viscosity_matrix_def} (the pressure tensor is the opposite of the stress tensor).
When $\omega_{\text{B}} \to 0$, we recover that the shear viscosity is $\eta(\omega_{\text{B}} = 0) = p \tau$.
Taking $\omega_{\text{B}} = \alpha/\tau$ to match Eq.~\eqref{SRTA_infinitesimal}, we find
\begin{equation}
\label{viscosity_magnetic_field_again}
\eta
= 
\begin{pmatrix} 
	\eta & \eta^{\text{odd}} \\
	-\eta^{\text{odd}} & \eta
\end{pmatrix}
=
\frac{p \tau}{1+ 4 \alpha^2}
\begin{pmatrix}
	1 & 2 \alpha \\
  - 2 \alpha & 1
\end{pmatrix}
\end{equation}
which is consistent with Eq.~\eqref{viscosity_matrix_srta} at first order in $\alpha$.

\medskip
\appendix

\medskip
\begin{center}
{\bfseries APPENDICES}
\end{center}

\let\theequation\oldtheequation

\section{The ambiguity of transport coefficients in bulk hydrodynamic equations}
\label{bulk_ambiguity}

In this Appendix, we discuss the issue of the uniqueness of the stress tensor and its consequence on the viscosity tensor.
The Navier-Stokes equations only contain the divergence $f_i = \partial_j \sigma_{i j}$ of the stress tensor.
As a consequence, we can add any divergence free tensor to the stress without changing the force, and hence without changing the bulk equations of motion.
Namely, the replacement $\sigma_{i j} \to \sigma_{i j} + \delta \sigma_{i j}$ with any $\delta \sigma_{i j}$ satisfying $\partial_j \delta \sigma_{i j} = 0$ (say, $\delta \sigma_{i j} = \epsilon_{j k} \partial_k \chi_i$ where $\chi_i$ is an arbitrary function) does change the contribution $f_i$ to the Navier-Stokes equations, because $\partial_j \delta \sigma_{i j} = \epsilon_{j k} \partial_j \partial_k \chi_i = 0$. Nevertheless, the stress tensor itself \emph{does} change. 
From the point of view of the bulk theory, this divergence-free part is akin to gauge degrees of freedom: $\sigma$ is defined modulo transformations $\sigma \to \sigma + \delta \sigma$ with any $\delta \sigma$ satisfying $\text{Div}(\delta \sigma) = 0$, in the same way as the vector potential $A$ (related to the magnetic field $B = \text{curl}(A)$) is defined up to the addition of a term $\delta A$ such that $\text{curl}(\delta A) = 0$.
In summary, from the point of view of bulk hydrodynamics, the stress tensor is not an observable, because it is not gauge invariant.

This ambiguity can be traced back to microscopic considerations (see Ref.~\cite{Goldhirsch2010} and references therein). 
The stress tensor is usually expressed from microscopic quantities through the Irving-Kirkwood formula~\cite{Irving1950,Schofield1982,Goldhirsch2010}
\begin{equation}
\label{IK}
\begin{split}
	\sigma_{\alpha \beta}(t, r) = 
	&- \sum_i m_i v_i^\alpha v_i^\beta \delta(r - r_i)
	\\
	&- \frac{1}{2} \sum_{i \neq j} f_{i j}^{\alpha} r_{i j}^{\beta} \int_{0}^{1} \dd s \delta(r - r_i + s r_{i j}).
\end{split}
\end{equation}
The kinetic part of the stress (first term) is not problematic: the velocities of classical particles can in principle be measured; in practice, this is even relatively easy in current active matter experiments based, for instance, on colloidal systems.  
The virial part (second term) is however ambiguous. 
The integral over a straight line connecting $r_i$ and $r_j$ in Eq.~\eqref{IK} could be replaced with an arbitrary curve $C_{i j}$ having the same end-points without affecting the equations of motion.
However, this replacements produces a divergence-free contribution to the stress tensor~\cite{Schofield1982}. 
Several arguments have been proposed in favour of (and against) the uniqueness of a microscopic expression of stress tensor (see Refs.~\cite{Wajnryb1995,Goldhirsch2010,Chen2018,Admal2011,TorresSanchez2016,Shi2021} and references therein)~\footnote{
Note that there are also troubles in defining a unique stress-energy tensor in field theory~\cite{Tichy1998,Forger2004}.}. 
However, there is no apparent consensus in the literature, and it is not clear to what extent these arguments apply in general (e.g., out of equilibrium or when usual symmetries are broken). 

The ambiguity in defining the stress tensor means that the viscosity tensor can in general not be unambiguously determined from the bulk hydrodynamic equations~\cite{Rao2020,Rao2021}.
Assuming that $\eta_{i j k \ell}$ is uniform, the viscous stress $\sigma_{i j}^\text{vis} = \eta_{i j k \ell} \partial_\ell v_k$  contributes the term
\begin{equation}
	f_i = \eta_{i j k \ell} \partial_j \partial_\ell v_k
\end{equation}
in the Navier-Stokes equation.
By symmetry of the second derivatives, the same force is obtained after the replacement
\begin{equation}
	\eta_{i j k \ell} \to \eta_{i j k \ell} + \delta \eta_{i j k \ell}
\end{equation}
in which 
\begin{equation}
	\label{delta_eta}
	\delta \eta_{i j k \ell} = a_{i k} \epsilon_{j \ell}
\end{equation}
is a \enquote{pure gauge} contribution to the viscosity. 
Here, $a_{i k}$ is an arbitrary rank-2 tensor, and $\epsilon_{j \ell} = - \epsilon_{\ell j}$ is the Levi-Civita tensor.
Let us parameterize $a_{i k}$ as
\begin{equation}
	a = \begin{pmatrix}
		a_1 + a_2 & a_3 + a_4 \\
		a_3 - a_4 & a_1 - a_2
	\end{pmatrix}
\end{equation}
so that the pure gauge viscosity represented as a matrix is
\begin{equation}
	\delta \eta = \begin{pmatrix}
	a_{4} & -a_{1} & -a_{3} & a_{2} \\
	 a_{1} & a_{4} & a_{2} & a_{3} \\
	 a_{3} & -a_{2} & -a_{4} & a_{1} \\
	 -a_{2} & -a_{3} & -a_{1} & -a_{4} \\
	\end{pmatrix}
\end{equation}
Note that this disappears when the stress tensor is symmetric, as $a_1 = a_2 = a_3 = a_4 = 0$.

A similar issue occurs in the heat conductivity tensor: the Hall thermal conductivity completely vanishes from the bulk heat equation.
The corresponding ambiguity in the particle diffusion tensor is reviewed and analysed in Ref.~\cite{Hargus2021}.

\section{Symmetry of the transport tensors}
\label{symmetries_tensors}

In this Appendix, we discuss the constraints put by rotation invariance on transport coefficients.
Under a transformation $R \in O(d)$ [$d$ is the space dimension], the transport coefficients transform as tensors, namely
\begin{subequations}
\label{transformation_tensors}
\begin{align}
	\lambda_{i j} &\to R_{i i'} R_{j j'} \; \lambda_{i' j'} \\
	\alpha_{i j k} &\to R_{i i'} R_{j j'} R_{k k'} \; \alpha_{i' j' k'} \\
	\alpha_{i j k} &\to R_{i i'} R_{j j'} R_{k k'} \; \alpha'_{i' j' k'} \\
	\eta_{i j k \ell} &\to R_{i i'} R_{j j'} R_{k k'} R_{\ell \ell'} \; \eta_{i' j' k' \ell'}
\end{align}
\end{subequations}

Let us assume that these tensors are invariant under all rotations $R \in SO(2)$ (in $d=2$), namely, impose that the two sides of the arrows in Eq.~\eqref{transformation_tensors} are equal.
We find by a direct calculation that the most general tensors compatible with this symmetry are
\begin{subequations}
\begin{align}
	\lambda_{i j} &= \lambda \, \delta_{i j} + \lambda^{\text{odd}} \, \epsilon_{i j} \\
	\alpha_{i j k} &= 0 \\
	\alpha'_{i j k} &= 0 \\
	\eta_{abcd} &= \;\zeta \, \delta_{ab} \delta_{cd} -  \eta^\text{A} \epsilon_{ab} \delta_{cd} - \eta^\text{B} \delta_{ab} \epsilon_{cd}  + \eta^\text{R} \epsilon_{ab} \epsilon_{cd} \notag \\
     & \;\;+ \eta \, (\delta_{ac} \delta_{bd} + \delta_{ad} \delta_{bc} - \delta_{ab} \delta_{cd}) \\
     & \;\;+ \eta^\text{o} (\epsilon_{ac} \delta_{bd}+\epsilon_{bd} \delta_{ac}) \notag
\end{align}
\end{subequations}
in which $\delta_{i j}$ is the Kronecker symbol and $\epsilon_{i j}$ is the Levi-Civita symbol.

\section{Chapman-Enskog procedure}
\label{chapman_enskog}

In this Appendix, we apply the Chapman-Enskog method to an odd ideal gas.
In this method, the stress is directly computed (instead of its divergence). 
Hence, it is possible to obtain some more information from the microscopic system than available at the bulk hydrodynamic level (see Appendix~\ref{bulk_ambiguity}).

\subsection{Decomposition of the Boltzmann equation}

The distribution function is decomposed into successive orders
\begin{equation}
	\label{expansion_f_orders}
	f = \frac{1}{\varepsilon} \left[ f^{(0)} + \varepsilon f^{(1)} + \varepsilon^2 f^{(2)} + \cdots \right]
\end{equation}
and the time derivative in the Boltzmann equation is also decomposed as 
\begin{equation}
	\partial_t = \partial_t^{(0)} + \varepsilon \partial_t^{(1)} + \cdots.
\end{equation}
This can be seen as a poor man's renormalization scheme: we are trying to reduce the full kinetic theory to simplified theory describing the evolution of slow variables (see Refs.~\cite{Kunihiro2006,Hatta2002} for more details).

To use this decomposition in the Boltzmann equation
\begin{equation}
	\mathscr{D}(f) = \mathscr{C}(f)
\end{equation}
let us first note that the collision term is bilinear in $f$, and we should more appropriately write
\begin{equation}
	\mathscr{C}(f) = \mathscr{C}(f, f)
\end{equation}
in which
\begin{equation}
\label{collision_bilinear}
\begin{split}
	\mathscr{C}(f, g) \equiv \iiint \big[
	 &W(\vec{c_1'}, \vec{c_2'} \mid \vec{c_1}, \vec{c_2}) f_1' g_2' \\
	-&W(\vec{c_1}, \vec{c_2} \mid \vec{c_1'}, \vec{c_2'}) f_1 g_2
	\big] \dd \vec{c_2} \, \dd \vec{c_1'} \, \dd \vec{c_2'}.
\end{split}
\end{equation}
We then stipulate that the Boltzmann equation must hold order by order.
This yields at order $\varepsilon^0$
\begin{equation}
	\mathscr{C}(f^{(0)}, f^{(0)}) = 0.
\end{equation}
The solution to this equation is the Boltzmann distribution Eq.~\eqref{boltzmann_distribution}.

Let us immediately compute some derivatives that will be required momentarily:
\begin{equation}
	\frac{\partial f^\circ}{\partial n} = \frac{1}{n} f^\circ
\end{equation}

\begin{equation}
	\frac{\partial f^\circ}{\partial v_\mu} = \beta m (c_\mu - v_\mu) f^\circ = - \frac{\partial f^\circ}{\partial c_\mu}
\end{equation}

\begin{equation}
	\frac{\partial f^\circ}{\partial T} 
	= \frac{\partial \beta}{\partial T} \frac{\partial f^\circ}{\partial \beta}
	= - k \beta \, \frac{d - \beta m (c - v)^2 }{2} f^\circ
\end{equation}

We now wish to determine the distribution function at first order.
The first-order distribution function is constrained so that the meanings of $n$, $\vec{v}$ and $T$ do not change. 
Namely, we assume
\begin{equation}
	\label{conditions_f1_keep_meaning_hydro}
	\int f^{(1)} = 0
	\quad
	\int f^{(1)} U_\mu = 0
	\quad
	\int f^{(1)} U^2 = 0
\end{equation}
in which we have introduced the peculiar velocity
\begin{equation}
	\vec{U} = \vec{c} - \vec{v}.
\end{equation}

At order $\varepsilon^1$ the Boltzmann equation gives
\begin{equation}
	\label{linearized_boltzmann_first_order}
	\mathscr{D}^{(0)}(f^{(0)}) = \mathscr{L}^{(0)}(f^{(1)})
\end{equation}
in which we have defined \begin{equation}
	\label{linearized_boltzmann_L0}
	\mathscr{L}^{(0)}(f^{(1)}) = \mathscr{C}(f^{(0)}, f^{(1)}) + \mathscr{C}(f^{(1)}, f^{(0)})
\end{equation}
as well as
\begin{equation}
	\mathscr{D}^{(0)}(f^{(0)}) \equiv \partial_t^{(0)} f^{(0)} + c_\mu \frac{\partial f^{(0)}}{\partial r_\mu} + b_\mu \frac{\partial f^{(0)}}{\partial c_\mu}.
\end{equation}
Using the notations of the main text, $f^{(1)} = f^{(0)} \phi$, and we also define the linearized collision operator $L$ by
\begin{equation}
	L \phi = \frac{1}{f^{(0)}} \, \mathscr{L}^{(0)}(f^{(0)} \phi).
\end{equation}

In the Chapman-Enskog procedure, we assume that $f$ depends on $t$ and $\vec{r}$ only through the local quantities $n(t,\vec{r})$, $\vec{v}(t,\vec{r})$, $T(t,\vec{r})$. Namely, we replace
\begin{equation}
	f(t, \vec{r}, \vec{c}) \to f(t, n(t, \vec{r}), \vec{v}(t, \vec{r}), T(t, \vec{r}))
\end{equation}

Correspondingly, we use the replacements
\begin{equation}
	\frac{\partial f}{\partial t} \to 
	\frac{\partial f}{\partial n} \, \frac{n}{\partial t} +
	\frac{\partial f}{\partial v_\mu} \, \frac{v_\mu}{\partial t} +
	\frac{\partial f}{\partial T} \, \frac{T}{\partial t}
\end{equation}
and
\begin{equation}
	\frac{\partial f}{\partial r_\mu} \to 
	\frac{\partial f}{\partial n} \, \frac{n}{\partial r_\mu} +
	\frac{\partial f}{\partial v_\mu} \, \frac{v_\mu}{\partial r_\mu} +
	\frac{\partial f}{\partial T} \, \frac{T}{\partial r_\mu}.
\end{equation}
The spatial derivatives of the fields $(n,\vec{v}, T)$ are left untouched; they are simply gradients of our physical fields.
Their time derivatives are evaluated using the zeroth-order balance equations of motion Eqs.~\eqref{conservation_eqs} (in which we use the zeroth-order quantities are given by Eqs.~(\ref{ideal_gas_law},\ref{q_zeroth_order},\ref{ekin_zeroth_order})).

We obtain
\begin{equation}
\label{linearized_boltzmann_D0}
\begin{split}
	\mathscr{D}^{(0)}(f^{(0)}) = f^{(0)} \Bigg[ \frac{m}{k T} \left( U_i U_j - \frac{1}{d} U^2 \delta_{i j} \right) \frac{\partial v_j}{\partial r_i} \\
	+ \frac{1}{T} \left( \frac{m}{2 k T} U^2 - \frac{d+2}{2}\right) U_i \, \frac{\partial T}{\partial r_i}
	 \Bigg]
\end{split}
\end{equation}
in which we have assumed $n_{\text{dof}} = d$ as we do not explicitly describe any internal degrees of freedom.

The linearized Boltzmann equation then becomes
\begin{equation}
	\label{general_form_linearized_boltzmann_D0}
	\bigg[ 
	\frac{\partial n}{\partial r_\mu} \, Y^{n}_{\mu}(\vec{U})
	+
	\frac{\partial v_\nu}{\partial r_\mu} \, Y^{v_\nu}_{\mu}(\vec{U})
	+
	\frac{\partial T}{\partial r_\mu} \, Y^{T}_{\mu}(\vec{U})
	\bigg] = L \phi
\end{equation}
in which we have defined
\begin{subequations}
\begin{align}
Y_i^{v_j}(\vec{U}) &= \frac{m}{k T} \left( U_i U_j - \frac{1}{d} U^2 \delta_{i j} \right) \\ Y_i^{T}(\vec{U}) &= \frac{1}{T} \left( \frac{m}{2 k T} U^2 - \frac{d+2}{2}\right) U_i \\ Y^{n}(\vec{U}) &= 0.
\end{align}
\end{subequations}
Equation~\eqref{general_form_linearized_boltzmann_D0} has the general form
\begin{equation}
	\label{general_form_LBE_L}
	\sum_{I} G^I \, \ket{Y^I_{\mu}} = L \ket{\phi}
\end{equation}
in which $G^I$ represent the gradients, and kets $\ket{\phi}$ represent functions $U \mapsto \phi(\vec{U})$ of the peculiar velocity $\vec{U}$.
We seek an expression of the distribution function in terms of the gradients of the form
\begin{equation}
	\label{expansion_phi_X}
	\ket{\phi} = \sum_{I} \left[ \frac{\partial \phi^I}{\partial r_\mu} \right] \, \ket{X^I_{\mu}}
\end{equation}
If $L$ was invertible, this would imply
\begin{equation}
	\label{LX_Y}
	L \ket{X^I_{\mu}} = \ket{Y^I_{\mu}}.
\end{equation}
as the gradients are arbitrary, and therefore
\begin{equation}
	\label{bad_X_LinvY}
	\ket{X^I_{\mu}} = L^{-1} \ket{Y^I_{\mu}}.
\end{equation}
Unfortunately, $L$ is not invertible. 
However, it is still possible to obtain a solution using Fredholm theory under certain hypotheses on $Y^I_{\mu}$, see for instance Ref.~\cite{Groetsch1993}. In practice, the solution consists in working in the orthogonal complement to the nullspace of $L$.

\subsection{General expression of transport coefficients}

The expansion Eq.~\eqref{expansion_f_orders} of the distribution function induces a similar expansion of everything, including the pressure tensor
\begin{equation}
	\label{pressure_tensor}
	P_{\mu \nu} = m \int f U_\mu U_\nu \dd^d \vec{c}.
\end{equation}
It is expanded as
\begin{equation}
	P_{\mu \nu} = P_{\mu \nu}^{(0)} + P_{\mu \nu}^{(1)} + \dots
\end{equation}
that are defined by replacing $f$ by $f^{(n)}$ in the equation above (the expansion parameter is already set to one).
Here, $P^{(0)}$ is the equilibrium/steady-state pressure tensor, that we have already computed in Eq.~\eqref{ideal_gas_law}, while the viscous response should be contained in
\begin{equation}
	P^{(1)}_{\mu \nu} = m \int f^{(0)} \phi U_\mu U_\nu \dd^d \vec{U}.
\end{equation}

Starting from Eq.~\eqref{expansion_phi_X}, we obtain
\begin{equation}
	P^{(1)}_{\mu \nu} = - \eta_{\mu \nu \rho \sigma} \, \frac{\partial v_\rho}{\partial r_\sigma} - \alpha_{\mu \nu \rho} \frac{\partial T}{\partial r_\rho}
\end{equation}
in which we have defined the viscosity tensor
\begin{equation}
	\eta_{\mu \nu \rho \sigma} = \left[ - m \int f^{(0)}(\vec{U}) U_\mu U_\nu X_{\sigma}^{v_\rho}(\vec{U}) \dd^d \vec{U} \right].
\end{equation}
This can be written as
\begin{equation}
	\eta_{\mu \nu \rho \sigma} = - m \; (B_{\mu \nu}, X_{\sigma}^{v_\rho})
\end{equation}
in which we have defined $B_{\mu \nu}(\vec{U}) = U_\mu U_\nu$.
The coefficient $\alpha$ is defined in a similar way as
\begin{equation}
	\alpha_{\mu \nu \sigma} = - m \; (B_{\mu \nu}, X_{\sigma}^{T}).
\end{equation}

In the same way, we compute the first-order heat flux from \eqref{eqdef}
\begin{equation}
	Q^{(1)} = - \lambda_{\mu \sigma} \partial_\sigma T  - \alpha'_{\mu \rho \sigma} \partial_\sigma v_\rho 
\end{equation}
in which the heat conductivity is
\begin{equation}
	\lambda_{\mu \sigma} = - \frac{m}{2} \; (A_{\mu}, X_{\sigma}^{T})
\end{equation}
with $A_{\mu}(\vec{U}) = U^2 U_\mu$ and
\begin{equation}
	\alpha'_{\mu \rho \sigma} = - \frac{m}{2} \; (A_{\mu}, X_{\sigma}^{v_\rho}).
\end{equation}

\section{Symmetries of the collision operator}
\label{symmetries_collisions}

In this Appendix, we discuss the symmetries of the collision operator. 
For the most part, we follow Ref.~\cite{Waldmann1958}.

\subsection{Symmetries and conservation laws}

For classical (distinguishable) particles, the function $W$ must satisfy~\footnote{
If the particles were indistinguishable (quantum), then we would also have
\begin{equation*}
	\!\!W(\vec{c_1},\vec{c_2} | \vec{c_1'},\vec{c_2'}) = W(\vec{c_2},\vec{c_1} | \vec{c_1'},\vec{c_2'})	= W(\vec{c_1},\vec{c_2} | \vec{c_2'},\vec{c_1'}) \end{equation*}
but we will not consider this case here.
}
\begin{equation}
	\label{exchange_symmetry}
	W(\vec{c_1},\vec{c_2} | \vec{c_1'},\vec{c_2'}) = W(\vec{c_2},\vec{c_1} | \vec{c_2'},\vec{c_1'})	
\end{equation}

Note that a change of variable $\vec{c_1'} \leftrightarrow \vec{c_2'}$ in the collision integral Eq.~\eqref{collision_integral} is equivalent to the replacement
\begin{equation}
	W(\vec{c_1},\vec{c_2} | \vec{c_1'},\vec{c_2'}) \to W(\vec{c_1},\vec{c_2} | \vec{c_2'},\vec{c_1'})
\end{equation}
in Eq.~\eqref{collision_integral}. 
Hence, as far as we are only interested in $\mathscr{C}(f)$, we can assume that $W$ has an additional exchange symmetry (it can also be combined with Eq.~\eqref{exchange_symmetry} to permute $\vec{c_1}$ with $\vec{c_2}$ without changing the primed velocities).

Galilean invariance imposes
\begin{equation}
	\label{galilean_invariance}
	W(\vec{c_1}, \vec{c_2} | \vec{c_1'}, \vec{c_2'}) = W(\vec{c_1} - \vec{c}, \vec{c_2} - \vec{c} | \vec{c_1'} - \vec{c}, \vec{c_2'} - \vec{c})
\end{equation}

Time-reversal changes the order of the collision, and also inverts the velocities ($\vec{c} \to -\vec{c}$), so it requires
\begin{equation}
	W(\vec{c_1}, \vec{c_2} | \vec{c_1'}, \vec{c_2'}) = W(-\vec{c_1'}, -\vec{c_2'} | -\vec{c_1}, -\vec{c_2})
\end{equation}
Spatial inversion ($\vec{c} \to -\vec{c}$) leads to 
\begin{equation}
	W(\vec{c_1}, \vec{c_2} | \vec{c_1'}, \vec{c_2'}) = W(-\vec{c_1}, -\vec{c_2} | -\vec{c_1'}, -\vec{c_2'})
\end{equation}
When it is combined with time-reversal, we obtain space-time inversion, which imposes 
\begin{equation}
	W(\vec{c_1}, \vec{c_2} | \vec{c_1'}, \vec{c_2'}) = W(\vec{c_1'}, \vec{c_2'} | \vec{c_1}, \vec{c_2})
\end{equation}

Conservation of linear momentum leads to
\begin{align}
	W \propto \delta^{(d)}\left( \tfrac{1}{2} (\vec{c_1} + \vec{c_2}) - \tfrac{1}{2} (\vec{c_1'} + \vec{c_2'}) \right) 
	\label{momentum_conservation_W}
\end{align}
while conservation of energy leads to
\begin{align}
	W \propto \delta^{(1)}\left(\tfrac{1}{2} (g^2 - g'^{\,2})\right)
	\label{energy_conservation_W}
\end{align}

\subsection{Collisional invariants}

Each conserved quantity corresponds to a collisional invariant, defined as a quantity $\chi(\vec{c})$ conserved by collisions as
\begin{equation}
	 \int \chi(\vec{c_1}) \mathscr{C}(f_1) \dd \vec{c_1} = 0
\end{equation}
Using the Boltzmann equation $\mathscr{D}(f_1) = \mathscr{C}(f_1)$, this leads to a conservation law of the form
\begin{equation}
	 \int \chi(\vec{c_1}) \mathscr{D}(f_1) \dd \vec{c_1} = 0
\end{equation}
that can be remixed into the usual local conservation equations.

First, consider conservation of mass:
\begin{equation}
\begin{split}
&\int \mathscr{C}(f_1) \dd \vec{c_1} \\
=
&\begin{aligned}[t]
	\int \big[
	 &W(\vec{c_1'}, \vec{c_2'} \mid \vec{c_1}, \vec{c_2}) f_1' f_2' \\
	-&W(\vec{c_1}, \vec{c_2} \mid \vec{c_1'}, \vec{c_2'}) f_1 f_2
	\big] \dd \vec{c_1} \dd \vec{c_2} \, \dd \vec{c_1'} \, \dd \vec{c_2'} = 0
\end{aligned}
\end{split}
\end{equation}
by a change of variables $(\vec{c_1}, \vec{c_2}) \leftrightarrow (\vec{c_1'}, \vec{c_2'})$.

Now, let us move on to the first moment
\begin{equation}
	\int \vec{c_1} \mathscr{C}(f_1) \dd \vec{c_1}.
\end{equation}
In this case, we will see that the RHS vanishes under the assumption of momentum conservation.
Indeed, a few changes of variables ($\vec{c_1} \leftrightarrow \vec{c_2}$ and $\vec{c_1'} \leftrightarrow \vec{c_2'}$) as well as the symmetry Eq.~\eqref{exchange_symmetry} always allow to rewrite
\begin{equation}
	\int \vec{c_1} \mathscr{C}(f_1) \dd \vec{c_1}
	=
	\int \frac{1}{2} \left( \vec{c_1} + \vec{c_2} \right) \mathscr{C}(f_1) \dd \vec{c_1}
\end{equation}
Using momentum conservation Eq.~\eqref{momentum_conservation_W} allows to proceed further and write
\begin{equation}
\begin{split}
	&\int \frac{1}{2} \left( \vec{c_1} + \vec{c_2} \right) \mathscr{C}(f_1) \dd \vec{c_1} \\
	= &
	\begin{aligned}[t]
	\int \big[
	 &W(\vec{c_1'}, \vec{c_2'} \mid \vec{c_1}, \vec{c_2}) \, f_1' f_2' \, \frac{1}{2} \left( \vec{c_1'} + \vec{c_2'} \right) \\
	-&W(\vec{c_1}, \vec{c_2} \mid \vec{c_1'}, \vec{c_2'}) \, f_1 f_2  \,  \frac{1}{2} \left( \vec{c_1} + \vec{c_2} \right)
	\big] \dd \vec{c_1} \dd \vec{c_2} \, \dd \vec{c_1'} \, \dd \vec{c_2'}
	\end{aligned}
	 \\
	= & \, 0
	\raisetag{2.15cm}
\end{split}
\end{equation}
in which the last equality comes again from a change of variables $(\vec{c_1}, \vec{c_2}) \leftrightarrow (\vec{c_1'}, \vec{c_2'})$ in one of the integrals.

Finally, conservation of energy Eq.~\eqref{energy_conservation_W} allows to show 
\begin{equation}
	\int c_1^2 \mathscr{C}(f_1) \dd \vec{c_1} = 0.
\end{equation}

\subsection{Simplified collision operator}

\def\ttilde#1{\ensuremath{\tilde{\raisebox{0pt}[0.85\height]{$\tilde{#1}$}}}}
\def\tttilde#1{\ensuremath{\tilde{\raisebox{0pt}[0.85\height]{$\ttilde{#1}$}}}}

We now derived the simplified collision operator used in the main text, under the assumption of mass, linear momentum, and energy conservation combined with Galilean invariance and rotation invariance (defined as invariance under $SO(d)$).
Because of Galilean invariance, $W$ only depends on three variables, and we can choose to define~\footnote{Set $c = c_1$ so that $W$ does depend only on the three variables $a_1 = c_2 - c_1$, $a_2 = \vec{c_1'} - c_1$ and $a_3 = \vec{c_2'} - c_1$.
We can apply a linear transformation to get three other independent variables $b_1 = a_1$, $b_2 = a_3 - a_2$ and $b_3 = a_1 - a_2 - a_3$ which gives the result.
}
\begin{equation}
	\label{galilean_invariance_reduction}
	W(\vec{c_1}, \vec{c_2} | \vec{c_1'}, \vec{c_2'}) = \tilde{W}(\vec{c_2} - \vec{c_1}, \vec{c_2'} - \vec{c_1'}, (\vec{c_1} + \vec{c_2}) - (\vec{c_1'} + \vec{c_2'}))
\end{equation}
(The choice of these variables will be convenient later when imposing conservation laws.)

Using conservation of momentum, we note that the third argument of $\tilde{W}$ is constrained to vanish, and further write
\begin{equation}
\label{conservation_momentum_Wtt}
\begin{split}
	\tilde{W}(\vec{g}, \vec{g'}, \vec{c_1} + \vec{c_2} - (\vec{c_1'} + \vec{c_2'})) \\
	= \ttilde{W}(\vec{g}, \vec{g'}) \;  \delta^{(d)}\left( \frac{(\vec{c_1} + \vec{c_2}) - (\vec{c_1'} + \vec{c_2'})}{2} \right)
\end{split}
\end{equation}

\let\para\sslash

We can now simplify the collision term under these assumptions by writing
\begin{equation}
\begin{split}
	\mathscr{C}(f) = &
	\int \big[
	 W(\vec{c_1'}, \vec{c_2'} \mid \vec{c_1}, \vec{c_2}) f_1' f_2' \\
	&-W(\vec{c_1}, \vec{c_2} \mid \vec{c_1'}, \vec{c_2'}) f_1 f_2
	\big] \dd^d \vec{c_2} \, \dd^d \vec{c_1'} \, \dd^d \vec{c_2'} \\
	=
	&\int \big[
	  \ttilde{W}(\vec{g}', \vec{g}) f_1' f_2' 
	- \ttilde{W}(\vec{g}, \vec{g'}) f_1 f_2
	\big] \dd^d \vec{c_2} \, \dd^d \vec{g'} \\
\end{split}
\end{equation}
in which we have used Galilean invariance Eq.~\eqref{galilean_invariance} and momentum conservation Eq.~\eqref{conservation_momentum_Wtt}.
Here, $\vec{c_1}'$ and $\vec{c_1}'$ are now function of the other variables through $\vec{c_{2(1)}}' = (\pm \vec{g'} + \vec{c_1} + \vec{c_2})/2$, etc.

Further simplifications from rotation invariance depend on the dimension, as we will now see.
In any case, we want to use the invariance
\begin{equation}
	\ttilde{W}(\vec{g}', \vec{g}) = \ttilde{W}(R \vec{g}', R \vec{g})
\end{equation}
under $R \in SO(d)$.
This reflects the invariance under rotation of the collision (not the fluid).
In principle, we need to compute the quotient of $\mathbb{R}^{d} \times \mathbb{R}^d$ by the equivalence relation $(\vec{g'}, \vec{g}) \sim (R \vec{g'}, R \vec{g})$ ($SO(d)$ acting diagonally), and consider functions on this quotient.
In practice, however, we only need to compute functions $\ttilde{W}(\vec{g}', \vec{g})$ that are invariant under the action of $SO(d)$.
This is at least doable if we focus on polynomial invariants: in this case, we only need to know fundamental invariants $I^\text{SO}_1, \dots, I^\text{SO}_n$ that form a Hilbert basis of the ring of invariant (these are polynomials $I^\text{SO}_n(g_1, \dots, g_d, g'_1, \dots, g'_d)$ of the components of the vectors).
We can then write
\begin{equation}
	\ttilde{W}(\vec{g}', \vec{g}) = \tttilde{W}(I^\text{SO}_1, \dots, I^\text{SO}_n)
\end{equation}

As energy is a scalar, it is invariant under rotation. Hence (as we shall see explicitly), we will always be able to choose $I^\text{SO}_1 = g'^2/2$ and $I^\text{SO}_2 = g^2/2$ and use conservation of energy to write
\begin{equation}
	\tttilde{W}(I^\text{SO}_1, \dots, I^\text{SO}_n) = \tilde{\sigma}(I^\text{SO}_2, I^\text{SO}_3, \dots) \; \delta^{(1)}\left( I^\text{SO}_1 - I^\text{SO}_2 \right).
\end{equation}
Below, we will use $I^\text{SO}_1 = g$ and $I^\text{SO}_2 = g'$ because it is more practical.

\strong{Three dimensional systems.} The invariants always include $\braket{\vec{g},\vec{g}}$, $\braket{\vec{g'},\vec{g'}}$, $\braket{\vec{g},\vec{g'}}$. 
In 3D, there is no other fundamental invariant (Ref.~\cite{Procesi2006}, chapter 11, \S~2.1 page 390). It will be convenient to use the combinations of invariants
\begin{equation}
	g = \lVert \vec{g} \rVert
	\qquad
	g' = \lVert \vec{g'} \rVert
	\qquad
	s = \frac{\braket{\vec{g},\vec{g'}}}{g \, g'}.
\end{equation}
to write
\begin{equation}
	\ttilde{W}(\vec{g}, \vec{g'}) = \tttilde{W}(g,g',s).
\end{equation}
We can then use conservation of energy to write
\begin{equation}
	\tttilde{W}(g,g',s) = \tilde{\sigma}(g, s) \; \delta^{(1)}\left(\frac{g^2 - g'^2}{2}\right).
\end{equation}

Here, time-reversal provides no additional constraint~\cite{DeGroot1973,Waldmann1958} because the action of time-reversal on the $SO(3)$-invariants is $(g,g',s) \to (g',g,s)$ and we already have $\tttilde{W}(g,g',s) = \tttilde{W}(g',g,s)$.

To write down the collision integral in a usable form, it is convenient to use spherical coordinates $(g,\chi,\epsilon)$ in 3D, and write
\begin{equation}
	(g_x, g_y, g_z) = g (\sin\chi \, \cos\epsilon, \sin\chi \, \sin\epsilon, \cos\chi)
\end{equation}
so that
\begin{equation}
	\dd^3 \vec{g} = \dd^2 \Omega \, g^2 \dd g
	\quad
	\text{with}
	\quad
	\dd^2 \Omega = \sin(\chi) \dd \chi \, \dd \epsilon.
\end{equation}
Here $g \geq 0$ is the radius, $\chi$ is the inclination with $0 \leq \chi \leq  \pi$, while $\epsilon$ is the azimuth and $0 \leq \epsilon < 2\pi$~\footnote{
The relation with Cartesian coordinates $(g_x,g_y,g_z)$ is $g_x = g \sin\chi \, \cos\epsilon$, $g_y = g \sin\chi \, \sin\epsilon$, $g_z = g \cos\chi$.}.
We do the same thing for $\vec{g'}$.

Because of rotation invariance we can always choose a coordinate system in which $\vec{g} = (0,0,1)$ (in Cartesian coordinates), so that $\chi = 0$ (and $\epsilon$ is indeterminate).
In this case, $s = \cos(\chi')$ so $\tilde{\sigma}(g, s) = \tilde{\sigma}(g, \cos(\chi'))$ and to connect with usual notations (see Refs.~\cite{Friedrich2015,Harris2004,ChapmanCowling}), we define the differential cross section
\begin{equation}
	\sigma(g, \chi') = \tilde{\sigma}(g, \cos(\chi'))
\end{equation}

We then use the identity
\begin{equation}
	\int_{0}^{\infty} \delta^{(1)}\left(\frac{g^2 - g'^2}{2}\right) g'^2 \dd^1 g' = g.
\end{equation}
to get 
\begin{equation}
	\mathscr{C}(f) = \int g \sigma\left(g, \frac{\braket{g',g}}{g' \, g} \right) 
	\big[ f_1' f_2' -  f_1 f_2 \big] \dd^3 \vec{c_2} \, \dd^{2} \Omega'
\end{equation}
in which we have used the fact that $\braket{g',g} = \braket{g,g'}$ to factor the scattering cross-section.
Hence, we can write
\begin{equation}
\label{simplified_collision_integral_3D}
\boxed{
	\mathscr{C}(f) = \int g \, \sigma(g, \chi') 
	\big[ f_1' f_2' -  f_1 f_2 \big] \dd^3 \vec{c_2} \, \dd^{2} \Omega'
	}
\end{equation}
Note that all the variables in the equation should be seen as functions of $(t, \vec{r}, \vec{c_1})$ (arguments of the LHS) and $(\vec{c_2}, \chi', \epsilon')$ (integration variables), defined by the conservation laws, etc.

\strong{Two-dimensional systems.} 
In 2D, we still have the invariants $\braket{\vec{g},\vec{g}}$, $\braket{\vec{g'},\vec{g'}}$, $\braket{\vec{g},\vec{g'}}$, but there is an additional fundamental invariant for $SO(2)$, namely
\begin{equation}
\vec{g} \times \vec{g'} \equiv \epsilon_{i j} g_i g'_j	
\end{equation}
We can therefore write
\begin{equation}
	\ttilde{W}(\vec{g}, \vec{g'}) = \tttilde{W}\left(g, g', \frac{\vec{g} \cdot \vec{g'}}{g \, g'}, \frac{\vec{g} \times \vec{g'}}{g \, g'}\right).
\end{equation}
The action of time-reversal is then
\begin{equation}
	\tttilde{W}(g, g', s_1, s_2) \to \tttilde{W}(g', g, s_1, -s_2)
\end{equation}

\let\polarangle\theta
We again move to polar coordinates. By rotation invariance, we can always assume that $\vec{g} = (1,0)$ (in Cartesian components)~\footnote{This has to be done separately for the two parts of the integral.}.
We then write
\begin{equation}
	\vec{g'} = g' (\cos(\polarangle'),\sin(\polarangle'))
\end{equation}
so that
\begin{equation}
	\dd^2 \vec{g'} = g' \dd g' \, \dd \polarangle'
\end{equation}
and 
\begin{equation}
	\frac{\vec{g} \cdot \vec{g'}}{g \, g'} = \cos(\polarangle')
	\quad
	\text{and}
	\quad
	\frac{\vec{g} \times \vec{g'}}{g \, g'} = \sin(\polarangle').
\end{equation}
The additional fundamental invariant gives us a tiny bit of information!
We can therefore write
\begin{equation}
	\ttilde{W}(\vec{g}, \vec{g'}) = \tttilde{W}(g, g', \cos(\polarangle'), \sin(\polarangle')).
\end{equation}

Again, we use conservation of energy to write
\begin{equation}
	\tttilde{W}(g,g',\cos(\polarangle'),\sin(\polarangle')) = g \, \sigma(g, -\polarangle') \; \delta^{(1)}\left(\frac{g^2 - g'^2}{2}\right)
\end{equation}
in which we have added a factor $g$ and introduced the differential cross-section $\sigma(g, -\polarangle')$ (we have introduced a minus sign in the definition for convenience).

Finally, we use the identity
\begin{equation}
	\int_{0}^{\infty} \delta^{(1)}\left(\frac{g^2 - g'^2}{2}\right) g' \dd^1 g' = 1
\end{equation}
to get
\begin{equation}
	\mathscr{C}(f) = 
	\int g \, \big[
	  \sigma(g, \polarangle') f_1' f_2' - \sigma(g, -\polarangle') f_1 f_2
	\big] \dd^2 \vec{c_2} \, \dd \polarangle'
\end{equation}
Again, all the variables in the equation should be seen as functions of $(t, \vec{r}, \vec{c_1}, \vec{c_2}, \polarangle')$ defined by the conservation laws, etc.

The identity
\begin{equation}
	\int_{-\pi}^{\pi} \sigma(g, \polarangle') \dd \polarangle' = \int_{-\pi}^{\pi} \sigma(g, -\polarangle') \dd \polarangle'
\end{equation}
expresses the bilateral normalization condition, that is trivially satisified in our simple model.
(The equation above is just a change of variable, not a physical condition.)
Using this, we can finally rewrite the collision integral as
\begin{equation}
\label{simplified_collision_integral_2D}
\boxed{
	\mathscr{C}(f) = 
	\int 
	  g \, \sigma(g, \polarangle') \big[ f_1' f_2' - f_1 f_2 \big] 
	  \dd^2 \vec{c_2} \, \dd \polarangle'
	  }
\end{equation}

With the notations of Eq.~\eqref{simplified_collision_integral_3D} and Eq.~\eqref{simplified_collision_integral_2D}, the cross-section $\sigma$ has the dimension $[\sigma] = \text{\textsf{L}}^{d-1}$ (an area in 3D, a distance in 2D).

We can permute $\vec{c_1} \leftrightarrow \vec{c_2}$ (without touching the primed ones) and independently $\vec{c_1'} \leftrightarrow \vec{c_2'}$.
Hence, everything should be invariant under $\polarangle' \to \polarangle' + \pi$.
The differential cross-section itself doesn't need to be invariant under this transformation (and usually isn't), but only the symmetrized part $[\sigma(g,\polarangle')+\sigma(g,\polarangle'+\pi)]/2$ should contribute to the collision integral.

\begin{figure*}
    \includegraphics[width=\textwidth]{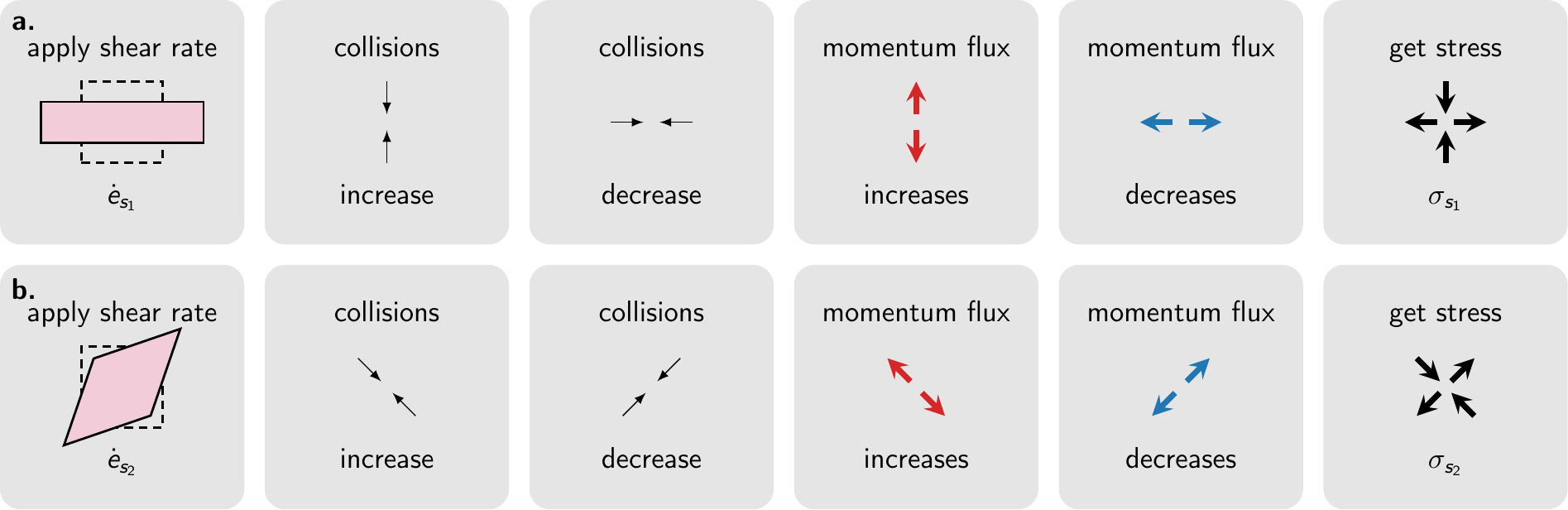}
    \caption{\label{figure_mechanism_normal_viscosity}
    \textbf{Schematic mechanism leading to normal viscosity.}
    The contributions to the pressure tensor leading to normal shear viscosity were not included in \mainTextFigureMechanismOddViscosity{} of the main text for simplicity.
    We display them here, see the caption of \mainTextFigureMechanismOddViscosity{} for details.
    }
\end{figure*}

\section{Relation between the dependent variables in the integrals}
\label{appendix_dependent_variables}

It is convenient to first write
\begin{align}
	\vec{c_1} &= c_1 (\cos(\varphi_1), \sin(\varphi_1)) \\
	\vec{c_2} &= c_2 (\cos(\varphi_2), \sin(\varphi_2)).
\end{align}
We now express the various quantities appearing in the integrals in terms of the independent variables $(c_1,c_2,\varphi_1,\varphi_2,\theta')$ as follows
\begin{align}
	\vec{g} &= \vec{c_2} - \vec{c_1} \\
	g &= \lVert \vec{g} \rVert \\
	\vec{g'} &= R(\theta') \vec{g}
\end{align}
in which $R(\theta')$ is the rotation matrix by an angle $\theta'$, and
\begin{align}
	\vec{c_1'} &= \frac{1}{2} \left( \vec{c_1} + \vec{c_2} - \vec{g'} \right) \\
	\vec{c_2'} &= \frac{1}{2} \left( \vec{c_1} + \vec{c_2} + \vec{g'} \right)
\end{align}


\begin{thebibliography}{155}\makeatletter
\providecommand \@ifxundefined [1]{\@ifx{#1\undefined}
}\providecommand \@ifnum [1]{\ifnum #1\expandafter \@firstoftwo
 \else \expandafter \@secondoftwo
 \fi
}\providecommand \@ifx [1]{\ifx #1\expandafter \@firstoftwo
 \else \expandafter \@secondoftwo
 \fi
}\providecommand \natexlab [1]{#1}\providecommand \enquote  [1]{``#1''}\providecommand \bibnamefont  [1]{#1}\providecommand \bibfnamefont [1]{#1}\providecommand \citenamefont [1]{#1}\providecommand \href@noop [0]{\@secondoftwo}\providecommand \href [0]{\begingroup \@sanitize@url \@href}\providecommand \@href[1]{\@@startlink{#1}\@@href}\providecommand \@@href[1]{\endgroup#1\@@endlink}\providecommand \@sanitize@url [0]{\catcode `\\12\catcode `\$12\catcode
  `\&12\catcode `\#12\catcode `\^12\catcode `\_12\catcode `\%12\relax}\providecommand \@@startlink[1]{}\providecommand \@@endlink[0]{}\providecommand \url  [0]{\begingroup\@sanitize@url \@url }\providecommand \@url [1]{\endgroup\@href {#1}{\urlprefix }}\providecommand \urlprefix  [0]{URL }\providecommand \Eprint [0]{\href }\providecommand \doibase [0]{https://doi.org/}\providecommand \selectlanguage [0]{\@gobble}\providecommand \bibinfo  [0]{\@secondoftwo}\providecommand \bibfield  [0]{\@secondoftwo}\providecommand \translation [1]{[#1]}\providecommand \BibitemOpen [0]{}\providecommand \bibitemStop [0]{}\providecommand \bibitemNoStop [0]{.\EOS\space}\providecommand \EOS [0]{\spacefactor3000\relax}\providecommand \BibitemShut  [1]{\csname bibitem#1\endcsname}\let\auto@bib@innerbib\@empty
\bibitem [{\citenamefont {Marchetti}\ \emph {et~al.}(2013)\citenamefont
  {Marchetti}, \citenamefont {Joanny}, \citenamefont {Ramaswamy}, \citenamefont
  {Liverpool}, \citenamefont {Prost}, \citenamefont {Rao},\ and\ \citenamefont
  {Simha}}]{Marchetti2013}\BibitemOpen
  \bibfield  {author} {\bibinfo {author} {\bibfnamefont {M.~C.}\ \bibnamefont
  {Marchetti}}, \bibinfo {author} {\bibfnamefont {J.~F.}\ \bibnamefont
  {Joanny}}, \bibinfo {author} {\bibfnamefont {S.}~\bibnamefont {Ramaswamy}},
  \bibinfo {author} {\bibfnamefont {T.~B.}\ \bibnamefont {Liverpool}}, \bibinfo
  {author} {\bibfnamefont {J.}~\bibnamefont {Prost}}, \bibinfo {author}
  {\bibfnamefont {M.}~\bibnamefont {Rao}},\ and\ \bibinfo {author}
  {\bibfnamefont {R.~A.}\ \bibnamefont {Simha}},\ }\bibfield  {title} {\bibinfo
  {title} {Hydrodynamics of soft active matter},\ }\href
  {https://doi.org/10.1103/revmodphys.85.1143} {\bibfield  {journal} {\bibinfo
  {journal} {Reviews of Modern Physics}\ }\textbf {\bibinfo {volume} {85}},\
  \bibinfo {pages} {1143–1189} (\bibinfo {year} {2013})}\BibitemShut
  {NoStop}\bibitem [{\citenamefont {Doostmohammadi}\ \emph {et~al.}(2018)\citenamefont
  {Doostmohammadi}, \citenamefont {Ignés-Mullol}, \citenamefont {Yeomans},\
  and\ \citenamefont {Sagués}}]{Doostmohammadi2018}\BibitemOpen
  \bibfield  {author} {\bibinfo {author} {\bibfnamefont {A.}~\bibnamefont
  {Doostmohammadi}}, \bibinfo {author} {\bibfnamefont {J.}~\bibnamefont
  {Ignés-Mullol}}, \bibinfo {author} {\bibfnamefont {J.~M.}\ \bibnamefont
  {Yeomans}},\ and\ \bibinfo {author} {\bibfnamefont {F.}~\bibnamefont
  {Sagués}},\ }\bibfield  {title} {\bibinfo {title} {Active nematics},\
  }\bibfield  {journal} {\bibinfo  {journal} {Nature Communications}\ }\textbf
  {\bibinfo {volume} {9}},\ \href {https://doi.org/10.1038/s41467-018-05666-8}
  {10.1038/s41467-018-05666-8} (\bibinfo {year} {2018})\BibitemShut {NoStop}\bibitem [{\citenamefont {Prost}\ \emph {et~al.}(2015)\citenamefont {Prost},
  \citenamefont {Jülicher},\ and\ \citenamefont {Joanny}}]{Prost2015}\BibitemOpen
  \bibfield  {author} {\bibinfo {author} {\bibfnamefont {J.}~\bibnamefont
  {Prost}}, \bibinfo {author} {\bibfnamefont {F.}~\bibnamefont {Jülicher}},\
  and\ \bibinfo {author} {\bibfnamefont {J.-F.}\ \bibnamefont {Joanny}},\
  }\bibfield  {title} {\bibinfo {title} {Active gel physics},\ }\href
  {https://doi.org/10.1038/nphys3224} {\bibfield  {journal} {\bibinfo
  {journal} {Nature Physics}\ }\textbf {\bibinfo {volume} {11}},\ \bibinfo
  {pages} {111–117} (\bibinfo {year} {2015})}\BibitemShut {NoStop}\bibitem [{\citenamefont {Shankar}\ \emph {et~al.}(2020)\citenamefont
  {Shankar}, \citenamefont {Souslov}, \citenamefont {Bowick}, \citenamefont
  {Marchetti},\ and\ \citenamefont {Vitelli}}]{Shankar2020}\BibitemOpen
  \bibfield  {author} {\bibinfo {author} {\bibfnamefont {S.}~\bibnamefont
  {Shankar}}, \bibinfo {author} {\bibfnamefont {A.}~\bibnamefont {Souslov}},
  \bibinfo {author} {\bibfnamefont {M.~J.}\ \bibnamefont {Bowick}}, \bibinfo
  {author} {\bibfnamefont {M.~C.}\ \bibnamefont {Marchetti}},\ and\ \bibinfo
  {author} {\bibfnamefont {V.}~\bibnamefont {Vitelli}},\ }\href@noop {}
  {\bibinfo {title} {Topological active matter}} (\bibinfo {year} {2020}),\
  \Eprint {https://arxiv.org/abs/2010.00364} {arXiv:2010.00364} \BibitemShut
  {NoStop}\bibitem [{\citenamefont {Colen}\ \emph {et~al.}(2021)\citenamefont {Colen},
  \citenamefont {Han}, \citenamefont {Zhang}, \citenamefont {Redford},
  \citenamefont {Lemma}, \citenamefont {Morgan}, \citenamefont {Ruijgrok},
  \citenamefont {Adkins}, \citenamefont {Bryant}, \citenamefont {Dogic},\ and\
  \citenamefont {et~al.}}]{Colen2021}\BibitemOpen
  \bibfield  {author} {\bibinfo {author} {\bibfnamefont {J.}~\bibnamefont
  {Colen}}, \bibinfo {author} {\bibfnamefont {M.}~\bibnamefont {Han}}, \bibinfo
  {author} {\bibfnamefont {R.}~\bibnamefont {Zhang}}, \bibinfo {author}
  {\bibfnamefont {S.~A.}\ \bibnamefont {Redford}}, \bibinfo {author}
  {\bibfnamefont {L.~M.}\ \bibnamefont {Lemma}}, \bibinfo {author}
  {\bibfnamefont {L.}~\bibnamefont {Morgan}}, \bibinfo {author} {\bibfnamefont
  {P.~V.}\ \bibnamefont {Ruijgrok}}, \bibinfo {author} {\bibfnamefont
  {R.}~\bibnamefont {Adkins}}, \bibinfo {author} {\bibfnamefont
  {Z.}~\bibnamefont {Bryant}}, \bibinfo {author} {\bibfnamefont
  {Z.}~\bibnamefont {Dogic}},\ and\ \bibinfo {author} {\bibnamefont {et~al.}},\
  }\bibfield  {title} {\bibinfo {title} {Machine learning active-nematic
  hydrodynamics},\ }\href {https://doi.org/10.1073/pnas.2016708118} {\bibfield
  {journal} {\bibinfo  {journal} {Proceedings of the National Academy of
  Sciences}\ }\textbf {\bibinfo {volume} {118}},\ \bibinfo {pages}
  {e2016708118} (\bibinfo {year} {2021})}\BibitemShut {NoStop}\bibitem [{\citenamefont {Lucas}\ and\ \citenamefont {Fong}(2018)}]{Lucas2018}\BibitemOpen
  \bibfield  {author} {\bibinfo {author} {\bibfnamefont {A.}~\bibnamefont
  {Lucas}}\ and\ \bibinfo {author} {\bibfnamefont {K.~C.}\ \bibnamefont
  {Fong}},\ }\bibfield  {title} {\bibinfo {title} {Hydrodynamics of electrons
  in graphene},\ }\href {https://doi.org/10.1088/1361-648x/aaa274} {\bibfield
  {journal} {\bibinfo  {journal} {Journal of Physics: Condensed Matter}\
  }\textbf {\bibinfo {volume} {30}},\ \bibinfo {pages} {053001} (\bibinfo
  {year} {2018})}\BibitemShut {NoStop}\bibitem [{\citenamefont {Cook}\ and\ \citenamefont {Lucas}(2019)}]{Cook2019}\BibitemOpen
  \bibfield  {author} {\bibinfo {author} {\bibfnamefont {C.~Q.}\ \bibnamefont
  {Cook}}\ and\ \bibinfo {author} {\bibfnamefont {A.}~\bibnamefont {Lucas}},\
  }\bibfield  {title} {\bibinfo {title} {Electron hydrodynamics with a
  polygonal fermi surface},\ }\href
  {https://doi.org/10.1103/physrevb.99.235148} {\bibfield  {journal} {\bibinfo
  {journal} {Physical Review B}\ }\textbf {\bibinfo {volume} {99}},\ \bibinfo
  {pages} {235148} (\bibinfo {year} {2019})}\BibitemShut {NoStop}\bibitem [{\citenamefont {Sulpizio}\ \emph {et~al.}(2019)\citenamefont
  {Sulpizio}, \citenamefont {Ella}, \citenamefont {Rozen}, \citenamefont
  {Birkbeck}, \citenamefont {Perello}, \citenamefont {Dutta}, \citenamefont
  {Ben-Shalom}, \citenamefont {Taniguchi}, \citenamefont {Watanabe},
  \citenamefont {Holder},\ and\ \citenamefont {et~al.}}]{Sulpizio2019}\BibitemOpen
  \bibfield  {author} {\bibinfo {author} {\bibfnamefont {J.~A.}\ \bibnamefont
  {Sulpizio}}, \bibinfo {author} {\bibfnamefont {L.}~\bibnamefont {Ella}},
  \bibinfo {author} {\bibfnamefont {A.}~\bibnamefont {Rozen}}, \bibinfo
  {author} {\bibfnamefont {J.}~\bibnamefont {Birkbeck}}, \bibinfo {author}
  {\bibfnamefont {D.~J.}\ \bibnamefont {Perello}}, \bibinfo {author}
  {\bibfnamefont {D.}~\bibnamefont {Dutta}}, \bibinfo {author} {\bibfnamefont
  {M.}~\bibnamefont {Ben-Shalom}}, \bibinfo {author} {\bibfnamefont
  {T.}~\bibnamefont {Taniguchi}}, \bibinfo {author} {\bibfnamefont
  {K.}~\bibnamefont {Watanabe}}, \bibinfo {author} {\bibfnamefont
  {T.}~\bibnamefont {Holder}},\ and\ \bibinfo {author} {\bibnamefont
  {et~al.}},\ }\bibfield  {title} {\bibinfo {title} {Visualizing poiseuille
  flow of hydrodynamic electrons},\ }\href
  {https://doi.org/10.1038/s41586-019-1788-9} {\bibfield  {journal} {\bibinfo
  {journal} {Nature}\ }\textbf {\bibinfo {volume} {576}},\ \bibinfo {pages}
  {75–79} (\bibinfo {year} {2019})}\BibitemShut {NoStop}\bibitem [{\citenamefont {Ku}\ \emph {et~al.}(2020)\citenamefont {Ku},
  \citenamefont {Zhou}, \citenamefont {Li}, \citenamefont {Shin}, \citenamefont
  {Shi}, \citenamefont {Burch}, \citenamefont {Anderson}, \citenamefont
  {Pierce}, \citenamefont {Xie}, \citenamefont {Hamo},\ and\ \citenamefont
  {et~al.}}]{Ku2020}\BibitemOpen
  \bibfield  {author} {\bibinfo {author} {\bibfnamefont {M.~J.~H.}\
  \bibnamefont {Ku}}, \bibinfo {author} {\bibfnamefont {T.~X.}\ \bibnamefont
  {Zhou}}, \bibinfo {author} {\bibfnamefont {Q.}~\bibnamefont {Li}}, \bibinfo
  {author} {\bibfnamefont {Y.~J.}\ \bibnamefont {Shin}}, \bibinfo {author}
  {\bibfnamefont {J.~K.}\ \bibnamefont {Shi}}, \bibinfo {author} {\bibfnamefont
  {C.}~\bibnamefont {Burch}}, \bibinfo {author} {\bibfnamefont {L.~E.}\
  \bibnamefont {Anderson}}, \bibinfo {author} {\bibfnamefont {A.~T.}\
  \bibnamefont {Pierce}}, \bibinfo {author} {\bibfnamefont {Y.}~\bibnamefont
  {Xie}}, \bibinfo {author} {\bibfnamefont {A.}~\bibnamefont {Hamo}},\ and\
  \bibinfo {author} {\bibnamefont {et~al.}},\ }\bibfield  {title} {\bibinfo
  {title} {Imaging viscous flow of the dirac fluid in graphene},\ }\href
  {https://doi.org/10.1038/s41586-020-2507-2} {\bibfield  {journal} {\bibinfo
  {journal} {Nature}\ }\textbf {\bibinfo {volume} {583}},\ \bibinfo {pages}
  {537–541} (\bibinfo {year} {2020})}\BibitemShut {NoStop}\bibitem [{\citenamefont {Policastro}\ \emph {et~al.}(2001)\citenamefont
  {Policastro}, \citenamefont {Son},\ and\ \citenamefont
  {Starinets}}]{Policastro2001}\BibitemOpen
  \bibfield  {author} {\bibinfo {author} {\bibfnamefont {G.}~\bibnamefont
  {Policastro}}, \bibinfo {author} {\bibfnamefont {D.~T.}\ \bibnamefont
  {Son}},\ and\ \bibinfo {author} {\bibfnamefont {A.~O.}\ \bibnamefont
  {Starinets}},\ }\bibfield  {title} {\bibinfo {title} {Shear viscosity of
  strongly coupled $n=4$ supersymmetric yang-mills plasma},\ }\href
  {https://doi.org/10.1103/physrevlett.87.081601} {\bibfield  {journal}
  {\bibinfo  {journal} {Physical Review Letters}\ }\textbf {\bibinfo {volume}
  {87}},\ \bibinfo {pages} {081601} (\bibinfo {year} {2001})}\BibitemShut
  {NoStop}\bibitem [{\citenamefont {Kovtun}\ \emph {et~al.}(2005)\citenamefont {Kovtun},
  \citenamefont {Son},\ and\ \citenamefont {Starinets}}]{Kovtun2005}\BibitemOpen
  \bibfield  {author} {\bibinfo {author} {\bibfnamefont {P.~K.}\ \bibnamefont
  {Kovtun}}, \bibinfo {author} {\bibfnamefont {D.~T.}\ \bibnamefont {Son}},\
  and\ \bibinfo {author} {\bibfnamefont {A.~O.}\ \bibnamefont {Starinets}},\
  }\bibfield  {title} {\bibinfo {title} {Viscosity in strongly interacting
  quantum field theories from black hole physics},\ }\href
  {https://doi.org/10.1103/physrevlett.94.111601} {\bibfield  {journal}
  {\bibinfo  {journal} {Physical Review Letters}\ }\textbf {\bibinfo {volume}
  {94}},\ \bibinfo {pages} {111601} (\bibinfo {year} {2005})}\BibitemShut
  {NoStop}\bibitem [{\citenamefont {Cen}(1992)}]{Cen1992}\BibitemOpen
  \bibfield  {author} {\bibinfo {author} {\bibfnamefont {R.}~\bibnamefont
  {Cen}},\ }\bibfield  {title} {\bibinfo {title} {A hydrodynamic approach to
  cosmology - methodology},\ }\href@noop {} {\bibfield  {journal} {\bibinfo
  {journal} {The Astrophysical Journal Supplement Series}\ }\textbf {\bibinfo
  {volume} {78}},\ \bibinfo {pages} {341} (\bibinfo {year} {1992})}\BibitemShut
  {NoStop}\bibitem [{\citenamefont {Andersson}\ and\ \citenamefont
  {Comer}(2021)}]{Andersson2021}\BibitemOpen
  \bibfield  {author} {\bibinfo {author} {\bibfnamefont {N.}~\bibnamefont
  {Andersson}}\ and\ \bibinfo {author} {\bibfnamefont {G.~L.}\ \bibnamefont
  {Comer}},\ }\bibfield  {title} {\bibinfo {title} {Relativistic fluid
  dynamics: physics for many different scales},\ }\bibfield  {journal}
  {\bibinfo  {journal} {Living Reviews in Relativity}\ }\textbf {\bibinfo
  {volume} {24}},\ \href {https://doi.org/10.1007/s41114-021-00031-6}
  {10.1007/s41114-021-00031-6} (\bibinfo {year} {2021})\BibitemShut {NoStop}\bibitem [{\citenamefont {Forster}(1975)}]{Forster1975}\BibitemOpen
  \bibfield  {author} {\bibinfo {author} {\bibfnamefont {D.}~\bibnamefont
  {Forster}},\ }\href@noop {} {\emph {\bibinfo {title} {Hydrodynamic
  Fluctuations, Broken Symmetry, and Correlation Functions}}},\ Advanced Books
  Classics\ (\bibinfo  {publisher} {Taylor \& Francis Group},\ \bibinfo {year}
  {1975})\BibitemShut {NoStop}\bibitem [{\citenamefont {Chaikin}\ and\ \citenamefont
  {Lubensky}(2000)}]{Chaikin2000}\BibitemOpen
  \bibfield  {author} {\bibinfo {author} {\bibfnamefont {P.}~\bibnamefont
  {Chaikin}}\ and\ \bibinfo {author} {\bibfnamefont {T.}~\bibnamefont
  {Lubensky}},\ }\href@noop {} {\emph {\bibinfo {title} {Principles of
  Condensed Matter Physics}}}\ (\bibinfo  {publisher} {Cambridge University
  Press},\ \bibinfo {year} {2000})\BibitemShut {NoStop}\bibitem [{\citenamefont {Schäfer}(2014)}]{Schafer2014}\BibitemOpen
  \bibfield  {author} {\bibinfo {author} {\bibfnamefont {T.}~\bibnamefont
  {Schäfer}},\ }\bibfield  {title} {\bibinfo {title} {Fluid dynamics and
  viscosity in strongly correlated fluids},\ }\href
  {https://doi.org/10.1146/annurev-nucl-102313-025439} {\bibfield  {journal}
  {\bibinfo  {journal} {Annual Review of Nuclear and Particle Science}\
  }\textbf {\bibinfo {volume} {64}},\ \bibinfo {pages} {125–148} (\bibinfo
  {year} {2014})}\BibitemShut {NoStop}\bibitem [{\citenamefont {Landau}\ and\ \citenamefont
  {Lifshitz}(2013)}]{LandauVI}\BibitemOpen
  \bibfield  {author} {\bibinfo {author} {\bibfnamefont {L.}~\bibnamefont
  {Landau}}\ and\ \bibinfo {author} {\bibfnamefont {E.}~\bibnamefont
  {Lifshitz}},\ }\href@noop {} {\emph {\bibinfo {title} {Fluid Mechanics}}},\
  \bibinfo {number} {v. 6}\ (\bibinfo  {publisher} {Elsevier Science},\
  \bibinfo {year} {2013})\BibitemShut {NoStop}\bibitem [{\citenamefont {Dubovsky}\ \emph {et~al.}(2012)\citenamefont
  {Dubovsky}, \citenamefont {Hui}, \citenamefont {Nicolis},\ and\ \citenamefont
  {Son}}]{Dubovsky2012}\BibitemOpen
  \bibfield  {author} {\bibinfo {author} {\bibfnamefont {S.}~\bibnamefont
  {Dubovsky}}, \bibinfo {author} {\bibfnamefont {L.}~\bibnamefont {Hui}},
  \bibinfo {author} {\bibfnamefont {A.}~\bibnamefont {Nicolis}},\ and\ \bibinfo
  {author} {\bibfnamefont {D.~T.}\ \bibnamefont {Son}},\ }\bibfield  {title}
  {\bibinfo {title} {Effective field theory for hydrodynamics: Thermodynamics,
  and the derivative expansion},\ }\href
  {https://doi.org/10.1103/physrevd.85.085029} {\bibfield  {journal} {\bibinfo
  {journal} {Physical Review D}\ }\textbf {\bibinfo {volume} {85}},\ \bibinfo
  {pages} {085029} (\bibinfo {year} {2012})}\BibitemShut {NoStop}\bibitem [{\citenamefont {Haehl}\ \emph {et~al.}(2016)\citenamefont {Haehl},
  \citenamefont {Loganayagam},\ and\ \citenamefont {Rangamani}}]{Haehl2016}\BibitemOpen
  \bibfield  {author} {\bibinfo {author} {\bibfnamefont {F.~M.}\ \bibnamefont
  {Haehl}}, \bibinfo {author} {\bibfnamefont {R.}~\bibnamefont {Loganayagam}},\
  and\ \bibinfo {author} {\bibfnamefont {M.}~\bibnamefont {Rangamani}},\
  }\bibfield  {title} {\bibinfo {title} {The fluid manifesto: emergent
  symmetries, hydrodynamics, and black holes},\ }\bibfield  {journal} {\bibinfo
   {journal} {Journal of High Energy Physics}\ }\textbf {\bibinfo {volume}
  {2016}},\ \href {https://doi.org/10.1007/jhep01(2016)184}
  {10.1007/jhep01(2016)184} (\bibinfo {year} {2016})\BibitemShut {NoStop}\bibitem [{\citenamefont {Liu}\ and\ \citenamefont {Glorioso}(2018)}]{Liu2018}\BibitemOpen
  \bibfield  {author} {\bibinfo {author} {\bibfnamefont {H.}~\bibnamefont
  {Liu}}\ and\ \bibinfo {author} {\bibfnamefont {P.}~\bibnamefont {Glorioso}},\
  }\bibfield  {title} {\bibinfo {title} {Lectures on non-equilibrium effective
  field theories and fluctuating hydrodynamics},\ }\bibfield  {journal}
  {\bibinfo  {journal} {Proceedings of Theoretical Advanced Study Institute
  Summer School 2017 “Physics at the Fundamental Frontier” —
  PoS(TASI2017)}\ }\href {https://doi.org/10.22323/1.305.0008}
  {10.22323/1.305.0008} (\bibinfo {year} {2018})\BibitemShut {NoStop}\bibitem [{\citenamefont {Kaufman}(1960)}]{Kaufman1960}\BibitemOpen
  \bibfield  {author} {\bibinfo {author} {\bibfnamefont {A.~N.}\ \bibnamefont
  {Kaufman}},\ }\bibfield  {title} {\bibinfo {title} {Plasma viscosity in a
  magnetic field},\ }\href {https://doi.org/10.1063/1.1706096} {\bibfield
  {journal} {\bibinfo  {journal} {Physics of Fluids}\ }\textbf {\bibinfo
  {volume} {3}},\ \bibinfo {pages} {610} (\bibinfo {year} {1960})}\BibitemShut
  {NoStop}\bibitem [{\citenamefont {Chapman}\ \emph {et~al.}(1990)\citenamefont
  {Chapman}, \citenamefont {Cowling}, \citenamefont {Burnett},\ and\
  \citenamefont {Cercignani}}]{ChapmanCowling}\BibitemOpen
  \bibfield  {author} {\bibinfo {author} {\bibfnamefont {S.}~\bibnamefont
  {Chapman}}, \bibinfo {author} {\bibfnamefont {T.}~\bibnamefont {Cowling}},
  \bibinfo {author} {\bibfnamefont {D.}~\bibnamefont {Burnett}},\ and\ \bibinfo
  {author} {\bibfnamefont {C.}~\bibnamefont {Cercignani}},\ }\href@noop {}
  {\emph {\bibinfo {title} {The Mathematical Theory of Non-uniform Gases: An
  Account of the Kinetic Theory of Viscosity, Thermal Conduction and Diffusion
  in Gases}}},\ Cambridge Mathematical Library\ (\bibinfo  {publisher}
  {Cambridge University Press},\ \bibinfo {year} {1990})\BibitemShut {NoStop}\bibitem [{\citenamefont {{Braginskii}}(1958)}]{Braginskii1958}\BibitemOpen
  \bibfield  {author} {\bibinfo {author} {\bibfnamefont {S.~I.}\ \bibnamefont
  {{Braginskii}}},\ }\bibfield  {title} {\bibinfo {title} {{Transport Phenomena
  in a Completely Ionized Two-Temperature Plasma}},\ }\href
  {http://www.jetp.ras.ru/cgi-bin/dn/e_006_02_0358.pdf} {\bibfield  {journal}
  {\bibinfo  {journal} {Soviet Physics JETP}\ }\textbf {\bibinfo {volume}
  {6}},\ \bibinfo {pages} {358} (\bibinfo {year} {1958})}\BibitemShut {NoStop}\bibitem [{\citenamefont {{Braginskii}}(1965)}]{Braginskii1965}\BibitemOpen
  \bibfield  {author} {\bibinfo {author} {\bibfnamefont {S.~I.}\ \bibnamefont
  {{Braginskii}}},\ }\bibfield  {title} {\bibinfo {title} {Transport processes
  in a plasma},\ }\href@noop {} {\bibfield  {journal} {\bibinfo  {journal}
  {Reviews of plasma physics}\ }\textbf {\bibinfo {volume} {1}} (\bibinfo
  {year} {1965})}\BibitemShut {NoStop}\bibitem [{\citenamefont {Korving}\ \emph {et~al.}(1966)\citenamefont
  {Korving}, \citenamefont {Hulsman}, \citenamefont {Knaap},\ and\
  \citenamefont {Beenakker}}]{Korving1966}\BibitemOpen
  \bibfield  {author} {\bibinfo {author} {\bibfnamefont {J.}~\bibnamefont
  {Korving}}, \bibinfo {author} {\bibfnamefont {H.}~\bibnamefont {Hulsman}},
  \bibinfo {author} {\bibfnamefont {H.}~\bibnamefont {Knaap}},\ and\ \bibinfo
  {author} {\bibfnamefont {J.}~\bibnamefont {Beenakker}},\ }\bibfield  {title}
  {\bibinfo {title} {Transverse momentum transport in viscous flow of diatomic
  gases in a magnetic field},\ }\href
  {https://doi.org/https://doi.org/10.1016/0031-9163(66)91315-1} {\bibfield
  {journal} {\bibinfo  {journal} {Physics Letters}\ }\textbf {\bibinfo {volume}
  {21}},\ \bibinfo {pages} {5 } (\bibinfo {year} {1966})}\BibitemShut {NoStop}\bibitem [{\citenamefont {Korving}\ \emph {et~al.}(1967)\citenamefont
  {Korving}, \citenamefont {Hulsman}, \citenamefont {Scoles}, \citenamefont
  {Knaap},\ and\ \citenamefont {Beenakker}}]{Korving1967}\BibitemOpen
  \bibfield  {author} {\bibinfo {author} {\bibfnamefont {J.}~\bibnamefont
  {Korving}}, \bibinfo {author} {\bibfnamefont {H.}~\bibnamefont {Hulsman}},
  \bibinfo {author} {\bibfnamefont {G.}~\bibnamefont {Scoles}}, \bibinfo
  {author} {\bibfnamefont {H.}~\bibnamefont {Knaap}},\ and\ \bibinfo {author}
  {\bibfnamefont {J.}~\bibnamefont {Beenakker}},\ }\bibfield  {title} {\bibinfo
  {title} {The influence of a magnetic field on the transport properties of
  gases of polyatomic molecules: Part i, viscosity},\ }\href
  {https://doi.org/10.1016/0031-8914(67)90243-1} {\bibfield  {journal}
  {\bibinfo  {journal} {Physica}\ }\textbf {\bibinfo {volume} {36}},\ \bibinfo
  {pages} {177} (\bibinfo {year} {1967})}\BibitemShut {NoStop}\bibitem [{\citenamefont {Yu.~Kagan}(1962{\natexlab{a}})}]{Kagan1962}\BibitemOpen
  \bibfield  {author} {\bibinfo {author} {\bibfnamefont {L.~M.}\ \bibnamefont
  {Yu.~Kagan}},\ }\bibfield  {title} {\bibinfo {title} {Transport phenomena in
  a paramagnetic gas},\ }\href
  {http://jetp.ras.ru/cgi-bin/e/index/e/14/3/p604?a=list} {\bibfield  {journal}
  {\bibinfo  {journal} {JETP Lett.}\ }\textbf {\bibinfo {volume} {14}},\
  \bibinfo {pages} {604} (\bibinfo {year} {1962}{\natexlab{a}})},\ \bibinfo
  {note} {[Pis'ma Zh. Eksp. Teor. Fiz. , Vol. 41, No. 3, p. 842, March
  1962]}\BibitemShut {NoStop}\bibitem [{\citenamefont {Yu.~Kagan}(1962{\natexlab{b}})}]{Kagan1962b}\BibitemOpen
  \bibfield  {author} {\bibinfo {author} {\bibfnamefont {A.~A.}\ \bibnamefont
  {Yu.~Kagan}},\ }\bibfield  {title} {\bibinfo {title} {On the kinetic theory
  of gases with rotational degrees of freedom},\ }\href
  {http://jetp.ras.ru/cgi-bin/e/index/e/14/5/p1096?a=list} {\bibfield
  {journal} {\bibinfo  {journal} {JETP Lett.}\ }\textbf {\bibinfo {volume}
  {14}},\ \bibinfo {pages} {1096} (\bibinfo {year} {1962}{\natexlab{b}})},\
  \bibinfo {note} {[Pis'ma Zh. Eksp. Teor. Fiz. , Vol. 41, No. 5, p. 1536, May
  1962]}\BibitemShut {NoStop}\bibitem [{\citenamefont {Yu.~Kagan}(1967)}]{Kagan1967}\BibitemOpen
  \bibfield  {author} {\bibinfo {author} {\bibfnamefont {L.~M.}\ \bibnamefont
  {Yu.~Kagan}},\ }\bibfield  {title} {\bibinfo {title} {Kinetic theory of gases
  taking into account rotational degrees of freedom in an external field},\
  }\href {http://jetp.ras.ru/cgi-bin/e/index/e/24/6/p1272?a=list} {\bibfield
  {journal} {\bibinfo  {journal} {JETP Lett.}\ }\textbf {\bibinfo {volume}
  {24}},\ \bibinfo {pages} {1893} (\bibinfo {year} {1967})},\ \bibinfo {note}
  {[Pis'ma Zh. Eksp. Teor. Fiz. , Vol. 51, No. 6, p. 1893, June
  1967]}\BibitemShut {NoStop}\bibitem [{\citenamefont {Moraal}\ \emph {et~al.}(1969)\citenamefont {Moraal},
  \citenamefont {McCourt},\ and\ \citenamefont {H.F.P.}}]{Moraal1969}\BibitemOpen
  \bibfield  {author} {\bibinfo {author} {\bibfnamefont {H.}~\bibnamefont
  {Moraal}}, \bibinfo {author} {\bibfnamefont {F.}~\bibnamefont {McCourt}},\
  and\ \bibinfo {author} {\bibfnamefont {K.}~\bibnamefont {H.F.P.}},\
  }\bibfield  {title} {\bibinfo {title} {The senftleben-beenakker effects for a
  gas of rough spherical molecules},\ }\href
  {https://doi.org/10.1016/0031-8914(69)90273-0} {\bibfield  {journal}
  {\bibinfo  {journal} {Physica}\ }\textbf {\bibinfo {volume} {45}},\ \bibinfo
  {pages} {455–468} (\bibinfo {year} {1969})}\BibitemShut {NoStop}\bibitem [{\citenamefont {McCourt}\ \emph {et~al.}(1969)\citenamefont
  {McCourt}, \citenamefont {Knaap},\ and\ \citenamefont
  {Moraal}}]{McCourt1969}\BibitemOpen
  \bibfield  {author} {\bibinfo {author} {\bibfnamefont {F.}~\bibnamefont
  {McCourt}}, \bibinfo {author} {\bibfnamefont {H.}~\bibnamefont {Knaap}},\
  and\ \bibinfo {author} {\bibfnamefont {H.}~\bibnamefont {Moraal}},\
  }\bibfield  {title} {\bibinfo {title} {The senftleben-beenakker effects for a
  gas of rough spherical molecules},\ }\href
  {https://doi.org/10.1016/0031-8914(69)90181-5} {\bibfield  {journal}
  {\bibinfo  {journal} {Physica}\ }\textbf {\bibinfo {volume} {43}},\ \bibinfo
  {pages} {485–512} (\bibinfo {year} {1969})}\BibitemShut {NoStop}\bibitem [{\citenamefont {Hulsman}\ \emph {et~al.}(1970)\citenamefont
  {Hulsman}, \citenamefont {Van~Waasdijk}, \citenamefont {Burgmans},
  \citenamefont {Knaap},\ and\ \citenamefont {Beenakker}}]{Hulsman1970}\BibitemOpen
  \bibfield  {author} {\bibinfo {author} {\bibfnamefont {H.}~\bibnamefont
  {Hulsman}}, \bibinfo {author} {\bibfnamefont {E.}~\bibnamefont
  {Van~Waasdijk}}, \bibinfo {author} {\bibfnamefont {A.}~\bibnamefont
  {Burgmans}}, \bibinfo {author} {\bibfnamefont {H.}~\bibnamefont {Knaap}},\
  and\ \bibinfo {author} {\bibfnamefont {J.}~\bibnamefont {Beenakker}},\
  }\bibfield  {title} {\bibinfo {title} {Transverse momentum transport in
  polyatomic gases under the influence of a magnetic field},\ }\href
  {https://doi.org/10.1016/0031-8914(70)90053-4} {\bibfield  {journal}
  {\bibinfo  {journal} {Physica}\ }\textbf {\bibinfo {volume} {50}},\ \bibinfo
  {pages} {53–76} (\bibinfo {year} {1970})}\BibitemShut {NoStop}\bibitem [{\citenamefont {Knaap}\ and\ \citenamefont
  {Beenakker}(1967)}]{Knaap1967}\BibitemOpen
  \bibfield  {author} {\bibinfo {author} {\bibfnamefont {H.}~\bibnamefont
  {Knaap}}\ and\ \bibinfo {author} {\bibfnamefont {J.}~\bibnamefont
  {Beenakker}},\ }\bibfield  {title} {\bibinfo {title} {Heat conductivity and
  viscosity of a gas of non-spherical molecules in a magnetic field},\ }\href
  {https://doi.org/10.1016/0031-8914(67)90209-1} {\bibfield  {journal}
  {\bibinfo  {journal} {Physica}\ }\textbf {\bibinfo {volume} {33}},\ \bibinfo
  {pages} {643–670} (\bibinfo {year} {1967})}\BibitemShut {NoStop}\bibitem [{\citenamefont {McCourt}\ and\ \citenamefont
  {Snider}(1967)}]{McCourt1967}\BibitemOpen
  \bibfield  {author} {\bibinfo {author} {\bibfnamefont {F.~R.}\ \bibnamefont
  {McCourt}}\ and\ \bibinfo {author} {\bibfnamefont {R.~F.}\ \bibnamefont
  {Snider}},\ }\bibfield  {title} {\bibinfo {title} {Senftleben—beenakker
  effect for the viscosity of a dilute gas of diamagnetic diatomic molecules},\
  }\href {https://doi.org/10.1063/1.1701586} {\bibfield  {journal} {\bibinfo
  {journal} {The Journal of Chemical Physics}\ }\textbf {\bibinfo {volume}
  {47}},\ \bibinfo {pages} {4117–4128} (\bibinfo {year} {1967})}\BibitemShut
  {NoStop}\bibitem [{\citenamefont {Levi}\ and\ \citenamefont
  {McCourt}(1968)}]{Levi1968}\BibitemOpen
  \bibfield  {author} {\bibinfo {author} {\bibfnamefont {A.}~\bibnamefont
  {Levi}}\ and\ \bibinfo {author} {\bibfnamefont {F.}~\bibnamefont {McCourt}},\
  }\bibfield  {title} {\bibinfo {title} {Odd terms in angular momentum and
  transport properties of polyatomic gases in a field},\ }\href
  {https://doi.org/10.1016/0031-8914(68)90108-0} {\bibfield  {journal}
  {\bibinfo  {journal} {Physica}\ }\textbf {\bibinfo {volume} {38}},\ \bibinfo
  {pages} {415–437} (\bibinfo {year} {1968})}\BibitemShut {NoStop}\bibitem [{\citenamefont {Waldmann}(1958{\natexlab{a}})}]{Waldmann1958b}\BibitemOpen
  \bibfield  {author} {\bibinfo {author} {\bibfnamefont {L.}~\bibnamefont
  {Waldmann}},\ }\bibfield  {title} {\bibinfo {title} {Die boltzmann-gleichung
  für gase aus spinteilchen},\ }\href {https://doi.org/10.1515/zna-1958-0803}
  {\bibfield  {journal} {\bibinfo  {journal} {Zeitschrift für Naturforschung
  A}\ }\textbf {\bibinfo {volume} {13}},\ \bibinfo {pages} {609} (\bibinfo
  {year} {1958}{\natexlab{a}})}\BibitemShut {NoStop}\bibitem [{\citenamefont {Beenakker}\ and\ \citenamefont
  {McCourt}(1970)}]{Beenakker1970}\BibitemOpen
  \bibfield  {author} {\bibinfo {author} {\bibfnamefont {J.~J.~M.}\
  \bibnamefont {Beenakker}}\ and\ \bibinfo {author} {\bibfnamefont {F.~R.}\
  \bibnamefont {McCourt}},\ }\bibfield  {title} {\bibinfo {title} {Magnetic and
  electric effects on transport properties},\ }\href
  {https://doi.org/10.1146/annurev.pc.21.100170.000403} {\bibfield  {journal}
  {\bibinfo  {journal} {Annual Review of Physical Chemistry}\ }\textbf
  {\bibinfo {volume} {21}},\ \bibinfo {pages} {47} (\bibinfo {year}
  {1970})}\BibitemShut {NoStop}\bibitem [{\citenamefont {McCourt}(1990)}]{Mccourt1990}\BibitemOpen
  \bibfield  {author} {\bibinfo {author} {\bibfnamefont {F.}~\bibnamefont
  {McCourt}},\ }\href@noop {} {\emph {\bibinfo {title} {Nonequilibrium
  phenomena in polyatomic gases}}}\ (\bibinfo  {publisher} {Clarendon Press
  Oxford University Press},\ \bibinfo {address} {Oxford New York},\ \bibinfo
  {year} {1990})\BibitemShut {NoStop}\bibitem [{\citenamefont {Hess}(2003)}]{Hess2003}\BibitemOpen
  \bibfield  {author} {\bibinfo {author} {\bibfnamefont {S.}~\bibnamefont
  {Hess}},\ }\bibfield  {title} {\bibinfo {title} {In memoriam ludwig
  waldmann},\ }\href {https://doi.org/10.1515/zna-2003-5-602} {\bibfield
  {journal} {\bibinfo  {journal} {Zeitschrift für Naturforschung A}\ }\textbf
  {\bibinfo {volume} {58}},\ \bibinfo {pages} {269} (\bibinfo {year}
  {2003})}\BibitemShut {NoStop}\bibitem [{\citenamefont {Furusawa}\ \emph {et~al.}(2021)\citenamefont
  {Furusawa}, \citenamefont {Fujii},\ and\ \citenamefont
  {Nishida}}]{Furusawa2021}\BibitemOpen
  \bibfield  {author} {\bibinfo {author} {\bibfnamefont {T.}~\bibnamefont
  {Furusawa}}, \bibinfo {author} {\bibfnamefont {K.}~\bibnamefont {Fujii}},\
  and\ \bibinfo {author} {\bibfnamefont {Y.}~\bibnamefont {Nishida}},\
  }\bibfield  {title} {\bibinfo {title} {Hall viscosity in the a phase of
  superfluid he3},\ }\href {https://doi.org/10.1103/physrevb.103.064506}
  {\bibfield  {journal} {\bibinfo  {journal} {Physical Review B}\ }\textbf
  {\bibinfo {volume} {103}},\ \bibinfo {pages} {064506} (\bibinfo {year}
  {2021})}\BibitemShut {NoStop}\bibitem [{\citenamefont {Fujii}\ and\ \citenamefont
  {Nishida}(2018)}]{Fujii2018}\BibitemOpen
  \bibfield  {author} {\bibinfo {author} {\bibfnamefont {K.}~\bibnamefont
  {Fujii}}\ and\ \bibinfo {author} {\bibfnamefont {Y.}~\bibnamefont
  {Nishida}},\ }\bibfield  {title} {\bibinfo {title} {Low-energy effective
  field theory of superfluid 3he-b and its gyromagnetic and hall responses},\
  }\href {https://doi.org/10.1016/j.aop.2018.06.003} {\bibfield  {journal}
  {\bibinfo  {journal} {Annals of Physics}\ }\textbf {\bibinfo {volume}
  {395}},\ \bibinfo {pages} {170} (\bibinfo {year} {2018})}\BibitemShut
  {NoStop}\bibitem [{\citenamefont {Wiegmann}\ and\ \citenamefont
  {Abanov}(2014)}]{Wiegmann2014}\BibitemOpen
  \bibfield  {author} {\bibinfo {author} {\bibfnamefont {P.}~\bibnamefont
  {Wiegmann}}\ and\ \bibinfo {author} {\bibfnamefont {A.~G.}\ \bibnamefont
  {Abanov}},\ }\bibfield  {title} {\bibinfo {title} {Anomalous hydrodynamics of
  two-dimensional vortex fluids},\ }\href
  {https://doi.org/10.1103/physrevlett.113.034501} {\bibfield  {journal}
  {\bibinfo  {journal} {Physical Review Letters}\ }\textbf {\bibinfo {volume}
  {113}},\ \bibinfo {pages} {034501} (\bibinfo {year} {2014})}\BibitemShut
  {NoStop}\bibitem [{\citenamefont {Bogatskiy}\ and\ \citenamefont
  {Wiegmann}(2019)}]{Bogatskiy2019}\BibitemOpen
  \bibfield  {author} {\bibinfo {author} {\bibfnamefont {A.}~\bibnamefont
  {Bogatskiy}}\ and\ \bibinfo {author} {\bibfnamefont {P.}~\bibnamefont
  {Wiegmann}},\ }\bibfield  {title} {\bibinfo {title} {Edge wave and boundary
  layer of vortex matter},\ }\href
  {https://doi.org/10.1103/physrevlett.122.214505} {\bibfield  {journal}
  {\bibinfo  {journal} {Physical Review Letters}\ }\textbf {\bibinfo {volume}
  {122}},\ \bibinfo {pages} {214505} (\bibinfo {year} {2019})}\BibitemShut
  {NoStop}\bibitem [{\citenamefont {Tokatly}\ and\ \citenamefont
  {Vignale}(2007)}]{Tokatly2007}\BibitemOpen
  \bibfield  {author} {\bibinfo {author} {\bibfnamefont {I.~V.}\ \bibnamefont
  {Tokatly}}\ and\ \bibinfo {author} {\bibfnamefont {G.}~\bibnamefont
  {Vignale}},\ }\bibfield  {title} {\bibinfo {title} {Lorentz shear modulus of
  a two-dimensional electron gas at high magnetic field},\ }\href
  {https://doi.org/10.1103/physrevb.76.161305} {\bibfield  {journal} {\bibinfo
  {journal} {Physical Review B}\ }\textbf {\bibinfo {volume} {76}},\ \bibinfo
  {pages} {161305} (\bibinfo {year} {2007})}\BibitemShut {NoStop}\bibitem [{\citenamefont {Alekseev}(2016)}]{Alekseev2016}\BibitemOpen
  \bibfield  {author} {\bibinfo {author} {\bibfnamefont {P.~S.}\ \bibnamefont
  {Alekseev}},\ }\bibfield  {title} {\bibinfo {title} {Negative
  magnetoresistance in viscous flow of two-dimensional electrons},\ }\href
  {https://doi.org/10.1103/physrevlett.117.166601} {\bibfield  {journal}
  {\bibinfo  {journal} {Physical Review Letters}\ }\textbf {\bibinfo {volume}
  {117}},\ \bibinfo {pages} {166601} (\bibinfo {year} {2016})}\BibitemShut
  {NoStop}\bibitem [{\citenamefont {Scaffidi}\ \emph {et~al.}(2017)\citenamefont
  {Scaffidi}, \citenamefont {Nandi}, \citenamefont {Schmidt}, \citenamefont
  {Mackenzie},\ and\ \citenamefont {Moore}}]{Scaffidi2017}\BibitemOpen
  \bibfield  {author} {\bibinfo {author} {\bibfnamefont {T.}~\bibnamefont
  {Scaffidi}}, \bibinfo {author} {\bibfnamefont {N.}~\bibnamefont {Nandi}},
  \bibinfo {author} {\bibfnamefont {B.}~\bibnamefont {Schmidt}}, \bibinfo
  {author} {\bibfnamefont {A.~P.}\ \bibnamefont {Mackenzie}},\ and\ \bibinfo
  {author} {\bibfnamefont {J.~E.}\ \bibnamefont {Moore}},\ }\bibfield  {title}
  {\bibinfo {title} {Hydrodynamic electron flow and hall viscosity},\ }\href
  {https://doi.org/10.1103/physrevlett.118.226601} {\bibfield  {journal}
  {\bibinfo  {journal} {Physical Review Letters}\ }\textbf {\bibinfo {volume}
  {118}},\ \bibinfo {pages} {226601} (\bibinfo {year} {2017})}\BibitemShut
  {NoStop}\bibitem [{\citenamefont {Berdyugin}\ \emph {et~al.}(2019)\citenamefont
  {Berdyugin}, \citenamefont {Xu}, \citenamefont {Pellegrino}, \citenamefont
  {Kumar}, \citenamefont {Principi}, \citenamefont {Torre}, \citenamefont
  {Shalom}, \citenamefont {Taniguchi}, \citenamefont {Watanabe}, \citenamefont
  {Grigorieva}, \citenamefont {Polini}, \citenamefont {Geim},\ and\
  \citenamefont {Bandurin}}]{Berdyugin2019}\BibitemOpen
  \bibfield  {author} {\bibinfo {author} {\bibfnamefont {A.~I.}\ \bibnamefont
  {Berdyugin}}, \bibinfo {author} {\bibfnamefont {S.~G.}\ \bibnamefont {Xu}},
  \bibinfo {author} {\bibfnamefont {F.~M.~D.}\ \bibnamefont {Pellegrino}},
  \bibinfo {author} {\bibfnamefont {R.~K.}\ \bibnamefont {Kumar}}, \bibinfo
  {author} {\bibfnamefont {A.}~\bibnamefont {Principi}}, \bibinfo {author}
  {\bibfnamefont {I.}~\bibnamefont {Torre}}, \bibinfo {author} {\bibfnamefont
  {M.~B.}\ \bibnamefont {Shalom}}, \bibinfo {author} {\bibfnamefont
  {T.}~\bibnamefont {Taniguchi}}, \bibinfo {author} {\bibfnamefont
  {K.}~\bibnamefont {Watanabe}}, \bibinfo {author} {\bibfnamefont {I.~V.}\
  \bibnamefont {Grigorieva}}, \bibinfo {author} {\bibfnamefont
  {M.}~\bibnamefont {Polini}}, \bibinfo {author} {\bibfnamefont {A.~K.}\
  \bibnamefont {Geim}},\ and\ \bibinfo {author} {\bibfnamefont {D.~A.}\
  \bibnamefont {Bandurin}},\ }\bibfield  {title} {\bibinfo {title} {Measuring
  hall viscosity of graphene's electron fluid},\ }\href
  {https://doi.org/10.1126/science.aau0685} {\bibfield  {journal} {\bibinfo
  {journal} {Science}\ ,\ \bibinfo {pages} {eaau0685}} (\bibinfo {year}
  {2019})}\BibitemShut {NoStop}\bibitem [{\citenamefont {van Zuiden}\ \emph {et~al.}(2016)\citenamefont {van
  Zuiden}, \citenamefont {Paulose}, \citenamefont {Irvine}, \citenamefont
  {Bartolo},\ and\ \citenamefont {Vitelli}}]{vanZuiden2016}\BibitemOpen
  \bibfield  {author} {\bibinfo {author} {\bibfnamefont {B.~C.}\ \bibnamefont
  {van Zuiden}}, \bibinfo {author} {\bibfnamefont {J.}~\bibnamefont {Paulose}},
  \bibinfo {author} {\bibfnamefont {W.~T.~M.}\ \bibnamefont {Irvine}}, \bibinfo
  {author} {\bibfnamefont {D.}~\bibnamefont {Bartolo}},\ and\ \bibinfo {author}
  {\bibfnamefont {V.}~\bibnamefont {Vitelli}},\ }\bibfield  {title} {\bibinfo
  {title} {Spatiotemporal order and emergent edge currents in active spinner
  materials},\ }\href {https://doi.org/10.1073/pnas.1609572113} {\bibfield
  {journal} {\bibinfo  {journal} {Proc. Natl. Acad. Sci. U.S.A.}\ }\textbf
  {\bibinfo {volume} {113}},\ \bibinfo {pages} {12919} (\bibinfo {year}
  {2016})}\BibitemShut {NoStop}\bibitem [{\citenamefont {Tsai}\ \emph {et~al.}(2005)\citenamefont {Tsai},
  \citenamefont {Ye}, \citenamefont {Rodriguez}, \citenamefont {Gollub},\ and\
  \citenamefont {Lubensky}}]{Tsai2005}\BibitemOpen
  \bibfield  {author} {\bibinfo {author} {\bibfnamefont {J.-C.}\ \bibnamefont
  {Tsai}}, \bibinfo {author} {\bibfnamefont {F.}~\bibnamefont {Ye}}, \bibinfo
  {author} {\bibfnamefont {J.}~\bibnamefont {Rodriguez}}, \bibinfo {author}
  {\bibfnamefont {J.~P.}\ \bibnamefont {Gollub}},\ and\ \bibinfo {author}
  {\bibfnamefont {T.}~\bibnamefont {Lubensky}},\ }\bibfield  {title} {\bibinfo
  {title} {A chiral granular gas},\ }\href
  {https://doi.org/10.1103/PhysRevLett.94.214301} {\bibfield  {journal}
  {\bibinfo  {journal} {Phys. Rev. Lett.}\ }\textbf {\bibinfo {volume} {94}},\
  \bibinfo {pages} {214301} (\bibinfo {year} {2005})}\BibitemShut {NoStop}\bibitem [{\citenamefont {Banerjee}\ \emph {et~al.}(2017)\citenamefont
  {Banerjee}, \citenamefont {Souslov}, \citenamefont {Abanov},\ and\
  \citenamefont {Vitelli}}]{Banerjee2017}\BibitemOpen
  \bibfield  {author} {\bibinfo {author} {\bibfnamefont {D.}~\bibnamefont
  {Banerjee}}, \bibinfo {author} {\bibfnamefont {A.}~\bibnamefont {Souslov}},
  \bibinfo {author} {\bibfnamefont {A.~G.}\ \bibnamefont {Abanov}},\ and\
  \bibinfo {author} {\bibfnamefont {V.}~\bibnamefont {Vitelli}},\ }\bibfield
  {title} {\bibinfo {title} {Odd viscosity in chiral active fluids},\
  }\href@noop {} {\bibfield  {journal} {\bibinfo  {journal} {Nat. Commun.}\
  }\textbf {\bibinfo {volume} {8}},\ \bibinfo {pages} {1573} (\bibinfo {year}
  {2017})}\BibitemShut {NoStop}\bibitem [{\citenamefont {Soni}\ \emph {et~al.}(2019)\citenamefont {Soni},
  \citenamefont {Bililign}, \citenamefont {Magkiriadou}, \citenamefont
  {Sacanna}, \citenamefont {Bartolo}, \citenamefont {Shelley},\ and\
  \citenamefont {Irvine}}]{Soni2019}\BibitemOpen
  \bibfield  {author} {\bibinfo {author} {\bibfnamefont {V.}~\bibnamefont
  {Soni}}, \bibinfo {author} {\bibfnamefont {E.~S.}\ \bibnamefont {Bililign}},
  \bibinfo {author} {\bibfnamefont {S.}~\bibnamefont {Magkiriadou}}, \bibinfo
  {author} {\bibfnamefont {S.}~\bibnamefont {Sacanna}}, \bibinfo {author}
  {\bibfnamefont {D.}~\bibnamefont {Bartolo}}, \bibinfo {author} {\bibfnamefont
  {M.~J.}\ \bibnamefont {Shelley}},\ and\ \bibinfo {author} {\bibfnamefont
  {W.~T.~M.}\ \bibnamefont {Irvine}},\ }\bibfield  {title} {\bibinfo {title}
  {The odd free surface flows of a colloidal chiral fluid},\ }\href
  {https://doi.org/10.1038/s41567-019-0603-8} {\bibfield  {journal} {\bibinfo
  {journal} {Nat. Phys.}\ }\textbf {\bibinfo {volume} {15}},\ \bibinfo {pages}
  {1188} (\bibinfo {year} {2019})}\BibitemShut {NoStop}\bibitem [{\citenamefont {Han}\ \emph {et~al.}(2021)\citenamefont {Han},
  \citenamefont {Fruchart}, \citenamefont {Scheibner}, \citenamefont
  {Vaikuntanathan}, \citenamefont {de~Pablo},\ and\ \citenamefont
  {Vitelli}}]{Han2020}\BibitemOpen
  \bibfield  {author} {\bibinfo {author} {\bibfnamefont {M.}~\bibnamefont
  {Han}}, \bibinfo {author} {\bibfnamefont {M.}~\bibnamefont {Fruchart}},
  \bibinfo {author} {\bibfnamefont {C.}~\bibnamefont {Scheibner}}, \bibinfo
  {author} {\bibfnamefont {S.}~\bibnamefont {Vaikuntanathan}}, \bibinfo
  {author} {\bibfnamefont {J.~J.}\ \bibnamefont {de~Pablo}},\ and\ \bibinfo
  {author} {\bibfnamefont {V.}~\bibnamefont {Vitelli}},\ }\bibfield  {title}
  {\bibinfo {title} {Fluctuating hydrodynamics of chiral active fluids},\
  }\href {https://doi.org/10.1038/s41567-021-01360-7} {\bibfield  {journal}
  {\bibinfo  {journal} {Nature Physics}\ }\textbf {\bibinfo {volume} {17}},\
  \bibinfo {pages} {1260–1269} (\bibinfo {year} {2021})}\BibitemShut
  {NoStop}\bibitem [{\citenamefont {Bililign}\ \emph {et~al.}(2021)\citenamefont
  {Bililign}, \citenamefont {Balboa~Usabiaga}, \citenamefont {Ganan},
  \citenamefont {Poncet}, \citenamefont {Soni}, \citenamefont {Magkiriadou},
  \citenamefont {Shelley}, \citenamefont {Bartolo},\ and\ \citenamefont
  {Irvine}}]{Bililign2021}\BibitemOpen
  \bibfield  {author} {\bibinfo {author} {\bibfnamefont {E.~S.}\ \bibnamefont
  {Bililign}}, \bibinfo {author} {\bibfnamefont {F.}~\bibnamefont
  {Balboa~Usabiaga}}, \bibinfo {author} {\bibfnamefont {Y.~A.}\ \bibnamefont
  {Ganan}}, \bibinfo {author} {\bibfnamefont {A.}~\bibnamefont {Poncet}},
  \bibinfo {author} {\bibfnamefont {V.}~\bibnamefont {Soni}}, \bibinfo {author}
  {\bibfnamefont {S.}~\bibnamefont {Magkiriadou}}, \bibinfo {author}
  {\bibfnamefont {M.~J.}\ \bibnamefont {Shelley}}, \bibinfo {author}
  {\bibfnamefont {D.}~\bibnamefont {Bartolo}},\ and\ \bibinfo {author}
  {\bibfnamefont {W.~T.~M.}\ \bibnamefont {Irvine}},\ }\bibfield  {title}
  {\bibinfo {title} {Motile dislocations knead odd crystals into whorls},\
  }\bibfield  {journal} {\bibinfo  {journal} {Nature Physics}\ }\href
  {https://doi.org/10.1038/s41567-021-01429-3} {10.1038/s41567-021-01429-3}
  (\bibinfo {year} {2021})\BibitemShut {NoStop}\bibitem [{\citenamefont {Yamauchi}\ \emph {et~al.}(2020)\citenamefont
  {Yamauchi}, \citenamefont {Hayata}, \citenamefont {Uwamichi}, \citenamefont
  {Ozawa},\ and\ \citenamefont {Kawaguchi}}]{Yamauchi2020}\BibitemOpen
  \bibfield  {author} {\bibinfo {author} {\bibfnamefont {L.}~\bibnamefont
  {Yamauchi}}, \bibinfo {author} {\bibfnamefont {T.}~\bibnamefont {Hayata}},
  \bibinfo {author} {\bibfnamefont {M.}~\bibnamefont {Uwamichi}}, \bibinfo
  {author} {\bibfnamefont {T.}~\bibnamefont {Ozawa}},\ and\ \bibinfo {author}
  {\bibfnamefont {K.}~\bibnamefont {Kawaguchi}},\ }\href@noop {} {\bibinfo
  {title} {Chirality-driven edge flow and non-hermitian topology in active
  nematic cells}} (\bibinfo {year} {2020}),\ \Eprint
  {https://arxiv.org/abs/2008.10852} {arXiv:2008.10852} \BibitemShut {NoStop}\bibitem [{\citenamefont {Hosaka}\ \emph
  {et~al.}(2021{\natexlab{a}})\citenamefont {Hosaka}, \citenamefont {Komura},\
  and\ \citenamefont {Andelman}}]{Hosaka2021}\BibitemOpen
  \bibfield  {author} {\bibinfo {author} {\bibfnamefont {Y.}~\bibnamefont
  {Hosaka}}, \bibinfo {author} {\bibfnamefont {S.}~\bibnamefont {Komura}},\
  and\ \bibinfo {author} {\bibfnamefont {D.}~\bibnamefont {Andelman}},\
  }\bibfield  {title} {\bibinfo {title} {Nonreciprocal response of a
  two-dimensional fluid with odd viscosity},\ }\href
  {https://doi.org/10.1103/physreve.103.042610} {\bibfield  {journal} {\bibinfo
   {journal} {Physical Review E}\ }\textbf {\bibinfo {volume} {103}},\ \bibinfo
  {pages} {042610} (\bibinfo {year} {2021}{\natexlab{a}})}\BibitemShut
  {NoStop}\bibitem [{\citenamefont {Reichhardt}\ and\ \citenamefont
  {Reichhardt}(2021)}]{Reichhardt2021}\BibitemOpen
  \bibfield  {author} {\bibinfo {author} {\bibfnamefont {C.~J.~O.}\
  \bibnamefont {Reichhardt}}\ and\ \bibinfo {author} {\bibfnamefont
  {C.}~\bibnamefont {Reichhardt}},\ }\href@noop {} {\bibinfo {title} {Active
  rheology in odd viscosity systems}} (\bibinfo {year} {2021}),\ \Eprint
  {https://arxiv.org/abs/2106.15719v1} {arXiv:2106.15719v1} \BibitemShut
  {NoStop}\bibitem [{\citenamefont {Kogan}(2016)}]{Kogan2016}\BibitemOpen
  \bibfield  {author} {\bibinfo {author} {\bibfnamefont {E.}~\bibnamefont
  {Kogan}},\ }\bibfield  {title} {\bibinfo {title} {Lift force due to odd hall
  viscosity},\ }\href {https://doi.org/10.1103/physreve.94.043111} {\bibfield
  {journal} {\bibinfo  {journal} {Physical Review E}\ }\textbf {\bibinfo
  {volume} {94}},\ \bibinfo {pages} {043111} (\bibinfo {year}
  {2016})}\BibitemShut {NoStop}\bibitem [{\citenamefont {Khain}\ \emph {et~al.}(2022)\citenamefont {Khain},
  \citenamefont {Scheibner}, \citenamefont {Fruchart},\ and\ \citenamefont
  {Vitelli}}]{Khain2022}\BibitemOpen
  \bibfield  {author} {\bibinfo {author} {\bibfnamefont {T.}~\bibnamefont
  {Khain}}, \bibinfo {author} {\bibfnamefont {C.}~\bibnamefont {Scheibner}},
  \bibinfo {author} {\bibfnamefont {M.}~\bibnamefont {Fruchart}},\ and\
  \bibinfo {author} {\bibfnamefont {V.}~\bibnamefont {Vitelli}},\ }\bibfield
  {title} {\bibinfo {title} {Stokes flows in three-dimensional fluids with odd
  and parity-violating viscosities},\ }\bibfield  {journal} {\bibinfo
  {journal} {Journal of Fluid Mechanics}\ }\textbf {\bibinfo {volume} {934}},\
  \href {https://doi.org/10.1017/jfm.2021.1079} {10.1017/jfm.2021.1079}
  (\bibinfo {year} {2022}),\ \Eprint {https://arxiv.org/abs/2011.07681}
  {arXiv:2011.07681} \BibitemShut {NoStop}\bibitem [{\citenamefont {Hosaka}\ \emph
  {et~al.}(2021{\natexlab{b}})\citenamefont {Hosaka}, \citenamefont {Komura},\
  and\ \citenamefont {Andelman}}]{Hosaka2021b}\BibitemOpen
  \bibfield  {author} {\bibinfo {author} {\bibfnamefont {Y.}~\bibnamefont
  {Hosaka}}, \bibinfo {author} {\bibfnamefont {S.}~\bibnamefont {Komura}},\
  and\ \bibinfo {author} {\bibfnamefont {D.}~\bibnamefont {Andelman}},\
  }\href@noop {} {\bibinfo {title} {Hydrodynamic lift of a two-dimensional
  liquid domain with odd viscosity}} (\bibinfo {year} {2021}{\natexlab{b}}),\
  \Eprint {https://arxiv.org/abs/2109.02321v2} {arXiv:2109.02321v2}
  \BibitemShut {NoStop}\bibitem [{\citenamefont {Yang}\ \emph {et~al.}(2021)\citenamefont {Yang},
  \citenamefont {Zhu}, \citenamefont {Liu}, \citenamefont {Liu}, \citenamefont
  {Shi}, \citenamefont {Chen}, \citenamefont {Zheng}, \citenamefont {Ye},\ and\
  \citenamefont {Yang}}]{Yang2021}\BibitemOpen
  \bibfield  {author} {\bibinfo {author} {\bibfnamefont {Q.}~\bibnamefont
  {Yang}}, \bibinfo {author} {\bibfnamefont {H.}~\bibnamefont {Zhu}}, \bibinfo
  {author} {\bibfnamefont {P.}~\bibnamefont {Liu}}, \bibinfo {author}
  {\bibfnamefont {R.}~\bibnamefont {Liu}}, \bibinfo {author} {\bibfnamefont
  {Q.}~\bibnamefont {Shi}}, \bibinfo {author} {\bibfnamefont {K.}~\bibnamefont
  {Chen}}, \bibinfo {author} {\bibfnamefont {N.}~\bibnamefont {Zheng}},
  \bibinfo {author} {\bibfnamefont {F.}~\bibnamefont {Ye}},\ and\ \bibinfo
  {author} {\bibfnamefont {M.}~\bibnamefont {Yang}},\ }\bibfield  {title}
  {\bibinfo {title} {Topologically protected transport of cargo in a chiral
  active fluid aided by odd-viscosity-enhanced depletion interactions},\ }\href
  {https://doi.org/10.1103/physrevlett.126.198001} {\bibfield  {journal}
  {\bibinfo  {journal} {Physical Review Letters}\ }\textbf {\bibinfo {volume}
  {126}},\ \bibinfo {pages} {198001} (\bibinfo {year} {2021})}\BibitemShut
  {NoStop}\bibitem [{\citenamefont {Hargus}\ \emph {et~al.}(2020)\citenamefont {Hargus},
  \citenamefont {Klymko}, \citenamefont {Epstein},\ and\ \citenamefont
  {Mandadapu}}]{Hargus2020}\BibitemOpen
  \bibfield  {author} {\bibinfo {author} {\bibfnamefont {C.}~\bibnamefont
  {Hargus}}, \bibinfo {author} {\bibfnamefont {K.}~\bibnamefont {Klymko}},
  \bibinfo {author} {\bibfnamefont {J.~M.}\ \bibnamefont {Epstein}},\ and\
  \bibinfo {author} {\bibfnamefont {K.~K.}\ \bibnamefont {Mandadapu}},\
  }\bibfield  {title} {\bibinfo {title} {Time reversal symmetry breaking and
  odd viscosity in active fluids: Green–kubo and nemd results},\ }\href
  {https://doi.org/10.1063/5.0006441} {\bibfield  {journal} {\bibinfo
  {journal} {The Journal of Chemical Physics}\ }\textbf {\bibinfo {volume}
  {152}},\ \bibinfo {pages} {201102} (\bibinfo {year} {2020})}\BibitemShut
  {NoStop}\bibitem [{\citenamefont {Epstein}\ and\ \citenamefont
  {Mandadapu}(2020)}]{Epstein2020}\BibitemOpen
  \bibfield  {author} {\bibinfo {author} {\bibfnamefont {J.~M.}\ \bibnamefont
  {Epstein}}\ and\ \bibinfo {author} {\bibfnamefont {K.~K.}\ \bibnamefont
  {Mandadapu}},\ }\bibfield  {title} {\bibinfo {title} {Time-reversal symmetry
  breaking in two-dimensional nonequilibrium viscous fluids},\ }\href
  {https://doi.org/10.1103/physreve.101.052614} {\bibfield  {journal} {\bibinfo
   {journal} {Physical Review E}\ }\textbf {\bibinfo {volume} {101}},\ \bibinfo
  {pages} {052614} (\bibinfo {year} {2020})}\BibitemShut {NoStop}\bibitem [{\citenamefont {Lapa}\ and\ \citenamefont {Hughes}(2014)}]{Lapa2014}\BibitemOpen
  \bibfield  {author} {\bibinfo {author} {\bibfnamefont {M.~F.}\ \bibnamefont
  {Lapa}}\ and\ \bibinfo {author} {\bibfnamefont {T.~L.}\ \bibnamefont
  {Hughes}},\ }\bibfield  {title} {\bibinfo {title} {Swimming at low reynolds
  number in fluids with odd, or hall, viscosity},\ }\href
  {https://doi.org/10.1103/physreve.89.043019} {\bibfield  {journal} {\bibinfo
  {journal} {Physical Review E}\ }\textbf {\bibinfo {volume} {89}},\ \bibinfo
  {pages} {043019} (\bibinfo {year} {2014})}\BibitemShut {NoStop}\bibitem [{\citenamefont {Jensen}\ \emph {et~al.}(2012)\citenamefont {Jensen},
  \citenamefont {Kaminski}, \citenamefont {Kovtun}, \citenamefont {Meyer},
  \citenamefont {Ritz},\ and\ \citenamefont {Yarom}}]{Jensen2012}\BibitemOpen
  \bibfield  {author} {\bibinfo {author} {\bibfnamefont {K.}~\bibnamefont
  {Jensen}}, \bibinfo {author} {\bibfnamefont {M.}~\bibnamefont {Kaminski}},
  \bibinfo {author} {\bibfnamefont {P.}~\bibnamefont {Kovtun}}, \bibinfo
  {author} {\bibfnamefont {R.}~\bibnamefont {Meyer}}, \bibinfo {author}
  {\bibfnamefont {A.}~\bibnamefont {Ritz}},\ and\ \bibinfo {author}
  {\bibfnamefont {A.}~\bibnamefont {Yarom}},\ }\bibfield  {title} {\bibinfo
  {title} {Parity-violating hydrodynamics in 2 + 1 dimensions},\ }\bibfield
  {journal} {\bibinfo  {journal} {Journal of High Energy Physics}\ }\textbf
  {\bibinfo {volume} {2012}},\ \href {https://doi.org/10.1007/jhep05(2012)102}
  {10.1007/jhep05(2012)102} (\bibinfo {year} {2012})\BibitemShut {NoStop}\bibitem [{\citenamefont {Kaminski}\ and\ \citenamefont
  {Moroz}(2014)}]{Kaminski2014}\BibitemOpen
  \bibfield  {author} {\bibinfo {author} {\bibfnamefont {M.}~\bibnamefont
  {Kaminski}}\ and\ \bibinfo {author} {\bibfnamefont {S.}~\bibnamefont
  {Moroz}},\ }\bibfield  {title} {\bibinfo {title} {Nonrelativistic
  parity-violating hydrodynamics in two spatial dimensions},\ }\href
  {https://doi.org/10.1103/physrevb.89.115418} {\bibfield  {journal} {\bibinfo
  {journal} {Physical Review B}\ }\textbf {\bibinfo {volume} {89}},\ \bibinfo
  {pages} {115418} (\bibinfo {year} {2014})}\BibitemShut {NoStop}\bibitem [{\citenamefont {Tauber}\ \emph {et~al.}(2019)\citenamefont {Tauber},
  \citenamefont {Delplace},\ and\ \citenamefont {Venaille}}]{Tauber2019}\BibitemOpen
  \bibfield  {author} {\bibinfo {author} {\bibfnamefont {C.}~\bibnamefont
  {Tauber}}, \bibinfo {author} {\bibfnamefont {P.}~\bibnamefont {Delplace}},\
  and\ \bibinfo {author} {\bibfnamefont {A.}~\bibnamefont {Venaille}},\
  }\bibfield  {title} {\bibinfo {title} {A bulk-interface correspondence for
  equatorial waves},\ }\bibfield  {journal} {\bibinfo  {journal} {Journal of
  Fluid Mechanics}\ }\textbf {\bibinfo {volume} {868}},\ \href
  {https://doi.org/10.1017/jfm.2019.233} {10.1017/jfm.2019.233} (\bibinfo
  {year} {2019})\BibitemShut {NoStop}\bibitem [{\citenamefont {Souslov}\ \emph {et~al.}(2019)\citenamefont
  {Souslov}, \citenamefont {Dasbiswas}, \citenamefont {Fruchart}, \citenamefont
  {Vaikuntanathan},\ and\ \citenamefont {Vitelli}}]{Souslov2019}\BibitemOpen
  \bibfield  {author} {\bibinfo {author} {\bibfnamefont {A.}~\bibnamefont
  {Souslov}}, \bibinfo {author} {\bibfnamefont {K.}~\bibnamefont {Dasbiswas}},
  \bibinfo {author} {\bibfnamefont {M.}~\bibnamefont {Fruchart}}, \bibinfo
  {author} {\bibfnamefont {S.}~\bibnamefont {Vaikuntanathan}},\ and\ \bibinfo
  {author} {\bibfnamefont {V.}~\bibnamefont {Vitelli}},\ }\bibfield  {title}
  {\bibinfo {title} {Topological waves in fluids with odd viscosity},\ }\href
  {https://doi.org/10.1103/physrevlett.122.128001} {\bibfield  {journal}
  {\bibinfo  {journal} {Physical Review Letters}\ }\textbf {\bibinfo {volume}
  {122}},\ \bibinfo {pages} {128001} (\bibinfo {year} {2019})}\BibitemShut
  {NoStop}\bibitem [{\citenamefont {Bao}\ and\ \citenamefont {Jian}(2021)}]{Bao2021}\BibitemOpen
  \bibfield  {author} {\bibinfo {author} {\bibfnamefont {G.}~\bibnamefont
  {Bao}}\ and\ \bibinfo {author} {\bibfnamefont {Y.}~\bibnamefont {Jian}},\
  }\bibfield  {title} {\bibinfo {title} {Odd-viscosity-induced instability of a
  falling thin film with an external electric field},\ }\href
  {https://doi.org/10.1103/physreve.103.013104} {\bibfield  {journal} {\bibinfo
   {journal} {Physical Review E}\ }\textbf {\bibinfo {volume} {103}},\ \bibinfo
  {pages} {013104} (\bibinfo {year} {2021})}\BibitemShut {NoStop}\bibitem [{\citenamefont {Kirkinis}\ and\ \citenamefont
  {Andreev}(2019)}]{Kirkinis2019}\BibitemOpen
  \bibfield  {author} {\bibinfo {author} {\bibfnamefont {E.}~\bibnamefont
  {Kirkinis}}\ and\ \bibinfo {author} {\bibfnamefont {A.~V.}\ \bibnamefont
  {Andreev}},\ }\bibfield  {title} {\bibinfo {title} {Odd-viscosity-induced
  stabilization of viscous thin liquid films},\ }\href
  {https://doi.org/10.1017/jfm.2019.644} {\bibfield  {journal} {\bibinfo
  {journal} {Journal of Fluid Mechanics}\ }\textbf {\bibinfo {volume} {878}},\
  \bibinfo {pages} {169–189} (\bibinfo {year} {2019})}\BibitemShut {NoStop}\bibitem [{\citenamefont {Ganeshan}\ and\ \citenamefont
  {Abanov}(2017)}]{Ganeshan2017}\BibitemOpen
  \bibfield  {author} {\bibinfo {author} {\bibfnamefont {S.}~\bibnamefont
  {Ganeshan}}\ and\ \bibinfo {author} {\bibfnamefont {A.~G.}\ \bibnamefont
  {Abanov}},\ }\bibfield  {title} {\bibinfo {title} {Odd viscosity in
  two-dimensional incompressible fluids},\ }\href
  {https://doi.org/10.1103/physrevfluids.2.094101} {\bibfield  {journal}
  {\bibinfo  {journal} {Physical Review Fluids}\ }\textbf {\bibinfo {volume}
  {2}},\ \bibinfo {pages} {094101} (\bibinfo {year} {2017})}\BibitemShut
  {NoStop}\bibitem [{\citenamefont {Monteiro}\ and\ \citenamefont
  {Ganeshan}(2021)}]{Monteiro2021}\BibitemOpen
  \bibfield  {author} {\bibinfo {author} {\bibfnamefont {G.~M.}\ \bibnamefont
  {Monteiro}}\ and\ \bibinfo {author} {\bibfnamefont {S.}~\bibnamefont
  {Ganeshan}},\ }\bibfield  {title} {\bibinfo {title} {Nonlinear shallow water
  dynamics with odd viscosity},\ }\href
  {https://doi.org/10.1103/physrevfluids.6.l092401} {\bibfield  {journal}
  {\bibinfo  {journal} {Physical Review Fluids}\ }\textbf {\bibinfo {volume}
  {6}},\ \bibinfo {pages} {l092401} (\bibinfo {year} {2021})}\BibitemShut
  {NoStop}\bibitem [{\citenamefont {Abanov}\ \emph {et~al.}(2018)\citenamefont {Abanov},
  \citenamefont {Can},\ and\ \citenamefont {Ganeshan}}]{Abanov2018}\BibitemOpen
  \bibfield  {author} {\bibinfo {author} {\bibfnamefont {A.}~\bibnamefont
  {Abanov}}, \bibinfo {author} {\bibfnamefont {T.}~\bibnamefont {Can}},\ and\
  \bibinfo {author} {\bibfnamefont {S.}~\bibnamefont {Ganeshan}},\ }\bibfield
  {title} {\bibinfo {title} {Odd surface waves in two-dimensional
  incompressible fluids},\ }\bibfield  {journal} {\bibinfo  {journal} {SciPost
  Physics}\ }\textbf {\bibinfo {volume} {5}},\ \href
  {https://doi.org/10.21468/scipostphys.5.1.010} {10.21468/scipostphys.5.1.010}
  (\bibinfo {year} {2018})\BibitemShut {NoStop}\bibitem [{\citenamefont {Abanov}\ and\ \citenamefont
  {Monteiro}(2019)}]{Abanov2019}\BibitemOpen
  \bibfield  {author} {\bibinfo {author} {\bibfnamefont {A.~G.}\ \bibnamefont
  {Abanov}}\ and\ \bibinfo {author} {\bibfnamefont {G.~M.}\ \bibnamefont
  {Monteiro}},\ }\bibfield  {title} {\bibinfo {title} {Free-surface variational
  principle for an incompressible fluid with odd viscosity},\ }\href
  {https://doi.org/10.1103/physrevlett.122.154501} {\bibfield  {journal}
  {\bibinfo  {journal} {Physical Review Letters}\ }\textbf {\bibinfo {volume}
  {122}},\ \bibinfo {pages} {154501} (\bibinfo {year} {2019})}\BibitemShut
  {NoStop}\bibitem [{\citenamefont {Landsteiner}\ \emph {et~al.}(2016)\citenamefont
  {Landsteiner}, \citenamefont {Liu},\ and\ \citenamefont
  {Sun}}]{Landsteiner2016}\BibitemOpen
  \bibfield  {author} {\bibinfo {author} {\bibfnamefont {K.}~\bibnamefont
  {Landsteiner}}, \bibinfo {author} {\bibfnamefont {Y.}~\bibnamefont {Liu}},\
  and\ \bibinfo {author} {\bibfnamefont {Y.-W.}\ \bibnamefont {Sun}},\
  }\bibfield  {title} {\bibinfo {title} {Odd viscosity in the quantum critical
  region of a holographic weyl semimetal},\ }\href
  {https://doi.org/10.1103/physrevlett.117.081604} {\bibfield  {journal}
  {\bibinfo  {journal} {Physical Review Letters}\ }\textbf {\bibinfo {volume}
  {117}},\ \bibinfo {pages} {081604} (\bibinfo {year} {2016})}\BibitemShut
  {NoStop}\bibitem [{\citenamefont {Lucas}\ and\ \citenamefont
  {Surówka}(2014)}]{Lucas2014}\BibitemOpen
  \bibfield  {author} {\bibinfo {author} {\bibfnamefont {A.}~\bibnamefont
  {Lucas}}\ and\ \bibinfo {author} {\bibfnamefont {P.}~\bibnamefont
  {Surówka}},\ }\bibfield  {title} {\bibinfo {title} {Phenomenology of
  nonrelativistic parity-violating hydrodynamics in 2+1 dimensions},\ }\href
  {https://doi.org/10.1103/physreve.90.063005} {\bibfield  {journal} {\bibinfo
  {journal} {Physical Review E}\ }\textbf {\bibinfo {volume} {90}},\ \bibinfo
  {pages} {063005} (\bibinfo {year} {2014})}\BibitemShut {NoStop}\bibitem [{\citenamefont {Holder}\ \emph {et~al.}(2019)\citenamefont {Holder},
  \citenamefont {Queiroz},\ and\ \citenamefont {Stern}}]{Holder2019}\BibitemOpen
  \bibfield  {author} {\bibinfo {author} {\bibfnamefont {T.}~\bibnamefont
  {Holder}}, \bibinfo {author} {\bibfnamefont {R.}~\bibnamefont {Queiroz}},\
  and\ \bibinfo {author} {\bibfnamefont {A.}~\bibnamefont {Stern}},\ }\bibfield
   {title} {\bibinfo {title} {Unified description of the classical hall
  viscosity},\ }\href {https://doi.org/10.1103/physrevlett.123.106801}
  {\bibfield  {journal} {\bibinfo  {journal} {Physical Review Letters}\
  }\textbf {\bibinfo {volume} {123}},\ \bibinfo {pages} {106801} (\bibinfo
  {year} {2019})}\BibitemShut {NoStop}\bibitem [{\citenamefont {Markovich}\ and\ \citenamefont
  {Lubensky}(2021)}]{Markovich2021}\BibitemOpen
  \bibfield  {author} {\bibinfo {author} {\bibfnamefont {T.}~\bibnamefont
  {Markovich}}\ and\ \bibinfo {author} {\bibfnamefont {T.~C.}\ \bibnamefont
  {Lubensky}},\ }\bibfield  {title} {\bibinfo {title} {Odd viscosity in active
  matter: Microscopic origin and 3d effects},\ }\href
  {https://doi.org/10.1103/physrevlett.127.048001} {\bibfield  {journal}
  {\bibinfo  {journal} {Physical Review Letters}\ }\textbf {\bibinfo {volume}
  {127}},\ \bibinfo {pages} {048001} (\bibinfo {year} {2021})}\BibitemShut
  {NoStop}\bibitem [{\citenamefont {Barabanov}\ \emph {et~al.}(2015)\citenamefont
  {Barabanov}, \citenamefont {Kagan}, \citenamefont {Maksimov}, \citenamefont
  {Mikheyenkov},\ and\ \citenamefont {Khabarova}}]{Barabanov2015}\BibitemOpen
  \bibfield  {author} {\bibinfo {author} {\bibfnamefont {A.~F.}\ \bibnamefont
  {Barabanov}}, \bibinfo {author} {\bibfnamefont {Y.~M.}\ \bibnamefont
  {Kagan}}, \bibinfo {author} {\bibfnamefont {L.~A.}\ \bibnamefont {Maksimov}},
  \bibinfo {author} {\bibfnamefont {A.~V.}\ \bibnamefont {Mikheyenkov}},\ and\
  \bibinfo {author} {\bibfnamefont {T.~V.}\ \bibnamefont {Khabarova}},\
  }\bibfield  {title} {\bibinfo {title} {The hall effect and its analogs},\
  }\href {https://doi.org/10.3367/ufne.0185.201505b.0479} {\bibfield  {journal}
  {\bibinfo  {journal} {Physics-Uspekhi}\ }\textbf {\bibinfo {volume} {58}},\
  \bibinfo {pages} {446–454} (\bibinfo {year} {2015})}\BibitemShut {NoStop}\bibitem [{\citenamefont {Maksimov}\ \emph {et~al.}(2017)\citenamefont
  {Maksimov}, \citenamefont {Mikheyenkov},\ and\ \citenamefont
  {Khabarova}}]{Maksimov2017}\BibitemOpen
  \bibfield  {author} {\bibinfo {author} {\bibfnamefont {L.~A.}\ \bibnamefont
  {Maksimov}}, \bibinfo {author} {\bibfnamefont {A.~V.}\ \bibnamefont
  {Mikheyenkov}},\ and\ \bibinfo {author} {\bibfnamefont {T.~V.}\ \bibnamefont
  {Khabarova}},\ }\bibfield  {title} {\bibinfo {title} {Nondiagonal
  cross-transport phenomena in a magnetic field},\ }\href
  {https://doi.org/10.3367/ufne.2017.02.038057} {\bibfield  {journal} {\bibinfo
   {journal} {Physics-Uspekhi}\ }\textbf {\bibinfo {volume} {60}},\ \bibinfo
  {pages} {623–627} (\bibinfo {year} {2017})}\BibitemShut {NoStop}\bibitem [{\citenamefont {Nakagawa}(1956)}]{Nakagawa1956}\BibitemOpen
  \bibfield  {author} {\bibinfo {author} {\bibfnamefont {Y.}~\bibnamefont
  {Nakagawa}},\ }\bibfield  {title} {\bibinfo {title} {The kinetic theory of
  gases for the rotating system.},\ }\href
  {https://doi.org/10.4294/jpe1952.4.105} {\bibfield  {journal} {\bibinfo
  {journal} {Journal of Physics of the Earth}\ }\textbf {\bibinfo {volume}
  {4}},\ \bibinfo {pages} {105–111} (\bibinfo {year} {1956})}\BibitemShut
  {NoStop}\bibitem [{\citenamefont {Morrison}\ \emph {et~al.}(2014)\citenamefont
  {Morrison}, \citenamefont {Lingam},\ and\ \citenamefont
  {Acevedo}}]{Morrison2014}\BibitemOpen
  \bibfield  {author} {\bibinfo {author} {\bibfnamefont {P.~J.}\ \bibnamefont
  {Morrison}}, \bibinfo {author} {\bibfnamefont {M.}~\bibnamefont {Lingam}},\
  and\ \bibinfo {author} {\bibfnamefont {R.}~\bibnamefont {Acevedo}},\
  }\bibfield  {title} {\bibinfo {title} {Hamiltonian and action formalisms for
  two-dimensional gyroviscous magnetohydrodynamics},\ }\href
  {https://doi.org/10.1063/1.4891321} {\bibfield  {journal} {\bibinfo
  {journal} {Physics of Plasmas}\ }\textbf {\bibinfo {volume} {21}},\ \bibinfo
  {pages} {082102} (\bibinfo {year} {2014})}\BibitemShut {NoStop}\bibitem [{\citenamefont {Morrison}\ \emph {et~al.}(1984)\citenamefont
  {Morrison}, \citenamefont {Caldas},\ and\ \citenamefont
  {Tasso}}]{Morrison1984}\BibitemOpen
  \bibfield  {author} {\bibinfo {author} {\bibfnamefont {P.~J.}\ \bibnamefont
  {Morrison}}, \bibinfo {author} {\bibfnamefont {I.~L.}\ \bibnamefont
  {Caldas}},\ and\ \bibinfo {author} {\bibfnamefont {H.}~\bibnamefont
  {Tasso}},\ }\bibfield  {title} {\bibinfo {title} {Hamiltonian formulation of
  two-dimensional gyroviscous mhd},\ }\href
  {https://doi.org/10.1515/zna-1984-1102} {\bibfield  {journal} {\bibinfo
  {journal} {Zeitschrift für Naturforschung A}\ }\textbf {\bibinfo {volume}
  {39}},\ \bibinfo {pages} {1023–1027} (\bibinfo {year} {1984})}\BibitemShut
  {NoStop}\bibitem [{\citenamefont {Lingam}\ \emph {et~al.}(2020)\citenamefont {Lingam},
  \citenamefont {Morrison},\ and\ \citenamefont {Wurm}}]{Lingam2020}\BibitemOpen
  \bibfield  {author} {\bibinfo {author} {\bibfnamefont {M.}~\bibnamefont
  {Lingam}}, \bibinfo {author} {\bibfnamefont {P.~J.}\ \bibnamefont
  {Morrison}},\ and\ \bibinfo {author} {\bibfnamefont {A.}~\bibnamefont
  {Wurm}},\ }\bibfield  {title} {\bibinfo {title} {A class of three-dimensional
  gyroviscous magnetohydrodynamic models},\ }\bibfield  {journal} {\bibinfo
  {journal} {Journal of Plasma Physics}\ }\textbf {\bibinfo {volume} {86}},\
  \href {https://doi.org/10.1017/s0022377820001038} {10.1017/s0022377820001038}
  (\bibinfo {year} {2020})\BibitemShut {NoStop}\bibitem [{\citenamefont {Monteiro}\ \emph {et~al.}(2021)\citenamefont
  {Monteiro}, \citenamefont {Abanov},\ and\ \citenamefont
  {Ganeshan}}]{Monteiro2021b}\BibitemOpen
  \bibfield  {author} {\bibinfo {author} {\bibfnamefont {G.~M.}\ \bibnamefont
  {Monteiro}}, \bibinfo {author} {\bibfnamefont {A.~G.}\ \bibnamefont
  {Abanov}},\ and\ \bibinfo {author} {\bibfnamefont {S.}~\bibnamefont
  {Ganeshan}},\ }\href@noop {} {\bibinfo {title} {Hamiltonian structure of 2d
  fluid dynamics with broken parity}} (\bibinfo {year} {2021}),\ \Eprint
  {https://arxiv.org/abs/2105.01655} {arXiv:2105.01655} \BibitemShut {NoStop}\bibitem [{\citenamefont {Tan}\ \emph {et~al.}(2021)\citenamefont {Tan},
  \citenamefont {Mietke}, \citenamefont {Higinbotham}, \citenamefont {Li},
  \citenamefont {Chen}, \citenamefont {Foster}, \citenamefont {Gokhale},
  \citenamefont {Dunkel},\ and\ \citenamefont {Fakhri}}]{Tan2021}\BibitemOpen
  \bibfield  {author} {\bibinfo {author} {\bibfnamefont {T.~H.}\ \bibnamefont
  {Tan}}, \bibinfo {author} {\bibfnamefont {A.}~\bibnamefont {Mietke}},
  \bibinfo {author} {\bibfnamefont {H.}~\bibnamefont {Higinbotham}}, \bibinfo
  {author} {\bibfnamefont {J.}~\bibnamefont {Li}}, \bibinfo {author}
  {\bibfnamefont {Y.}~\bibnamefont {Chen}}, \bibinfo {author} {\bibfnamefont
  {P.~J.}\ \bibnamefont {Foster}}, \bibinfo {author} {\bibfnamefont
  {S.}~\bibnamefont {Gokhale}}, \bibinfo {author} {\bibfnamefont
  {J.}~\bibnamefont {Dunkel}},\ and\ \bibinfo {author} {\bibfnamefont
  {N.}~\bibnamefont {Fakhri}},\ }\href@noop {} {\bibinfo {title} {Development
  drives dynamics of living chiral crystals}} (\bibinfo {year} {2021}),\
  \Eprint {https://arxiv.org/abs/2105.07507} {arXiv:2105.07507} \BibitemShut
  {NoStop}\bibitem [{\citenamefont {Hirschfelder}\ \emph {et~al.}(1964)\citenamefont
  {Hirschfelder}, \citenamefont {Curtiss},\ and\ \citenamefont
  {Bird}}]{Hirschfelder1964}\BibitemOpen
  \bibfield  {author} {\bibinfo {author} {\bibfnamefont {J.}~\bibnamefont
  {Hirschfelder}}, \bibinfo {author} {\bibfnamefont {C.}~\bibnamefont
  {Curtiss}},\ and\ \bibinfo {author} {\bibfnamefont {R.}~\bibnamefont
  {Bird}},\ }\href@noop {} {\emph {\bibinfo {title} {The Molecular Theory of
  Gases and Liquids}}},\ Molecular Theory of Gases and Liquids\ (\bibinfo
  {publisher} {Wiley},\ \bibinfo {year} {1964})\BibitemShut {NoStop}\bibitem [{\citenamefont {Waldmann}(1958{\natexlab{b}})}]{Waldmann1958}\BibitemOpen
  \bibfield  {author} {\bibinfo {author} {\bibfnamefont {L.}~\bibnamefont
  {Waldmann}},\ }\bibfield  {title} {\bibinfo {title} {Transporterscheinungen
  in gasen von mittlerem druck},\ }in\ \href
  {https://doi.org/10.1007/978-3-642-45892-7_4} {\emph {\bibinfo {booktitle}
  {Handbuch der Physik / Encyclopedia of Physics}}}\ (\bibinfo  {publisher}
  {Springer Berlin Heidelberg},\ \bibinfo {year} {1958})\ pp.\ \bibinfo {pages}
  {295--514}\BibitemShut {NoStop}\bibitem [{\citenamefont {Grad}(1958)}]{Grad1958}\BibitemOpen
  \bibfield  {author} {\bibinfo {author} {\bibfnamefont {H.}~\bibnamefont
  {Grad}},\ }\bibfield  {title} {\bibinfo {title} {Principles of the kinetic
  theory of gases},\ }in\ \href {https://doi.org/10.1007/978-3-642-45892-7_3}
  {\emph {\bibinfo {booktitle} {Handbuch der Physik / Encyclopedia of
  Physics}}}\ (\bibinfo  {publisher} {Springer Berlin Heidelberg},\ \bibinfo
  {year} {1958})\ pp.\ \bibinfo {pages} {205--294}\BibitemShut {NoStop}\bibitem [{\citenamefont {Harris}(2004)}]{Harris2004}\BibitemOpen
  \bibfield  {author} {\bibinfo {author} {\bibfnamefont {S.}~\bibnamefont
  {Harris}},\ }\href@noop {} {\emph {\bibinfo {title} {An Introduction to the
  Theory of the Boltzmann Equation}}},\ Dover books on physics\ (\bibinfo
  {publisher} {Dover Publications},\ \bibinfo {year} {2004})\BibitemShut
  {NoStop}\bibitem [{\citenamefont {Dorfman}\ \emph {et~al.}(2021)\citenamefont
  {Dorfman}, \citenamefont {van Beijeren},\ and\ \citenamefont
  {Kirkpatrick}}]{Dorfman2021}\BibitemOpen
  \bibfield  {author} {\bibinfo {author} {\bibfnamefont {J.}~\bibnamefont
  {Dorfman}}, \bibinfo {author} {\bibfnamefont {H.}~\bibnamefont {van
  Beijeren}},\ and\ \bibinfo {author} {\bibfnamefont {T.}~\bibnamefont
  {Kirkpatrick}},\ }\href@noop {} {\emph {\bibinfo {title} {Contemporary
  Kinetic Theory of Matter}}}\ (\bibinfo  {publisher} {Cambridge University
  Press},\ \bibinfo {year} {2021})\BibitemShut {NoStop}\bibitem [{\citenamefont {Reif}(2009)}]{Reif2009}\BibitemOpen
  \bibfield  {author} {\bibinfo {author} {\bibfnamefont {F.}~\bibnamefont
  {Reif}},\ }\href {https://books.google.fr/books?id=ObsbAAAAQBAJ} {\emph
  {\bibinfo {title} {Fundamentals of Statistical and Thermal Physics}}}\
  (\bibinfo  {publisher} {Waveland Press},\ \bibinfo {year} {2009})\BibitemShut
  {NoStop}\bibitem [{\citenamefont {Ashida}\ \emph {et~al.}(2020)\citenamefont {Ashida},
  \citenamefont {Gong},\ and\ \citenamefont {Ueda}}]{Ashida2020}\BibitemOpen
  \bibfield  {author} {\bibinfo {author} {\bibfnamefont {Y.}~\bibnamefont
  {Ashida}}, \bibinfo {author} {\bibfnamefont {Z.}~\bibnamefont {Gong}},\ and\
  \bibinfo {author} {\bibfnamefont {M.}~\bibnamefont {Ueda}},\ }\bibfield
  {title} {\bibinfo {title} {Non-hermitian physics},\ }\href
  {https://doi.org/10.1080/00018732.2021.1876991} {\bibfield  {journal}
  {\bibinfo  {journal} {Advances in Physics}\ }\textbf {\bibinfo {volume}
  {69}},\ \bibinfo {pages} {249–435} (\bibinfo {year} {2020})}\BibitemShut
  {NoStop}\bibitem [{\citenamefont {Lhuillier}\ and\ \citenamefont
  {Laloë}(1982)}]{Lhuillier1982}\BibitemOpen
  \bibfield  {author} {\bibinfo {author} {\bibfnamefont {C.}~\bibnamefont
  {Lhuillier}}\ and\ \bibinfo {author} {\bibfnamefont {F.}~\bibnamefont
  {Laloë}},\ }\bibfield  {title} {\bibinfo {title} {Transport properties in a
  spin polarized gas, i},\ }\href
  {https://doi.org/10.1051/jphys:01982004302019700} {\bibfield  {journal}
  {\bibinfo  {journal} {Journal de Physique}\ }\textbf {\bibinfo {volume}
  {43}},\ \bibinfo {pages} {197} (\bibinfo {year} {1982})}\BibitemShut
  {NoStop}\bibitem [{\citenamefont {Ah-Sam}\ \emph {et~al.}(1971)\citenamefont {Ah-Sam},
  \citenamefont {Jensen},\ and\ \citenamefont {Smith}}]{AhSam1971}\BibitemOpen
  \bibfield  {author} {\bibinfo {author} {\bibfnamefont {L.~E.~G.}\
  \bibnamefont {Ah-Sam}}, \bibinfo {author} {\bibfnamefont {H.~H.}\
  \bibnamefont {Jensen}},\ and\ \bibinfo {author} {\bibfnamefont
  {H.}~\bibnamefont {Smith}},\ }\bibfield  {title} {\bibinfo {title} {On the
  use of variational methods for solving boltzmann equations involving
  non-hermitian operators},\ }\href {https://doi.org/10.1007/bf01012184}
  {\bibfield  {journal} {\bibinfo  {journal} {Journal of Statistical Physics}\
  }\textbf {\bibinfo {volume} {3}},\ \bibinfo {pages} {17} (\bibinfo {year}
  {1971})}\BibitemShut {NoStop}\bibitem [{\citenamefont {Résibois}(1970)}]{Resibois1970}\BibitemOpen
  \bibfield  {author} {\bibinfo {author} {\bibfnamefont {P.}~\bibnamefont
  {Résibois}},\ }\bibfield  {title} {\bibinfo {title} {On linearized
  hydrodynamic modes in statistical physics},\ }\href
  {https://doi.org/10.1007/bf01009709} {\bibfield  {journal} {\bibinfo
  {journal} {Journal of Statistical Physics}\ }\textbf {\bibinfo {volume}
  {2}},\ \bibinfo {pages} {21} (\bibinfo {year} {1970})}\BibitemShut {NoStop}\bibitem [{\citenamefont {Zwanzig}(2001)}]{Zwanzig2001}\BibitemOpen
  \bibfield  {author} {\bibinfo {author} {\bibfnamefont {R.}~\bibnamefont
  {Zwanzig}},\ }\href@noop {} {\emph {\bibinfo {title} {Nonequilibrium
  Statistical Mechanics}}},\ \bibinfo {edition} {3rd}\ ed.\ (\bibinfo
  {publisher} {Oxford University Press},\ \bibinfo {year} {2001})\BibitemShut
  {NoStop}\bibitem [{\citenamefont {Mori}(1965)}]{Mori1965}\BibitemOpen
  \bibfield  {author} {\bibinfo {author} {\bibfnamefont {H.}~\bibnamefont
  {Mori}},\ }\bibfield  {title} {\bibinfo {title} {Transport, collective
  motion, and brownian motion},\ }\href {https://doi.org/10.1143/ptp.33.423}
  {\bibfield  {journal} {\bibinfo  {journal} {Prog. Theor. Exp. Phys.}\
  }\textbf {\bibinfo {volume} {33}},\ \bibinfo {pages} {423} (\bibinfo {year}
  {1965})}\BibitemShut {NoStop}\bibitem [{\citenamefont {Nakajima}(1958)}]{Nakajima1958}\BibitemOpen
  \bibfield  {author} {\bibinfo {author} {\bibfnamefont {S.}~\bibnamefont
  {Nakajima}},\ }\bibfield  {title} {\bibinfo {title} {On quantum theory of
  transport phenomena},\ }\href {https://doi.org/10.1143/ptp.20.948} {\bibfield
   {journal} {\bibinfo  {journal} {Prog. Theor. Exp. Phys.}\ }\textbf {\bibinfo
  {volume} {20}},\ \bibinfo {pages} {948} (\bibinfo {year} {1958})}\BibitemShut
  {NoStop}\bibitem [{\citenamefont {Zwanzig}(1960)}]{Zwanzig1960}\BibitemOpen
  \bibfield  {author} {\bibinfo {author} {\bibfnamefont {R.}~\bibnamefont
  {Zwanzig}},\ }\bibfield  {title} {\bibinfo {title} {Ensemble method in the
  theory of irreversibility},\ }\href {https://doi.org/10.1063/1.1731409}
  {\bibfield  {journal} {\bibinfo  {journal} {J. Chem. Phys.}\ }\textbf
  {\bibinfo {volume} {33}},\ \bibinfo {pages} {1338} (\bibinfo {year}
  {1960})}\BibitemShut {NoStop}\bibitem [{\citenamefont {Balescu}(1975)}]{Balescu1975}\BibitemOpen
  \bibfield  {author} {\bibinfo {author} {\bibfnamefont {R.}~\bibnamefont
  {Balescu}},\ }\href@noop {} {\emph {\bibinfo {title} {Equilibrium and
  Non-Equilibrium Statistical Mechanics}}},\ A Wiley interscience publication\
  (\bibinfo  {publisher} {Wiley},\ \bibinfo {year} {1975})\BibitemShut
  {NoStop}\bibitem [{\citenamefont {Ernst}(1970)}]{Ernst1970}\BibitemOpen
  \bibfield  {author} {\bibinfo {author} {\bibfnamefont {M.~H.}\ \bibnamefont
  {Ernst}},\ }\bibfield  {title} {\bibinfo {title} {The linearized boltzmann
  equation. navier-stokes and burnett transport coefficients},\ }\href
  {https://doi.org/10.1119/1.1976492} {\bibfield  {journal} {\bibinfo
  {journal} {American Journal of Physics}\ }\textbf {\bibinfo {volume} {38}},\
  \bibinfo {pages} {908} (\bibinfo {year} {1970})}\BibitemShut {NoStop}\bibitem [{\citenamefont {M.~Bixon}(1971)}]{Bixon1971}\BibitemOpen
  \bibfield  {author} {\bibinfo {author} {\bibfnamefont {K.~C.~M.}\
  \bibnamefont {M.~Bixon}, \bibfnamefont {J.~R.~Dorfman}},\ }\bibfield  {title}
  {\bibinfo {title} {General hydrodynamic equations from the linear boltzmann
  equation},\ }\href {https://doi.org/10.1063/1.1693563} {\bibfield  {journal}
  {\bibinfo  {journal} {The Physics of Fluids}\ }\textbf {\bibinfo {volume}
  {14}},\ \bibinfo {pages} {1049} (\bibinfo {year} {1971})}\BibitemShut
  {NoStop}\bibitem [{\citenamefont {Hauge}(1970)}]{Hauge1970}\BibitemOpen
  \bibfield  {author} {\bibinfo {author} {\bibfnamefont {E.~H.}\ \bibnamefont
  {Hauge}},\ }\bibfield  {title} {\bibinfo {title} {Exact and chapman-enskog
  solutions of the boltzmann equation for the lorentz model},\ }\href
  {https://doi.org/10.1063/1.1693050} {\bibfield  {journal} {\bibinfo
  {journal} {The Physics of Fluids}\ }\textbf {\bibinfo {volume} {13}},\
  \bibinfo {pages} {1201} (\bibinfo {year} {1970})}\BibitemShut {NoStop}\bibitem [{\citenamefont {Groot}\ and\ \citenamefont
  {Mazur}(1962)}]{DeGrootMazur}\BibitemOpen
  \bibfield  {author} {\bibinfo {author} {\bibfnamefont {S.~R.~D.}\
  \bibnamefont {Groot}}\ and\ \bibinfo {author} {\bibfnamefont
  {P.}~\bibnamefont {Mazur}},\ }\href@noop {} {\emph {\bibinfo {title}
  {{Non-Equilibrium Thermodynamics}}}}\ (\bibinfo  {publisher} {Dover
  Publications},\ \bibinfo {year} {1962})\BibitemShut {NoStop}\bibitem [{\citenamefont {de~Groot}(1973)}]{DeGroot1973}\BibitemOpen
  \bibfield  {author} {\bibinfo {author} {\bibfnamefont {S.~R.}\ \bibnamefont
  {de~Groot}},\ }\bibfield  {title} {\bibinfo {title} {The onsager relations;
  theoretical basis},\ }in\ \href {https://doi.org/10.1007/978-1-349-02235-9_9}
  {\emph {\bibinfo {booktitle} {Foundations of Continuum Thermodynamics}}},\
  \bibinfo {editor} {edited by\ \bibinfo {editor} {\bibfnamefont {J.~J.~D.}\
  \bibnamefont {Domingos}}, \bibinfo {editor} {\bibfnamefont {M.~N.~R.}\
  \bibnamefont {Nina}},\ and\ \bibinfo {editor} {\bibfnamefont {J.~H.}\
  \bibnamefont {Whitelaw}}}\ (\bibinfo  {publisher} {Macmillan Education UK},\
  \bibinfo {address} {London},\ \bibinfo {year} {1973})\ pp.\ \bibinfo {pages}
  {159--183}\BibitemShut {NoStop}\bibitem [{\citenamefont {Hahn}(2005)}]{Hahn2005}\BibitemOpen
  \bibfield  {author} {\bibinfo {author} {\bibfnamefont {T.}~\bibnamefont
  {Hahn}},\ }\bibfield  {title} {\bibinfo {title} {Cuba—a library for
  multidimensional numerical integration},\ }\href
  {https://doi.org/10.1016/j.cpc.2005.01.010} {\bibfield  {journal} {\bibinfo
  {journal} {Computer Physics Communications}\ }\textbf {\bibinfo {volume}
  {168}},\ \bibinfo {pages} {78–95} (\bibinfo {year} {2005})}\BibitemShut
  {NoStop}\bibitem [{\citenamefont {Bhatnagar}\ \emph {et~al.}(1954)\citenamefont
  {Bhatnagar}, \citenamefont {Gross},\ and\ \citenamefont
  {Krook}}]{Bhatnagar1954}\BibitemOpen
  \bibfield  {author} {\bibinfo {author} {\bibfnamefont {P.~L.}\ \bibnamefont
  {Bhatnagar}}, \bibinfo {author} {\bibfnamefont {E.~P.}\ \bibnamefont
  {Gross}},\ and\ \bibinfo {author} {\bibfnamefont {M.}~\bibnamefont {Krook}},\
  }\bibfield  {title} {\bibinfo {title} {A model for collision processes in
  gases. i. small amplitude processes in charged and neutral one-component
  systems},\ }\href {https://doi.org/10.1103/physrev.94.511} {\bibfield
  {journal} {\bibinfo  {journal} {Physical Review}\ }\textbf {\bibinfo {volume}
  {94}},\ \bibinfo {pages} {511} (\bibinfo {year} {1954})}\BibitemShut
  {NoStop}\bibitem [{\citenamefont {Hoffman}\ and\ \citenamefont
  {Dahler}(1969)}]{Hoffman1969}\BibitemOpen
  \bibfield  {author} {\bibinfo {author} {\bibfnamefont {D.~K.}\ \bibnamefont
  {Hoffman}}\ and\ \bibinfo {author} {\bibfnamefont {J.~S.}\ \bibnamefont
  {Dahler}},\ }\bibfield  {title} {\bibinfo {title} {The boltzmann equation for
  a polyatomic gas},\ }\href {https://doi.org/10.1007/bf01024129} {\bibfield
  {journal} {\bibinfo  {journal} {Journal of Statistical Physics}\ }\textbf
  {\bibinfo {volume} {1}},\ \bibinfo {pages} {521} (\bibinfo {year}
  {1969})}\BibitemShut {NoStop}\bibitem [{\citenamefont {Alexandre}\ and\ \citenamefont
  {Villani}(2001)}]{Alexandre2001}\BibitemOpen
  \bibfield  {author} {\bibinfo {author} {\bibfnamefont {R.}~\bibnamefont
  {Alexandre}}\ and\ \bibinfo {author} {\bibfnamefont {C.}~\bibnamefont
  {Villani}},\ }\bibfield  {title} {\bibinfo {title} {On the boltzmann equation
  for long-range interactions},\ }\href {https://doi.org/10.1002/cpa.10012}
  {\bibfield  {journal} {\bibinfo  {journal} {Communications on Pure and
  Applied Mathematics}\ }\textbf {\bibinfo {volume} {55}},\ \bibinfo {pages}
  {30} (\bibinfo {year} {2001})}\BibitemShut {NoStop}\bibitem [{\citenamefont {Campa}\ \emph {et~al.}(2009)\citenamefont {Campa},
  \citenamefont {Dauxois},\ and\ \citenamefont {Ruffo}}]{Campa2009}\BibitemOpen
  \bibfield  {author} {\bibinfo {author} {\bibfnamefont {A.}~\bibnamefont
  {Campa}}, \bibinfo {author} {\bibfnamefont {T.}~\bibnamefont {Dauxois}},\
  and\ \bibinfo {author} {\bibfnamefont {S.}~\bibnamefont {Ruffo}},\ }\bibfield
   {title} {\bibinfo {title} {Statistical mechanics and dynamics of solvable
  models with long-range interactions},\ }\href
  {https://doi.org/10.1016/j.physrep.2009.07.001} {\bibfield  {journal}
  {\bibinfo  {journal} {Physics Reports}\ }\textbf {\bibinfo {volume} {480}},\
  \bibinfo {pages} {57} (\bibinfo {year} {2009})}\BibitemShut {NoStop}\bibitem [{\citenamefont {Liao}\ \emph {et~al.}(2019)\citenamefont {Liao},
  \citenamefont {Han}, \citenamefont {Fruchart}, \citenamefont {Vitelli},\ and\
  \citenamefont {Vaikuntanathan}}]{Liao2019}\BibitemOpen
  \bibfield  {author} {\bibinfo {author} {\bibfnamefont {Z.}~\bibnamefont
  {Liao}}, \bibinfo {author} {\bibfnamefont {M.}~\bibnamefont {Han}}, \bibinfo
  {author} {\bibfnamefont {M.}~\bibnamefont {Fruchart}}, \bibinfo {author}
  {\bibfnamefont {V.}~\bibnamefont {Vitelli}},\ and\ \bibinfo {author}
  {\bibfnamefont {S.}~\bibnamefont {Vaikuntanathan}},\ }\bibfield  {title}
  {\bibinfo {title} {A mechanism for anomalous transport in chiral active
  liquids},\ }\href {https://doi.org/10.1063/1.5126962} {\bibfield  {journal}
  {\bibinfo  {journal} {The Journal of Chemical Physics}\ }\textbf {\bibinfo
  {volume} {151}},\ \bibinfo {pages} {194108} (\bibinfo {year}
  {2019})}\BibitemShut {NoStop}\bibitem [{Note1()}]{Note1}\BibitemOpen
  \bibinfo {note} {In Eq.~\protect \textup {\hbox {\mathsurround \z@ \protect
  \normalfont (\ignorespaces \ref {linearized_collision_operator}\unskip
  \@@italiccorr )}}, the bilinear function $\protect \mathscr {C}(f,g)$ is
  defined by Eq.~\protect \textup {\hbox {\mathsurround \z@ \protect
  \normalfont (\ignorespaces \ref {collision_integral}\unskip \@@italiccorr )}}
  in which $f_2$ is replaced by $g_2$ and $f_2'$ by $g_2'$.}\BibitemShut
  {Stop}\bibitem [{\citenamefont {Garzó}\ and\ \citenamefont
  {Dufty}(1999)}]{Garzo1999}\BibitemOpen
  \bibfield  {author} {\bibinfo {author} {\bibfnamefont {V.}~\bibnamefont
  {Garzó}}\ and\ \bibinfo {author} {\bibfnamefont {J.~W.}\ \bibnamefont
  {Dufty}},\ }\bibfield  {title} {\bibinfo {title} {Dense fluid transport for
  inelastic hard spheres},\ }\href {https://doi.org/10.1103/physreve.59.5895}
  {\bibfield  {journal} {\bibinfo  {journal} {Physical Review E}\ }\textbf
  {\bibinfo {volume} {59}},\ \bibinfo {pages} {5895–5911} (\bibinfo {year}
  {1999})}\BibitemShut {NoStop}\bibitem [{\citenamefont {Lutsko}(2005)}]{Lutsko2005}\BibitemOpen
  \bibfield  {author} {\bibinfo {author} {\bibfnamefont {J.~F.}\ \bibnamefont
  {Lutsko}},\ }\bibfield  {title} {\bibinfo {title} {Transport properties of
  dense dissipative hard-sphere fluids for arbitrary energy loss models},\
  }\href {https://doi.org/10.1103/physreve.72.021306} {\bibfield  {journal}
  {\bibinfo  {journal} {Physical Review E}\ }\textbf {\bibinfo {volume} {72}},\
  \bibinfo {pages} {021306} (\bibinfo {year} {2005})}\BibitemShut {NoStop}\bibitem [{\citenamefont {Garzó}(2019)}]{Garzo2019}\BibitemOpen
  \bibfield  {author} {\bibinfo {author} {\bibfnamefont {V.}~\bibnamefont
  {Garzó}},\ }\href {https://doi.org/10.1007/978-3-030-04444-2} {\emph
  {\bibinfo {title} {Granular Gaseous Flows}}}\ (\bibinfo  {publisher}
  {Springer International Publishing},\ \bibinfo {year} {2019})\BibitemShut
  {NoStop}\bibitem [{\citenamefont {Brilliantov}\ and\ \citenamefont
  {P{\"o}schel}(2004)}]{Brilliantov2004}\BibitemOpen
  \bibfield  {author} {\bibinfo {author} {\bibfnamefont {N.}~\bibnamefont
  {Brilliantov}}\ and\ \bibinfo {author} {\bibfnamefont {T.}~\bibnamefont
  {P{\"o}schel}},\ }\href@noop {} {\emph {\bibinfo {title} {Kinetic Theory of
  Granular Gases}}}\ (\bibinfo  {publisher} {Oxford University Press},\
  \bibinfo {year} {2004})\BibitemShut {NoStop}\bibitem [{\citenamefont {Bertin}\ \emph {et~al.}(2006)\citenamefont {Bertin},
  \citenamefont {Droz},\ and\ \citenamefont {Gr{\'{e}}goire}}]{Bertin2006}\BibitemOpen
  \bibfield  {author} {\bibinfo {author} {\bibfnamefont {E.}~\bibnamefont
  {Bertin}}, \bibinfo {author} {\bibfnamefont {M.}~\bibnamefont {Droz}},\ and\
  \bibinfo {author} {\bibfnamefont {G.}~\bibnamefont {Gr{\'{e}}goire}},\
  }\bibfield  {title} {\bibinfo {title} {Boltzmann and hydrodynamic description
  for self-propelled particles},\ }\href
  {https://doi.org/10.1103/physreve.74.022101} {\bibfield  {journal} {\bibinfo
  {journal} {Physical Review E}\ }\textbf {\bibinfo {volume} {74}},\ \bibinfo
  {pages} {022101} (\bibinfo {year} {2006})}\BibitemShut {NoStop}\bibitem [{\citenamefont {Bertin}\ \emph {et~al.}(2009)\citenamefont {Bertin},
  \citenamefont {Droz},\ and\ \citenamefont {Gr{\'{e}}goire}}]{Bertin2009}\BibitemOpen
  \bibfield  {author} {\bibinfo {author} {\bibfnamefont {E.}~\bibnamefont
  {Bertin}}, \bibinfo {author} {\bibfnamefont {M.}~\bibnamefont {Droz}},\ and\
  \bibinfo {author} {\bibfnamefont {G.}~\bibnamefont {Gr{\'{e}}goire}},\
  }\bibfield  {title} {\bibinfo {title} {Hydrodynamic equations for
  self-propelled particles: microscopic derivation and stability analysis},\
  }\href {https://doi.org/10.1088/1751-8113/42/44/445001} {\bibfield  {journal}
  {\bibinfo  {journal} {Journal of Physics A: Mathematical and Theoretical}\
  }\textbf {\bibinfo {volume} {42}},\ \bibinfo {pages} {445001} (\bibinfo
  {year} {2009})}\BibitemShut {NoStop}\bibitem [{\citenamefont {Ihle}(2011)}]{Ihle2011}\BibitemOpen
  \bibfield  {author} {\bibinfo {author} {\bibfnamefont {T.}~\bibnamefont
  {Ihle}},\ }\bibfield  {title} {\bibinfo {title} {Kinetic theory of flocking:
  Derivation of hydrodynamic equations},\ }\href
  {https://doi.org/10.1103/physreve.83.030901} {\bibfield  {journal} {\bibinfo
  {journal} {Physical Review E}\ }\textbf {\bibinfo {volume} {83}},\ \bibinfo
  {pages} {030901} (\bibinfo {year} {2011})}\BibitemShut {NoStop}\bibitem [{\citenamefont {Kremer}(2010)}]{Kremer2010}\BibitemOpen
  \bibfield  {author} {\bibinfo {author} {\bibfnamefont {G.~M.}\ \bibnamefont
  {Kremer}},\ }\href@noop {} {\emph {\bibinfo {title} {An introduction to the
  Boltzmann equation and transport processes in gases}}}\ (\bibinfo
  {publisher} {Springer Science \& Business Media},\ \bibinfo {year}
  {2010})\BibitemShut {NoStop}\bibitem [{Note2()}]{Note2}\BibitemOpen
  \bibinfo {note} {In general, some care is needed when working with non-normal
  operators, because they cannot be diagonalized in an orthonormal basis.
  Instead, one can use a biorthonormal system composed of right (usual)
  eigenvectors $\Phi _i$ and left eigenvectors $\protect \mathaccentV
  {tilde}07E{\Phi }_i$ that can be chosen so that $(\protect \mathaccentV
  {tilde}07E{\Phi }_i, \Phi _j) = \delta _{i j}$ provided that the matrix is
  diagonalizable. The standard functional calculus of normal matrices is then
  extended in a simple way (e.g., by replacing $|\Phi _i)(\Phi _i|$ with $|\Phi
  _i)(\protect \mathaccentV {tilde}07E{\Phi }_i|$ in Eq.~\protect \textup
  {\hbox {\mathsurround \z@ \protect \normalfont (\ignorespaces \ref
  {projector_kernel_L}\unskip \@@italiccorr )}}, see Refs.~\cite
  {Curtright2007,Brody2013,Riechers2018} and references therein. The (right)
  eigenvectors of $L$ in Eq.~\protect \textup {\hbox {\mathsurround \z@
  \protect \normalfont (\ignorespaces \ref
  {orthonormalized_basis_kernel}\unskip \@@italiccorr )}} are also the
  corresponding left eigenvectors ($\protect \mathaccentV {tilde}07E{\Phi }_i =
  \Phi _i$, $i=1,\protect \dots ,n_{\protect \text {cons}}$), so we do not need
  to distinguish them.}\BibitemShut {Stop}\bibitem [{\citenamefont {Zinn-Justin}(2021)}]{ZinnJustin2021}\BibitemOpen
  \bibfield  {author} {\bibinfo {author} {\bibfnamefont {J.}~\bibnamefont
  {Zinn-Justin}},\ }\href@noop {} {\emph {\bibinfo {title} {Quantum field
  theory and critical phenomena}}},\ Vol.\ \bibinfo {volume} {171}\ (\bibinfo
  {publisher} {Oxford university press},\ \bibinfo {year} {2021})\BibitemShut
  {NoStop}\bibitem [{\citenamefont {Kumar}(1966)}]{Kumar1966}\BibitemOpen
  \bibfield  {author} {\bibinfo {author} {\bibfnamefont {K.}~\bibnamefont
  {Kumar}},\ }\bibfield  {title} {\bibinfo {title} {Polynomial expansions in
  kinetic theory of gases},\ }\href
  {https://doi.org/10.1016/0003-4916(66)90280-6} {\bibfield  {journal}
  {\bibinfo  {journal} {Annals of Physics}\ }\textbf {\bibinfo {volume} {37}},\
  \bibinfo {pages} {113–141} (\bibinfo {year} {1966})}\BibitemShut {NoStop}\bibitem [{\citenamefont {Friedrich}(2015)}]{Friedrich2015}\BibitemOpen
  \bibfield  {author} {\bibinfo {author} {\bibfnamefont {H.}~\bibnamefont
  {Friedrich}},\ }\href@noop {} {\emph {\bibinfo {title} {Scattering Theory}}}\
  (\bibinfo  {publisher} {Springer Berlin Heidelberg},\ \bibinfo {year}
  {2015})\BibitemShut {NoStop}\bibitem [{\citenamefont {Monceau}\ \emph {et~al.}(2002)\citenamefont
  {Monceau}, \citenamefont {Szydlo},\ and\ \citenamefont
  {Valent}}]{Monceau2002}\BibitemOpen
  \bibfield  {author} {\bibinfo {author} {\bibfnamefont {P.}~\bibnamefont
  {Monceau}}, \bibinfo {author} {\bibfnamefont {T.}~\bibnamefont {Szydlo}},\
  and\ \bibinfo {author} {\bibfnamefont {G.}~\bibnamefont {Valent}},\
  }\bibfield  {title} {\bibinfo {title} {Screened coulomb scattering versus
  thomson scattering},\ }\href {https://doi.org/10.1088/0143-0807/24/1/307}
  {\bibfield  {journal} {\bibinfo  {journal} {European Journal of Physics}\
  }\textbf {\bibinfo {volume} {24}},\ \bibinfo {pages} {47–65} (\bibinfo
  {year} {2002})}\BibitemShut {NoStop}\bibitem [{\citenamefont {Everhart}\ \emph {et~al.}(1955)\citenamefont
  {Everhart}, \citenamefont {Stone},\ and\ \citenamefont
  {Carbone}}]{Everhart1955}\BibitemOpen
  \bibfield  {author} {\bibinfo {author} {\bibfnamefont {E.}~\bibnamefont
  {Everhart}}, \bibinfo {author} {\bibfnamefont {G.}~\bibnamefont {Stone}},\
  and\ \bibinfo {author} {\bibfnamefont {R.~J.}\ \bibnamefont {Carbone}},\
  }\bibfield  {title} {\bibinfo {title} {Classical calculation of differential
  cross section for scattering from a coulomb potential with exponential
  screening},\ }\href {https://doi.org/10.1103/physrev.99.1287} {\bibfield
  {journal} {\bibinfo  {journal} {Physical Review}\ }\textbf {\bibinfo {volume}
  {99}},\ \bibinfo {pages} {1287–1290} (\bibinfo {year} {1955})}\BibitemShut
  {NoStop}\bibitem [{\citenamefont {Lane}\ and\ \citenamefont
  {Everhart}(1960)}]{Lane1960}\BibitemOpen
  \bibfield  {author} {\bibinfo {author} {\bibfnamefont {G.~H.}\ \bibnamefont
  {Lane}}\ and\ \bibinfo {author} {\bibfnamefont {E.}~\bibnamefont
  {Everhart}},\ }\bibfield  {title} {\bibinfo {title} {Calculations of total
  cross sections for scattering from coulomb potentials with exponential
  screening},\ }\href {https://doi.org/10.1103/physrev.117.920} {\bibfield
  {journal} {\bibinfo  {journal} {Physical Review}\ }\textbf {\bibinfo {volume}
  {117}},\ \bibinfo {pages} {920–924} (\bibinfo {year} {1960})}\BibitemShut
  {NoStop}\bibitem [{\citenamefont {Scheibner}\ \emph {et~al.}(2020)\citenamefont
  {Scheibner}, \citenamefont {Souslov}, \citenamefont {Banerjee}, \citenamefont
  {Surówka}, \citenamefont {Irvine},\ and\ \citenamefont
  {Vitelli}}]{Scheibner2020}\BibitemOpen
  \bibfield  {author} {\bibinfo {author} {\bibfnamefont {C.}~\bibnamefont
  {Scheibner}}, \bibinfo {author} {\bibfnamefont {A.}~\bibnamefont {Souslov}},
  \bibinfo {author} {\bibfnamefont {D.}~\bibnamefont {Banerjee}}, \bibinfo
  {author} {\bibfnamefont {P.}~\bibnamefont {Surówka}}, \bibinfo {author}
  {\bibfnamefont {W.~T.~M.}\ \bibnamefont {Irvine}},\ and\ \bibinfo {author}
  {\bibfnamefont {V.}~\bibnamefont {Vitelli}},\ }\bibfield  {title} {\bibinfo
  {title} {Odd elasticity},\ }\href {https://doi.org/10.1038/s41567-020-0795-y}
  {\bibfield  {journal} {\bibinfo  {journal} {Nature Physics}\ }\textbf
  {\bibinfo {volume} {16}},\ \bibinfo {pages} {475–480} (\bibinfo {year}
  {2020})}\BibitemShut {NoStop}\bibitem [{\citenamefont {Sengers}(1969)}]{Sengers1969}\BibitemOpen
  \bibfield  {author} {\bibinfo {author} {\bibfnamefont {J.~V.}\ \bibnamefont
  {Sengers}},\ }\href {https://apps.dtic.mil/sti/citations/AD0684179} {\emph
  {\bibinfo {title} {Triple collision effects in the thermal conductivity and
  viscosity of moderately dense gases}}},\ \bibinfo {type} {Tech. Rep.}\
  (\bibinfo  {institution} {National Bureau Of Standards Gaithersburg Md},\
  \bibinfo {year} {1969})\BibitemShut {NoStop}\bibitem [{\citenamefont {Gass}(1971)}]{Gass1971}\BibitemOpen
  \bibfield  {author} {\bibinfo {author} {\bibfnamefont {D.~M.}\ \bibnamefont
  {Gass}},\ }\bibfield  {title} {\bibinfo {title} {Enskog theory for a rigid
  disk fluid},\ }\href {https://doi.org/10.1063/1.1675115} {\bibfield
  {journal} {\bibinfo  {journal} {The Journal of Chemical Physics}\ }\textbf
  {\bibinfo {volume} {54}},\ \bibinfo {pages} {1898–1902} (\bibinfo {year}
  {1971})}\BibitemShut {NoStop}\bibitem [{\citenamefont {Goldhirsch}(2010)}]{Goldhirsch2010}\BibitemOpen
  \bibfield  {author} {\bibinfo {author} {\bibfnamefont {I.}~\bibnamefont
  {Goldhirsch}},\ }\bibfield  {title} {\bibinfo {title} {Stress, stress
  asymmetry and couple stress: from discrete particles to continuous fields},\
  }\href {https://doi.org/10.1007/s10035-010-0181-z} {\bibfield  {journal}
  {\bibinfo  {journal} {Granular Matter}\ }\textbf {\bibinfo {volume} {12}},\
  \bibinfo {pages} {239} (\bibinfo {year} {2010})}\BibitemShut {NoStop}\bibitem [{\citenamefont {Irving}\ and\ \citenamefont
  {Kirkwood}(1950)}]{Irving1950}\BibitemOpen
  \bibfield  {author} {\bibinfo {author} {\bibfnamefont {J.~H.}\ \bibnamefont
  {Irving}}\ and\ \bibinfo {author} {\bibfnamefont {J.~G.}\ \bibnamefont
  {Kirkwood}},\ }\bibfield  {title} {\bibinfo {title} {The statistical
  mechanical theory of transport processes. {IV}. the equations of
  hydrodynamics},\ }\href {https://doi.org/10.1063/1.1747782} {\bibfield
  {journal} {\bibinfo  {journal} {The Journal of Chemical Physics}\ }\textbf
  {\bibinfo {volume} {18}},\ \bibinfo {pages} {817} (\bibinfo {year}
  {1950})}\BibitemShut {NoStop}\bibitem [{\citenamefont {Schofield}\ and\ \citenamefont
  {Henderson}(1982)}]{Schofield1982}\BibitemOpen
  \bibfield  {author} {\bibinfo {author} {\bibfnamefont {P.}~\bibnamefont
  {Schofield}}\ and\ \bibinfo {author} {\bibfnamefont {J.~R.}\ \bibnamefont
  {Henderson}},\ }\bibfield  {title} {\bibinfo {title} {Statistical mechanics
  of inhomogeneous fluids},\ }\href {https://doi.org/10.1098/rspa.1982.0015}
  {\bibfield  {journal} {\bibinfo  {journal} {Proceedings of the Royal Society
  of London. A. Mathematical and Physical Sciences}\ }\textbf {\bibinfo
  {volume} {379}},\ \bibinfo {pages} {231} (\bibinfo {year}
  {1982})}\BibitemShut {NoStop}\bibitem [{\citenamefont {Wajnryb}\ \emph {et~al.}(1995)\citenamefont
  {Wajnryb}, \citenamefont {Altenberger},\ and\ \citenamefont
  {Dahler}}]{Wajnryb1995}\BibitemOpen
  \bibfield  {author} {\bibinfo {author} {\bibfnamefont {E.}~\bibnamefont
  {Wajnryb}}, \bibinfo {author} {\bibfnamefont {A.~R.}\ \bibnamefont
  {Altenberger}},\ and\ \bibinfo {author} {\bibfnamefont {J.~S.}\ \bibnamefont
  {Dahler}},\ }\bibfield  {title} {\bibinfo {title} {Uniqueness of the
  microscopic stress tensor},\ }\href {https://doi.org/10.1063/1.469942}
  {\bibfield  {journal} {\bibinfo  {journal} {The Journal of Chemical Physics}\
  }\textbf {\bibinfo {volume} {103}},\ \bibinfo {pages} {9782} (\bibinfo {year}
  {1995})}\BibitemShut {NoStop}\bibitem [{\citenamefont {Chen}\ and\ \citenamefont {Diaz}(2018)}]{Chen2018}\BibitemOpen
  \bibfield  {author} {\bibinfo {author} {\bibfnamefont {Y.}~\bibnamefont
  {Chen}}\ and\ \bibinfo {author} {\bibfnamefont {A.}~\bibnamefont {Diaz}},\
  }\bibfield  {title} {\bibinfo {title} {Physical foundation and consistent
  formulation of atomic-level fluxes in transport processes},\ }\href
  {https://doi.org/10.1103/physreve.98.052113} {\bibfield  {journal} {\bibinfo
  {journal} {Physical Review E}\ }\textbf {\bibinfo {volume} {98}},\ \bibinfo
  {pages} {052113} (\bibinfo {year} {2018})}\BibitemShut {NoStop}\bibitem [{\citenamefont {Admal}\ and\ \citenamefont
  {Tadmor}(2011)}]{Admal2011}\BibitemOpen
  \bibfield  {author} {\bibinfo {author} {\bibfnamefont {N.~C.}\ \bibnamefont
  {Admal}}\ and\ \bibinfo {author} {\bibfnamefont {E.~B.}\ \bibnamefont
  {Tadmor}},\ }\bibfield  {title} {\bibinfo {title} {Stress and heat flux for
  arbitrary multibody potentials: A unified framework},\ }\href
  {https://doi.org/10.1063/1.3582905} {\bibfield  {journal} {\bibinfo
  {journal} {The Journal of Chemical Physics}\ }\textbf {\bibinfo {volume}
  {134}},\ \bibinfo {pages} {184106} (\bibinfo {year} {2011})}\BibitemShut
  {NoStop}\bibitem [{\citenamefont {Torres-Sánchez}\ \emph {et~al.}(2016)\citenamefont
  {Torres-Sánchez}, \citenamefont {Vanegas},\ and\ \citenamefont
  {Arroyo}}]{TorresSanchez2016}\BibitemOpen
  \bibfield  {author} {\bibinfo {author} {\bibfnamefont {A.}~\bibnamefont
  {Torres-Sánchez}}, \bibinfo {author} {\bibfnamefont {J.~M.}\ \bibnamefont
  {Vanegas}},\ and\ \bibinfo {author} {\bibfnamefont {M.}~\bibnamefont
  {Arroyo}},\ }\bibfield  {title} {\bibinfo {title} {Geometric derivation of
  the microscopic stress: A covariant central force decomposition},\ }\href
  {https://doi.org/10.1016/j.jmps.2016.03.006} {\bibfield  {journal} {\bibinfo
  {journal} {Journal of the Mechanics and Physics of Solids}\ }\textbf
  {\bibinfo {volume} {93}},\ \bibinfo {pages} {224–239} (\bibinfo {year}
  {2016})}\BibitemShut {NoStop}\bibitem [{\citenamefont {Shi}\ \emph {et~al.}(2021)\citenamefont {Shi},
  \citenamefont {Santiso},\ and\ \citenamefont {Gubbins}}]{Shi2021}\BibitemOpen
  \bibfield  {author} {\bibinfo {author} {\bibfnamefont {K.}~\bibnamefont
  {Shi}}, \bibinfo {author} {\bibfnamefont {E.~E.}\ \bibnamefont {Santiso}},\
  and\ \bibinfo {author} {\bibfnamefont {K.~E.}\ \bibnamefont {Gubbins}},\
  }\bibfield  {title} {\bibinfo {title} {Can we define a unique microscopic
  pressure in inhomogeneous fluids?},\ }\href
  {https://doi.org/10.1063/5.0044487} {\bibfield  {journal} {\bibinfo
  {journal} {The Journal of Chemical Physics}\ }\textbf {\bibinfo {volume}
  {154}},\ \bibinfo {pages} {084502} (\bibinfo {year} {2021})}\BibitemShut
  {NoStop}\bibitem [{Note3()}]{Note3}\BibitemOpen
  \bibinfo {note} {Note that there are also troubles in defining a unique
  stress-energy tensor in field theory~\cite
  {Tichy1998,Forger2004}.}\BibitemShut {Stop}\bibitem [{\citenamefont {Rao}\ and\ \citenamefont {Bradlyn}(2020)}]{Rao2020}\BibitemOpen
  \bibfield  {author} {\bibinfo {author} {\bibfnamefont {P.}~\bibnamefont
  {Rao}}\ and\ \bibinfo {author} {\bibfnamefont {B.}~\bibnamefont {Bradlyn}},\
  }\bibfield  {title} {\bibinfo {title} {Hall viscosity in quantum systems with
  discrete symmetry: Point group and lattice anisotropy},\ }\href
  {https://doi.org/10.1103/physrevx.10.021005} {\bibfield  {journal} {\bibinfo
  {journal} {Physical Review X}\ }\textbf {\bibinfo {volume} {10}},\ \bibinfo
  {pages} {021005} (\bibinfo {year} {2020})}\BibitemShut {NoStop}\bibitem [{\citenamefont {Rao}\ and\ \citenamefont {Bradlyn}(2021)}]{Rao2021}\BibitemOpen
  \bibfield  {author} {\bibinfo {author} {\bibfnamefont {P.}~\bibnamefont
  {Rao}}\ and\ \bibinfo {author} {\bibfnamefont {B.}~\bibnamefont {Bradlyn}},\
  }\href@noop {} {\bibinfo {title} {Resolving hall and dissipative viscosity
  ambiguities via boundary effects}} (\bibinfo {year} {2021}),\ \Eprint
  {https://arxiv.org/abs/2112.04545} {arXiv:2112.04545} \BibitemShut {NoStop}\bibitem [{\citenamefont {Hargus}\ \emph {et~al.}(2021)\citenamefont {Hargus},
  \citenamefont {Epstein},\ and\ \citenamefont {Mandadapu}}]{Hargus2021}\BibitemOpen
  \bibfield  {author} {\bibinfo {author} {\bibfnamefont {C.}~\bibnamefont
  {Hargus}}, \bibinfo {author} {\bibfnamefont {J.~M.}\ \bibnamefont
  {Epstein}},\ and\ \bibinfo {author} {\bibfnamefont {K.~K.}\ \bibnamefont
  {Mandadapu}},\ }\bibfield  {title} {\bibinfo {title} {Odd diffusivity of
  chiral random motion},\ }\href
  {https://doi.org/10.1103/physrevlett.127.178001} {\bibfield  {journal}
  {\bibinfo  {journal} {Physical Review Letters}\ }\textbf {\bibinfo {volume}
  {127}},\ \bibinfo {pages} {178001} (\bibinfo {year} {2021})}\BibitemShut
  {NoStop}\bibitem [{\citenamefont {Kunihiro}\ and\ \citenamefont
  {Tsumura}(2006)}]{Kunihiro2006}\BibitemOpen
  \bibfield  {author} {\bibinfo {author} {\bibfnamefont {T.}~\bibnamefont
  {Kunihiro}}\ and\ \bibinfo {author} {\bibfnamefont {K.}~\bibnamefont
  {Tsumura}},\ }\bibfield  {title} {\bibinfo {title} {Application of the
  renormalization-group method to the reduction of transport equations},\
  }\href {https://doi.org/10.1088/0305-4470/39/25/s20} {\bibfield  {journal}
  {\bibinfo  {journal} {Journal of Physics A: Mathematical and General}\
  }\textbf {\bibinfo {volume} {39}},\ \bibinfo {pages} {8089} (\bibinfo {year}
  {2006})}\BibitemShut {NoStop}\bibitem [{\citenamefont {Hatta}\ and\ \citenamefont
  {Kunihiro}(2002)}]{Hatta2002}\BibitemOpen
  \bibfield  {author} {\bibinfo {author} {\bibfnamefont {Y.}~\bibnamefont
  {Hatta}}\ and\ \bibinfo {author} {\bibfnamefont {T.}~\bibnamefont
  {Kunihiro}},\ }\bibfield  {title} {\bibinfo {title} {Renormalization group
  method applied to kinetic equations: Roles of initial values and time},\
  }\href {https://doi.org/10.1006/aphy.2002.6234} {\bibfield  {journal}
  {\bibinfo  {journal} {Annals of Physics}\ }\textbf {\bibinfo {volume}
  {298}},\ \bibinfo {pages} {24} (\bibinfo {year} {2002})}\BibitemShut
  {NoStop}\bibitem [{\citenamefont {Groetsch}(1993)}]{Groetsch1993}\BibitemOpen
  \bibfield  {author} {\bibinfo {author} {\bibfnamefont {C.~W.}\ \bibnamefont
  {Groetsch}},\ }\href {https://doi.org/10.1007/978-3-322-99202-4} {\emph
  {\bibinfo {title} {Inverse Problems in the Mathematical Sciences}}}\
  (\bibinfo  {publisher} {Vieweg+Teubner Verlag},\ \bibinfo {year}
  {1993})\BibitemShut {NoStop}\bibitem [{Note4()}]{Note4}\BibitemOpen
  \bibinfo {note} {If the particles were indistinguishable (quantum), then we
  would also have \begin {equation*} \protect \tmspace -\thinmuskip
  {.1667em}\protect \tmspace -\thinmuskip {.1667em}W(\protect \bm
  {c_1},\protect \bm {c_2} | \protect \bm {c_1'},\protect \bm {c_2'}) =
  W(\protect \bm {c_2},\protect \bm {c_1} | \protect \bm {c_1'},\protect \bm
  {c_2'}) = W(\protect \bm {c_1},\protect \bm {c_2} | \protect \bm
  {c_2'},\protect \bm {c_1'}) \end {equation*} but we will not consider this
  case here.}\BibitemShut {Stop}\bibitem [{Note5()}]{Note5}\BibitemOpen
  \bibinfo {note} {Set $c = c_1$ so that $W$ does depend only on the three
  variables $a_1 = c_2 - c_1$, $a_2 = \protect \bm {c_1'} - c_1$ and $a_3 =
  \protect \bm {c_2'} - c_1$. We can apply a linear transformation to get three
  other independent variables $b_1 = a_1$, $b_2 = a_3 - a_2$ and $b_3 = a_1 -
  a_2 - a_3$ which gives the result.}\BibitemShut {Stop}\bibitem [{\citenamefont {Procesi}(2006)}]{Procesi2006}\BibitemOpen
  \bibfield  {author} {\bibinfo {author} {\bibfnamefont {C.}~\bibnamefont
  {Procesi}},\ }\href@noop {} {\emph {\bibinfo {title} {Lie Groups: An Approach
  through Invariants and Representations}}},\ Universitext\ (\bibinfo
  {publisher} {Springer New York},\ \bibinfo {year} {2006})\BibitemShut
  {NoStop}\bibitem [{Note6()}]{Note6}\BibitemOpen
  \bibinfo {note} {The relation with Cartesian coordinates $(g_x,g_y,g_z)$ is
  $g_x = g \protect \qopname \relax o{sin}\chi \protect \tmspace +\thinmuskip
  {.1667em} \protect \qopname \relax o{cos}\epsilon $, $g_y = g \protect
  \qopname \relax o{sin}\chi \protect \tmspace +\thinmuskip {.1667em} \protect
  \qopname \relax o{sin}\epsilon $, $g_z = g \protect \qopname \relax
  o{cos}\chi $.}\BibitemShut {Stop}\bibitem [{Note7()}]{Note7}\BibitemOpen
  \bibinfo {note} {This has to be done separately for the two parts of the
  integral.}\BibitemShut {Stop}\bibitem [{\citenamefont {Curtright}\ and\ \citenamefont
  {Mezincescu}(2007)}]{Curtright2007}\BibitemOpen
  \bibfield  {author} {\bibinfo {author} {\bibfnamefont {T.}~\bibnamefont
  {Curtright}}\ and\ \bibinfo {author} {\bibfnamefont {L.}~\bibnamefont
  {Mezincescu}},\ }\bibfield  {title} {\bibinfo {title} {Biorthogonal quantum
  systems},\ }\href {https://doi.org/10.1063/1.2196243} {\bibfield  {journal}
  {\bibinfo  {journal} {Journal of Mathematical Physics}\ }\textbf {\bibinfo
  {volume} {48}},\ \bibinfo {pages} {092106} (\bibinfo {year}
  {2007})}\BibitemShut {NoStop}\bibitem [{\citenamefont {Brody}(2013)}]{Brody2013}\BibitemOpen
  \bibfield  {author} {\bibinfo {author} {\bibfnamefont {D.~C.}\ \bibnamefont
  {Brody}},\ }\bibfield  {title} {\bibinfo {title} {Biorthogonal quantum
  mechanics},\ }\href {https://doi.org/10.1088/1751-8113/47/3/035305}
  {\bibfield  {journal} {\bibinfo  {journal} {Journal of Physics A:
  Mathematical and Theoretical}\ }\textbf {\bibinfo {volume} {47}},\ \bibinfo
  {pages} {035305} (\bibinfo {year} {2013})}\BibitemShut {NoStop}\bibitem [{\citenamefont {Riechers}\ and\ \citenamefont
  {Crutchfield}(2018)}]{Riechers2018}\BibitemOpen
  \bibfield  {author} {\bibinfo {author} {\bibfnamefont {P.~M.}\ \bibnamefont
  {Riechers}}\ and\ \bibinfo {author} {\bibfnamefont {J.~P.}\ \bibnamefont
  {Crutchfield}},\ }\bibfield  {title} {\bibinfo {title} {Beyond the spectral
  theorem: Spectrally decomposing arbitrary functions of nondiagonalizable
  operators},\ }\href {https://doi.org/10.1063/1.5040705} {\bibfield  {journal}
  {\bibinfo  {journal} {{AIP} Advances}\ }\textbf {\bibinfo {volume} {8}},\
  \bibinfo {pages} {065305} (\bibinfo {year} {2018})}\BibitemShut {NoStop}\bibitem [{\citenamefont {Tichy}\ and\ \citenamefont
  {Flanagan}(1998)}]{Tichy1998}\BibitemOpen
  \bibfield  {author} {\bibinfo {author} {\bibfnamefont {W.}~\bibnamefont
  {Tichy}}\ and\ \bibinfo {author} {\bibfnamefont {{\'{E}}.~{\'{E}}.}\
  \bibnamefont {Flanagan}},\ }\bibfield  {title} {\bibinfo {title} {How unique
  is the expected stress-energy tensor of a massive scalar field?},\ }\href
  {https://doi.org/10.1103/physrevd.58.124007} {\bibfield  {journal} {\bibinfo
  {journal} {Physical Review D}\ }\textbf {\bibinfo {volume} {58}},\ \bibinfo
  {pages} {124007} (\bibinfo {year} {1998})}\BibitemShut {NoStop}\bibitem [{\citenamefont {Forger}\ and\ \citenamefont
  {Römer}(2004)}]{Forger2004}\BibitemOpen
  \bibfield  {author} {\bibinfo {author} {\bibfnamefont {M.}~\bibnamefont
  {Forger}}\ and\ \bibinfo {author} {\bibfnamefont {H.}~\bibnamefont
  {Römer}},\ }\bibfield  {title} {\bibinfo {title} {Currents and the
  energy-momentum tensor in classical field theory: a fresh look at an old
  problem},\ }\href {https://doi.org/10.1016/j.aop.2003.08.011} {\bibfield
  {journal} {\bibinfo  {journal} {Annals of Physics}\ }\textbf {\bibinfo
  {volume} {309}},\ \bibinfo {pages} {306} (\bibinfo {year}
  {2004})}\BibitemShut {NoStop}\end{thebibliography}
\end{document}